\documentclass[aps,pra,twocolumn,a4paper,superscriptaddress,longbibliography,nofootinbib]{revtex4-1}
\usepackage[english]{babel}
\usepackage{graphicx}
\usepackage{amssymb}
\usepackage{amsmath}
\usepackage{color}
\usepackage{comment}
\usepackage[caption = false]{subfig}
\usepackage[colorlinks]{hyperref}

\newcommand{\bk}{{\bf k}}
\newcommand{\br}{{\bf r}}
\newcommand{\ok}{\omega_k}
\newcommand{\kk}{\kappa_k}
\newcommand{\beq}{\begin{equation}}
\newcommand{\eeq}{\end{equation}}
\newcommand{\ave}[1]{\langle #1 \rangle}
\newcommand{\comm}[2]{\left[#1,#2\right]}

\newcommand{\diff}{{\mathrm d}}

\newcommand{\pie}[1]{{\color{black} #1}}

\newcommand{\mac}[1]{{\color{black} #1}}

\begin{document}

\title{Quantum Brownian Motion with Inhomogeneous Damping and Diffusion}

\author{Pietro Massignan}
        \affiliation{ICFO -- Institut de Ci\`{e}ncies Fot\`{o}niques, Av.\ C.F.\ Gauss, 3, E-08860 Castelldefels, Spain}
\author{Aniello Lampo}
        \affiliation{ICFO -- Institut de Ci\`{e}ncies Fot\`{o}niques, Av.\ C.F.\ Gauss, 3, E-08860 Castelldefels, Spain}
\author{Jan Wehr}
    \affiliation{Department of Mathematics, University of Arizona, Tucson, AZ 85721-0089, USA}
\author{Maciej Lewenstein}  \email{maciej.lewenstein@icfo.es}
    \affiliation{ICFO -- Institut de Ci\`{e}ncies Fot\`{o}niques, Av.\ C.F.\ Gauss, 3, E-08860 Castelldefels, Spain}
        \affiliation{ICREA -- Instituci{\'o} Catalana de Recerca i Estudis Avan\c{c}ats, Lluis Companys 23, E-08010 Barcelona, Spain}


\begin{abstract}
We analyze the microscopic model of quantum Brownian motion, describing a Brownian particle interacting with a bosonic bath through a coupling which is linear in the creation and annihilation operators of the bath, but may be a nonlinear function of the position of the  particle.
Physically, this corresponds to a configuration in which damping and diffusion are spatially inhomogeneous. 
We derive systematically the quantum master equation for the Brownian particle in the Born-Markov approximation and we discuss the appearance of novel terms, for various polynomials forms of the coupling.
We discuss the cases of linear and quadratic coupling in great detail and we derive, using Wigner function techniques, the stationary solutions of the master equation for a Brownian particle in a harmonic trapping potential.
We predict quite generally Gaussian stationary states, and we compute the aspect ratio and the spread of the distributions.
In particular, we find that these solutions may be squeezed (super-localized) with respect to the position of the Brownian particle.
We analyze various restrictions to the validity of our theory posed by non-Markovian effects and by the Heisenberg principle.
We further study the dynamical stability of the system, by applying a Gaussian approximation to the time dependent Wigner function, and we compute the decoherence rates of coherent quantum superpositions in position space.
Finally, we propose a possible experimental realization of the physics discussed here, by considering an impurity particle embedded in a degenerate quantum gas.
\end{abstract}

\date{\today}

\maketitle

\section{Introduction}

The theory of quantum Brownian motion (QBM) has been a subject of
studies for decades and belongs  nowadays to a standard textbook
material
\cite{GardinerBook,BreuerBook,SchlosshauerBook,WaldenfelsBook,Weiss}.
Nevertheless, there are some aspects of QBM that have not been, in
our opinion,  explored completely in the literature, and that is
what motivates our paper.

First, one should note that the vast majority of the work on QBM
is devoted to microscopic models in which the coupling of the
Brownian particle to the bosonic bath
 is linear in both in  bath creation and annihilation operators, and in position (or momentum) of the particle.
The case when such coupling is non-linear in either the bath or the system operators has been hardly studied -- unique exceptions to our knowledge provide the old works of Landauer \cite{Landauer1957}, 
who studied nonlinearity in bath operators, and Dykman and Krivoglaz \cite{Dykman1975}, Hu, Paz and Zhang \cite{Hu1993}, Brun \cite{Brun1993}, and Banerjee and Ghosh \cite{Banerjee2003}, who considered both cases.
  Physically, the case of a coupling which deviates from linearity in the system coordinates corresponds to a
 situation,  in which damping and diffusion are spatially inhomogeneous.
Obviously, such nonlinearity might have both classical and quantum consequences,
and as such deserves careful analysis.

Second, this type of inhomogeneity has been recently intensively studied in the context of classical Brownian motion (CBM) and other classical diffusive systems. 
 In particular, explicit formulae were derived for {\it noise-induced drifts} in the small-mass (Smoluchowski-Kramers \cite{Smoluchowski,Kramers}) and other limits \cite{Hottovy2012,Hottovy2014,McDaniel2014}. 
 Noise-induced drifts have been shown to appear in a general class of diffusive systems, including systems with time delay and systems driven by colored noise.  Applications include Brownian 
motion in diffusion gradient \cite{Volpe2010,Brettschneider2011}, noisy electrical circuits \cite{Pesce2012} and thermophoresis \cite{Hottovy2012a}.  In the first two cases the theoretical predictions have been
demonstrated to be in an excellent agreement with the experiment.
 Diffusion in
inhomogeneous and disordered media is presently
one of the fastest developing subjects in the theory of random walks  and CBM
\cite{Haus1987,Havlin87,BG1990,Klafter2011}, and finds vast applications in various areas
of science. There is a considerable interest in the studies of
various forms of anomalous diffusion and non-ergodicity
\cite{Metzler2004,Klafter2011,Hoefling2013,Metzler2014}, based
either on the theory of heavy-tailed continuous-time random walk
(CTRW)~\cite{MontrollWeiss65,ScherMontroll75}
 or on models characterized by a diffusivity
(\textit{i.e.,} a diffusion coefficient) that is inhomogeneous in
time~\cite{Saxton1993,*Saxton1997} or
space~\cite{Hottovy2012,Cherstvy2013a,*Cherstvy2014}.
Particularly impressive is the recent progress in single particle
imaging, for instance  in biophotonics (cf.
\cite{Tolic04,Golding06,Jeon2011,Weigel2011,Kusumi2012,Bakker2012,Cisse2013}
and references therein),  where the single particle trajectories
of, say, a receptor on a cell membrane can be traced. It is presently investigated how random walk and CBM models with inhomogeneous diffusion may be employed in the description of such phenomena
\cite{Massignan2014,Manzo2014}.

The examples mentioned above are strictly classical,
but the recent unprecedented progress in control, detection and
manipulation of ultracold atoms and ions \cite{Lewenstein2012} are giving us the possibility to perform similar kind of experiments (e.g., single particle tracking to monitor the real time dynamics
of given atoms) in the quantum regime \cite{Krinner2013}.
 Note that such
experiments were unthinkable, say, 20 years ago (see the
corresponding paragraphs about difficulties to observe QBM in Ref.\
\cite{GardinerBook}). Note also that ultracold set-ups will
naturally involve spatial inhomogeneities, due to the necessary
presence of trapping potentials and eventual stray fields. This is in fact the third
motivation of this paper: to formulate and study theory of QBM at low temperatures, and in the presence of spatially inhomogeneous damping and diffusion.

An immediate application of our theory concerns dilute impurities
embedded in an ultracold degenerate quantum gas.  Such problem has been intensively studied in the recent years in the context of polaron physics in strongly-interacting Fermi gases \cite{Schirotzek2009,Kohstall2012,Koschorreck2012,MassignanPolRev2014,Lan2014,Levinsen2014} and Bose gases \cite{Cote2002,Massignan2005,Cucchietti2006,Palzer2009,Catani2012,Rath2013,Fukuhara2013,Shashi2014,Grusdt2014a,Grusdt2014b}. Obviously, there is a vast amount of literature on the polaron problem, or more generally on electron-phonon interactions, in solid state systems (cf.\ \cite{Devreese2009,Alexandrov2009}). The theory of polarons has been also a subject of intensive studies in mathematical physics \cite{Lieb1958,Lieb1997,Frank2010, %Frank2012,Frank2013,
Anapolitanos2013}.
 In analogy to the
studies of classical stochastic processes \cite{Sancho1982,Freidlin2004,Cerrai2011,Freidlin2011,Shi2012,Hottovy2012,Hottovy2014}, the present work opens also the possibility of employing ultracold atoms to study the quantum Smoluchowski-Kramers limit of a very light Brownian particle, or correspondingly an over-damped Brownian motion.

Since the present paper revisits some of the handbook material,
part of the presentation reproduces known and well established
results. We include it here in order to make our further
argumentations and derivations self-contained. We start in Section \ref{sec:CLM}
 by presenting the microscopic model of QBM, known as
Caldeira-Leggett model \cite{Caldeira1983a,Caldeira1983b}, and we
derive the quantum master equation (QME) \pie{in the Born-Markov approximation}, following up to a certain point the \pie{standard weak-coupling treatment, i.e., by means of perturbation theory to second order in the bath-system coupling constant} \cite{GardinerBook}. 
\pie{
The resulting equation is systematic in the sense of Born expansion, and it takes a certain part of non-Markovian effects into account.
In its most common form, the QME is derived in the limit when the characteristic energy of the system  (i.e. the
Brownian particle) $\hbar\Omega$ is much smaller than the
cutoff energy $\hbar\Lambda$, and the latter is much smaller than the thermal energy $k_BT$ of the
bath -- in the following, we will refer to this regime as {\it the} Caldeira-Leggett limit.   However, in this paper we are interested to the regime where $k_BT$ becomes comparable to  $\hbar\Omega$. Section \ref{sec:QMElinear} deals with
the case of linear coupling, i.e. spatially homogeneous damping
and diffusion; although this case has been widely elaborated
previously \cite{BreuerBook,SchlosshauerBook}, we discuss carefully the non-standard modifications of
the generalized master equation appearing in the uncommon limit
$\hbar\Lambda\gg\hbar\Omega \sim k_BT$.} In Section \ref{sec:QMEquadratic} we present our results concerning \pie{a coupling which is quadratic in the position of the test particle, which yields a quadratic dependence of the damping and diffusion coefficients on the position of the Brownian particle}, and extract the corresponding position-space decoherence time.
 The stationary solutions of the QMEs and their properties for linear and quadratic coupling are discussed  in Section \ref{sec:WignerStationary}. We predict quite generally Gaussian
 stationary states which are asymmetric in the position and momentum variables, and that may be classified in terms of an effective cooling or heating, depending on whether the associated distribution is more or less spread out than the one of its quantum thermal Gibbs-Boltzmann counterpart.  The aspect ratio of the distribution can be so extreme, that the system may even become squeezed (super-localized) with respect to the position of the Brownian particle. The squeezing effect can be understood in terms of renormalization, or {\it Lamb-shift}, of the system frequency $\hbar\Omega$ due to virtual excitations by the non-resonant bath modes.
We analyze various restrictions on the validity of our theory imposed by Heisenberg principle and non-Markovian effects, and we stress the role and possibility of observation of quantum effects.
 In Section \ref{sec:nearEqDyn} we discuss
the near-equilibrium dynamics of the system by computing moments of the time dependent Wigner function.
 We conclude and present the outlook Section \ref{sec:conclusions}, where we comment on the experimental realization of the models described by our theory. There, we also comment on challenges of investigating the so-called Smoluchowski-Kramers limit using a quantum analog of classical homogenization theory (cf.\ \cite{Hottovy2014}). A number of more intricate issues are addressed in the Appendices.  
 In Appendix \ref{sec:genericCoupling} we discuss the most general QME for the
 case of \pie{a generical polynomial coupling in the system's position}, and Appendices \ref{sec:LaplaceTransforms} and \ref{sec:trigIdentities} deal with a rather technical point, the detailed calculation of the coefficients appearing in the generic QME.
In Appendix \ref{sec:asymptoticTable} we summarize the asymptotic behavior of the QME coefficients for the cases of a linear and quadratic coupling.
  In Appendix \ref{sec:highTlimitWithLeadingQuantumCorrections} we analyse a (somehow oversimplified) high-temperature limit of the QME, which includes however the leading quantum corrections.
Finally, Appendix \ref{sec:impurityInATrappedGas}
discusses challenges related to application of our theory to the
problem of an impurity in an ultracold quantum gas.

It is important to stress to which extent our paper goes beyond the results of the previously published work \cite{Hu1993,Brun1993,Banerjee2003}. In particular, the in-depth study of Hu, Paz and Zhang contains the derivations of time-dependent (Redfield) and time-independent master equation   
for the case of general system--bath coupling: linear or nonlinear in bath and system operators. 
%\pie{ These weak-coupling approaches, similarly to ours, are based on a systematic perturbation theory to second order in the bath-system coupling constant.}
In our paper we consider the case where the coupling is linear in bath operators and polynomial in the system position $x$, but in contrast to the earlier works we provide: i) a careful analysis of the parameter dependences of coefficients entering into the time independent master equation, obtained as a long time limit of the Redfield equation, and the various limits of the resulting equation; ii) a derivation and a detailed discussion of the properties of the stationary solutions, analyzing in particular their dynamical stability, classifying solutions in terms of an effective cooling or heating, and highlighting the presence of  quantum squeezed regimes; iii) a discussion of QBM in the context of physics of ultracold degenerate gases; in particular, the present paper provides a solid theoretical basis for further studies of quantum Brownian motion of an impurity atoms inside a Bose Einstein condensate.

\mac{Before turning to the body of the paper, let us clarify the use of the notions {\it nonlinear} or {\it nonlinear coupling} we will use in the following.  In this paper we limit ourselves to bosonic baths with effective bath Hamiltonians which are quadratic in bath creation and annihilation operators (i.e., the bath dynamics is linear here), although more general cases with bath Hamiltonians containing also quartic terms are also of physical interest (i.e., the bath dynamics is itself nonlinear there; cf.\ \cite{Shashi2014}). We limit ourselves also to system Hamiltonians quadratic in the system's position and momentum (i.e. the system dynamics is also linear here). Obviously, non-harmonic trapping or spatially periodic or even random potentials (leading to a nonlinear dynamics  for the system itself) are also of physical interest. 
When we refer in this paper to {\it linear coupling} between the system and the bath, we consider the situation when the coupling between the bath and the system has the form:  a bath operator linear in the bath creation and annihilation operator, times a system operator linear in the system's particle position, or momentum, or both. Note that if in such situation both the Hamiltonian of the bath and of the system are quadratic, the whole model corresponds to a system of coupled harmonic oscillators (i.e., the system plus bath  dynamics as a whole is linear), and the model is exactly solvable by standard methods (via, e.g., matrix diagonalization, or Fourier or Laplace techniques). We refer to concrete examples below. 
When instead we consider a {\it nonlinear coupling} between the system and the bath, we refer to the situation when the coupling between the bath and the system has the form:  a bath operator linear in the bath creation and annihilation operator times a system operator non-linear in the system's particle position, or momentum or both.    Of course, one may also consider a {\it nonlinear coupling} between a bath and a system which are themselves nonlinear (cf. \cite{Hu1993}).
}
 \pie{To conclude, the nonlinearity in the coupling (which is the main focus of the present work) should not be confused with the nonlinear order of the Born expansion in the coupling constant  $\kappa_k$. Models involving expansions to quartic and higher order in $\kappa_k$ have been discussed in detail elsewhere (see, e.g., \cite{Fleming2012} and references therein).}

\section{Caldeira-Leggett model and quantum master equation}
\label{sec:CLM}
\subsection{Caldeira-Leggett model}

The Caldeira-Leggett model (CLM) is one of many models describing
a (Brownian)  particle interacting with a bosonic
bath (for the models discussing interaction of an atom, or
ensemble of atoms, with a minimally coupled photon bath, see for
instance \cite{Rzazewski1976,Lewenstein1980}). Despite its simplicity, the CLM gained popularity in condensed matter physics due to its very general nature, and its ability to describe
quantum dissipation in the Ohmic, super- and sub-Ohmic limits.
  The model is defined by the  Hamiltonian
\beq
H=H_{\rm S}+ H_{\rm B} + H_{\rm I} ,
\eeq
where the system, bath and interaction Hamiltonians are respectively
 \beq
H_{\rm S}= H_{\rm sys} + V_c(x) = \frac{p^2}{2m} + V(x) + V_c(x),
\eeq
\beq
  H_{\rm B}= \sum_k \left(\frac{p_k^2}{2m_k} + \frac{m_k\omega_k^2 x_k^2}{2}\right)-E_0=\sum_k\hbar\ok g_k^{\dagger}g_k,
\eeq
\beq
  H_{\rm I}= -f(x)B= -\sum_k\kappa_k x_kf(x).
 \eeq 
 In the
above expressions $p$ is the  particle momentum, $m$ its mass,
 $V(x)$ the trapping potential, and the so called counter-term 
 \beq
  V_c(x)=
\sum_k\frac{\kappa_k^2}{2m_k\omega_k^2}f(x)^2, 
\eeq
will be needed in the following to remove unphysical divergent renormalizations of the trapping potential arising from the coupling to the bath. 
 The bath
bosons  have masses $m_k$ and  frequencies $\omega_k$, and their
momenta and position are denoted by $p_k$ and $x_k$, respectively.
Alternatively, we describe them with the help of annihilation and
creation operators, $g_k$ and $ g_k^{\dagger}$. From the bath Hamiltonian,
 we have removed the constant zero-point energy $E_0$.
The parameters
describing the coupling of the bath modes to the system are
denoted by $\kappa_k$. We consider here the case of a very general
position-dependent coupling, described by a function $f(x)$ of the particle position $x$. To keep notation as close as possible to the usual case of linear coupling, we take $f(x)$
to have dimension of length, i.e. we write it as $f(x)=a\tilde
f(x/a)$, with $\tilde f(x)$ being dimensionless, and $a$ denoting
a typical length scale on which  $f$ varies.  We will restrict our discussion in the
following to the one dimensional (1D) case, but generalizations to 2D or 3D are
straightforward.

Since in order to derive the QME we are going to use systematic
Born-Markov approximation, it is useful to identify orders of
magnitude of various terms with respect to the coupling. To this
aim we rewrite the Hamiltonian as \beq H=H_0+ H_1 + H_2 , \eeq
where $H_0=H_{\rm sys}+H_B$, 
$H_1=H_{\rm I}$, and $H_2= V_c(x)$.
 The Hamiltonian of the system+bath ensemble may be written as
\pie{
\beq
 H=H_{\rm sys} + V_c(x)+\sum_k\hbar\ok
g_k^{\dagger}g_k+\sqrt{\frac{\hbar\kk^2}{2m_k\ok}}\left(g_k+g_k^{\dagger}\right)f(x)
\eeq
}

The next steps consist in going to the interaction picture with
respect to $H_0$, writing the Liouville-von Neumann equation for
the total density matrix $\rho(t)$ of the system and bath \beq
\dot\rho(t) =-\frac{i}{\hbar}[H_{\rm I}(t),\rho], \label{LvN}\eeq
where $H_{\rm I}(t)$ is the interaction Hamiltonian in the
interaction picture. We solve the above equation  formally 
\beq
\rho(t)= \rho(0)-\frac{i}{\hbar}\int_0^t \diff s\, [H_{\rm I}(s), \rho(s)],
\eeq
 and insert the solution into \eqref{LvN}. Taking trace over the bath and assuming\footnote{This assumption is typically verified as a consequence of the symmetries; the initial state $\rho(0)$
 is often taken to be an even function of the bath modes' position and momentum operators,
 while the interaction Hamiltonian is an odd function. In any case, this condition may always be satisfied by suitably redefining the Hamiltonian.} that
 ${\rm tr_B}[H_{\rm I}(t),\rho(0)]=0$ we obtain
\beq \dot\rho_S(t) = -\frac{1}{\hbar^2}\int_0^t \diff s{\rm
Tr}_B [H_{\rm I}(t), [H_{\rm I}(s),\rho(s)]].\eeq

\subsection{Born-Markov approximation}

We assume also that initially the system and the bath were
uncorrelated, i.e. the initial density matrix was a simple tensor
product, $\rho_S(0)\otimes\rho_B(0)$. The first approximation that
we apply is the Born approximation: in a weak
coupling regime, we expect that the influence of the system on the bath
is negligible, and the state of the total system remains
approximately uncorrelated for all times,
 \beq \rho(t)\simeq \rho_S(t)\otimes\rho_B(0).\eeq 
 Under this standard approximation (cf.\ \cite{BreuerBook})  we obtain first 
\beq
\label{preRedfield}
\dot\rho_S(t) = -\frac{1}{\hbar^2}\int_0^t \diff s\, {\rm Tr}_B [H_{\rm I}(t), [H_{\rm I}(s),\rho_S(s)\otimes\rho_B(0)]].
\eeq

The next steps require more specific assumptions about the initial
state of the bath, and an explicit form of the bath parameters
$\kappa_k$, $m_k$, and $\omega_k$. We will assume a thermal state
of the bath, described by the density matrix
 \beq \rho_B(0)=\frac{\exp(-H_B/k_BT)}{{\rm Tr}_B [\exp(-H_B/k_BT)]}. \eeq
We will also introduce the spectral density, which contains all the relevant properties of the bath; it determines the analytical form of the coefficients of the QME, and therefore characterizes the main dissipation and decoherence processes occurring in the central system. The spectral density may be generally defined as \beq J(\omega)=\sum_k
\frac{\kappa_k^2}{2m_k\omega_k}\delta(\omega-\omega_k). \eeq
As we will see in the following, see Eq.\ \eqref{spectralDensityOhmLD}, the spectral density will be more specifically defined to be proportional to a damping constant $\gamma$, and will necessarily contain a UV momentum cut-off $\Lambda$.
As such, when taking the trace over the bath degrees of freedom, the bath correlation functions arising in Eq.\ \eqref{preRedfield} will decay on a fast characteristic time scale $\tau_B$, determined by $1/\Lambda$ and $\hbar/k_BT$. On the other hand, in presence of a weak coupling between the bath and the system, the interaction-picture system density matrix $\rho_S(t)$ will evolve only on a much slower time scale, set by $1/\gamma$.
\mac{We may thus safely shift
 $\rho_S(s)$ to $\rho_S(t)$ in Eq.\ \eqref{preRedfield}. Note that even if the
system exhibits at long times algebraic decay of the form
$C/t^\nu$ with some exponent $\nu$ of order 1, the shift from $s$
to $t$ for $|t-s|< \tau_B$ causes a relative error of order
$\nu\tau_B/t$, which is negligible at long times. Traditionally, this approximation is termed in the handbooks \cite{GardinerBook,BreuerBook,SchlosshauerBook,WaldenfelsBook,Weiss} Markov approximation, although the considered quantum stochastic process 
strictly speaking is non-Markovian. In the following, we will see this approximation actually is part of the systematic second order (weak-coupling) expansion in the coupling constant:   the shift from $s$
to $t$ will be accompanied by the corresponding zero-th order time translation (i.e., time translation for the system decoupled from the bath).} 

In this way we derive the, so called, Redfield equation
\cite{Redfield1957,Blum1981} for the reduced density matrix of the
systems. Going back to the Schr\"odinger picture, the latter reads
%\beq
\begin{multline}
\label{Redfield}
\dot\rho_S(t) =-\frac{i}{\hbar}[H_S,\rho_S]\\
-\frac{1}{\hbar^2}\int_0^t \diff\tau\, {\rm Tr}_B\,[H_{\rm
I}(0), [H_{\rm I}(-\tau),\rho_S(t)\otimes\rho_B(0)]].  
\end{multline}
%\eeq
\mac{Note that the Redfield  equation is in fact the systematically derived Master Equation in the second order of the expansion in coupling constant, also known as  weak-coupling master equation. It is explicitly time dependent, and as such it is capable of describing non-Markovian effects. This is discussed in some detail for the case of linear couplings in \cite{SchlosshauerBook}, and for the general nonlinear couplings in \cite{Hu1993}. To be more specific, the Redfield equation has a well defined long time limit, expected to describe correctly the long time
behavior, but it also describes the short time non-Markovian effects. In many cases these non-Markovian effects reduce to ``initial slips", i.e., rapid changes of the system density matrix before entering into the long time regime, and an ``adiabatic drag", when the systems during the slow, long time phase of the evolution ``drags" the bath  with itself  (cf.\ \cite{Haake-ZfP,MLFH,Reibold1}, and references therein)  
  The final  step of what is traditionally called Markov approximation consists in
taking the long time limit, extending the $\tau$ integration to infinity, obtaining in this
way a QME which is local in time,
\begin{align}
\label{BMQME}
 \dot\rho_S(t)
&=-\frac{i}{\hbar}[H_S,\rho_S]\\
 &-\frac{1}{\hbar^2}\int_0^\infty
\diff\tau\, {\rm Tr}_B [H_{\rm I}(0), [H_{\rm
I}(-\tau),\rho_S(t)\otimes\rho_B(0)]].
\nonumber
\end{align}
}
\pie{
We will refer to the latter as to the {\it Born-Markov Quantum Master Equation} (BM-QME), making explicit reference to the two key approximations performed to derive it.

At this point, two important issues are worth discussing. First of all, it should be noted that the BM-QME is not, strictly speaking, Markovian.}
\mac{A quantum stochastic process is a quantum Markov process only if it can be regarded as a quantum Langevin process with purely white noise, and if it is described by a time independent master equation of Lindblad form. If we treated our model solving, say, the Heisenberg equations of motion, we would see that: i) for the case of {\it linear coupling}, the quantum noise is additive, but by no means white: its correlations indeed would typically have finite (exponential decay) correlation time, and even small algebraic long time tails; ii) for the case of {\it non-linear coupling}, the quantum noise not only is coloured, but is multiplicative, which of course complicates the treatment even more.

\pie{
%In the following, we will study the As we will see below, the regime that we are studying appears to be the systematic second order in coupling constant, or in other words the weak-coupling master equation, as described by the time dependent Redfield equation. We will, however, study the long-time stationary, time-independent limit of this equation.
%, and treat it as if it was a Markovian QME. 
Moreover, the long-time limit taken, in Eq.~\eqref{BMQME}, during the Markov approximation, loosely speaking erases
% introduce above consists in taking this long time limit, and in some sense erasing 
the memory about the initial state.}  Such procedure is a frequent ``abuse" in quantum optics, 
%since it does not assure automatically the Lindblad form of the Markovian QME, 
leading to unphysical solutions in certain regimes of parameters \pie{(typically at very low temperatures)}.  Obviously, all these problems can be avoided in the case of {\it linear coupling} and harmonic trapping potential, when the exact QME is used \cite{Grabert1988,Fleming2011}. Unfortunately, the exact solutions are not known in the case of {\it nonlinear coupling}. In the latter case, the Markov-Born approximations quite naturally seem to be the method of choice to obtain novel results. Trying to improve them using a canonical perturbation theory \`a la Ref.~\cite{Fleming2012} is a very interesting challenge, which however goes beyond the scope of the present paper.} \pie{In order to obtain a fair comparison, we will compare the approximate solutions of the non-linear case with the results obtained for the linear coupling using the same Markov-Born approximation.}

\subsection{Caldeira-Leggett QME}

Following the notation of Ref.\ \cite{SchlosshauerBook},
we can express the environment self-correlation function as
$\mathcal{C}(\tau)=\langle B
(0)B(-\tau)\rangle_{B}=\nu(\tau)-i\eta(\tau)$, with the noise
kernel
\beq
 \nu(\tau)=\int_0^\infty \diff \omega\,J(\omega)\coth\left(\frac{\hbar\omega}{2 k_B T}\right)\cos(\omega \tau)
\eeq
 and the dissipation kernel
\beq
\eta(\tau)=\int_0^\infty  \diff \omega\,J(\omega)\sin(\omega \tau).
\eeq
 The master equation for the system density matrix $\rho(t)$ (we will skip in the following the subscript $S$) takes then the form
\begin{multline}
\dot\rho(t)=-\frac{i}{\hbar}[ H_S,\rho(t)]\\
-\frac{1}{\hbar^{2}}\int_0^\infty\diff
\tau\,\Big(\nu(\tau) [ f(x(0)),[f(x(-\tau)),\rho(t)]]  \\
- i
\eta(\tau) [ f(x(0)),\{f(x(-\tau)),\rho(t)\}]\Big)
\end{multline}

In the case, when the coupling is linear in the position of the particle and the environment is
Ohmic, Caldeira and Leggett in \cite{Caldeira1983a,Caldeira1983b}
showed that the reduced density matrix $\rho$ of a harmonic
oscillator of mass $m$ and frequency $\Omega$, obeys in the high temperature  limit $k_BT/\hbar\gg \Lambda\gg\Omega$
the following master equation (CLME):
\beq
\dot\rho=-\frac{i}{\hbar}[H_{\rm sys},\rho]
-\frac{i\gamma}{2\hbar}[x,\{p,\rho\}] -\frac{m\gamma k_B
T}{\hbar^{2}}[x,[x,\rho]].
\label{CaldeiraLeggettMasterEq}
\eeq
Here $\gamma=\eta/m$ is the characteristic damping rate of the
oscillator, and $\eta$ is the friction coefficient.

Similarly, as shown first in Ref.\ \cite{Hu1993}, in
the case of non-linear coupling $f(x)$ to the Ohmic environment
and $T\rightarrow\infty$, the evolution of the system is described
by a generalization of the Caldeira-Leggett Master Equation, which may be written as 
\begin{multline}
\dot\rho=-\frac{i}{\hbar}[H_{\rm sys},\rho]-\frac{i\gamma
m}{2\hbar}[f(x),\{\dot{f}(x),\rho\}] \\
 -\frac{m \gamma k_B
T}{\hbar^{2}}[f(x),[f(x),\rho]].
\label{NonLinearCaldeiraLeggettMasterEq}
\end{multline}
We have
introduced/defined here the ``dot" operator \beq \dot
f(x)=-\frac{i}{\hbar}[f(x),H_{\rm sys}]=\frac{pf'(x)+f'(x)p}{2m}.
\eeq

\mac{Our aim in the following  is to derive the generalizations of Eqs.\
\eqref{CaldeiraLeggettMasterEq} and \eqref{NonLinearCaldeiraLeggettMasterEq} to the situation in which $k_BT\simeq \hbar \Omega$, and the largest energy scale in the problem is the cutoff energy $\hbar \Lambda$. For the case of {\it linear coupling} the resulting master equation was derived in certain limits in Refs. \cite{BreuerBook,SchlosshauerBook}. One should stress once more that in the case 
when the Hamiltonians of both the bath and the system are quadratic, the whole model corresponds to a system of coupled harmonic oscillators (i.e. the system plus bath  dynamics as a whole is linear), and the model is exactly soluble by standard methods such as matrix diagonalization, Fourier or Laplace techniques (cf.\ \cite{Ullersma1,Ullersma2,Ullersma3,Ullersma4,Rzazewski1976,Reibold}). In such situation, the exact time-dependent Master Equation
(i.e., an exact analogue of the Redfield equation)  can be worked out rigorously for various spectral functions (cf.\ \cite{Grabert1988,Fleming2011}). The case of {\it nonlinear coupling} (in the system's variables, as discussed in the last paragraph of the Introduction), to our knowledge, has been only discussed in Refs.\ \cite{Hu1993,Brun1993,Banerjee2003}; however, explicit analytic expressions for the coefficients entering the master equation have generally not been discussed there.}

\section{BM-QME with linear coupling}
\label{sec:QMElinear}

In this work we will focus on the simplest case of a %perfectly
 harmonic potential $V(x)=m\Omega^2x^2/2$, where $\Omega$ denotes the oscillator frequency, and $m\Omega^2$ is the corresponding spring constant.
%In general a system might undergo itself non-linear dynamics due to the presence of anharmonicities in the potential, but we will leave the studies of the more general situations to further publications.
In the interaction picture, the position operator obeys $x(\tau)=x\cos(\Omega\tau)+(p/m\Omega)\sin(\Omega\tau)$, and the master equation may be written in the simple form
\begin{multline}
\dot\rho(t)=-\frac{i}{\hbar}\left[\hat H_S+C_x x^2,\rho(t)\right]
-\frac{iC_p}{\hbar m\Omega} [x,\{p,\rho(t)\}]
\\
-\frac{D_x}{\hbar}[x,[x,\rho(t)]]
-\frac{D_p}{\hbar m\Omega}[x,[p,\rho(t)]],
\label{master_equation_linear}
\end{multline}
where the frequency renormalization of the harmonic potential, the momentum damping coefficient, the normal diffusion coefficient, and the anomalous diffusion coefficient are respectively proportional to
\begin{align}
C_x&=-\int_0^\infty\diff \tau\,\eta(\tau)\cos(\Omega\tau)\\\nonumber
C_p&=\int_0^\infty\diff \tau\,\eta(\tau)\sin(\Omega\tau)\\\nonumber
D_x&=\int_0^\infty\diff \tau\,\nu(\tau)\cos(\Omega\tau)\\\nonumber
D_p&=-\int_0^\infty \diff \tau\,\nu(\tau)\sin(\Omega\tau)
\end{align}

For definiteness, in this paper we focus on the case where the spectral density is Ohmic (i.e., it is linear in $\omega$) and has a Lorentz-Drude (LD) cutoff,
\beq
J(\omega)=\frac{m\gamma}{\pi}\omega\frac{\Lambda^2}{\omega^2+\Lambda^2}.
\label{spectralDensityOhmLD}
\eeq
%We have checked that t
The specific choice of cutoff function yields minor quantitative changes to the QME coefficients, but as physically expected, it does not alter their asymptotic behaviour. Exploiting the Matsubara representation
\beq
\coth\left(\frac{\hbar\omega}{2k_BT}\right)=\frac{2k_B T}{\hbar\omega}\sum_{n=-\infty}^{\infty}\frac{1}{1+(\nu_n/\omega)^2}
\eeq
with bosonic frequencies $\nu_n=2\pi n k_B T/\hbar$, the noise and dissipation kernels may be evaluated analytically with the help of the Cauchy's residue theorem,
\begin{align}
\nu(\tau)=& \frac{m k_B T\gamma\Lambda^2}{\hbar} \sum_{n=-\infty}^\infty\frac{\Lambda e^{-\Lambda|\tau|}-|\nu_n| e^{-|\nu_n\tau|}}{\Lambda^2-\nu_n^2}\label{nu_Drude},\\
\eta(\tau)=&\frac{m\gamma\Lambda^2}{2}{\rm sign}(\tau)e^{-\Lambda|\tau|},
\label{eta_Drude}
\end{align}
and the coefficients can be evaluated as follows:
\begin{align}
C_x(\Omega)&=
%-\frac{1}{4i}\int_{-\infty}^\infty\diff \tau\,\int_{-\infty}^\infty\diff \omega\,J(\omega){\rm sign}(\tau)e^{i(\omega+\Omega)\tau}=
-\frac{m\gamma}{2\pi}\int_{-\infty}^\infty\diff \omega\,{\cal P}\left(\frac1{\omega+\Omega}\frac{\omega\Lambda^2}{\omega^2+\Lambda^2}\right)\\\nonumber
&=-\frac{m\gamma\Lambda^3}{2(\Omega^2+\Lambda^2)}\\\nonumber
C_p(\Omega)&
%=-\frac1{4}\int_{-\infty}^\infty\diff \tau\,\int_{-\infty}^\infty\diff \omega\,J(\omega)e^{i(\omega+\Omega)\tau}
=\frac{m\gamma\Omega\Lambda^2}{2(\Omega^2+\Lambda^2)}\\\nonumber
D_x(\Omega)&
%=\frac14\int_{-\infty}^\infty\diff \tau\,\int_{-\infty}^\infty\diff \omega\,J(\omega)\coth\left(\frac{\omega}{2 k_B T}\right)e^{i(\omega+\Omega)\tau}
=\frac{m\gamma\Omega\Lambda^2}{2(\Omega^2+\Lambda^2)}\coth\left(\frac{\hbar\Omega}{2 k_B T}\right),
\end{align}
In the first equation above we have used the identity
$2i\int_{0}^\infty \diff \tau\,
\sin(\omega\tau)=\int_{-\infty}^\infty \diff \tau\, {\rm sign}
(\tau)e^{i\omega\tau}=2i{\cal P}\left(\frac 1\omega\right)$, where
$\cal P$ denotes the principal value of the integral. 

\begin{figure}[t!]
\begin{center}
\includegraphics[width=\columnwidth]{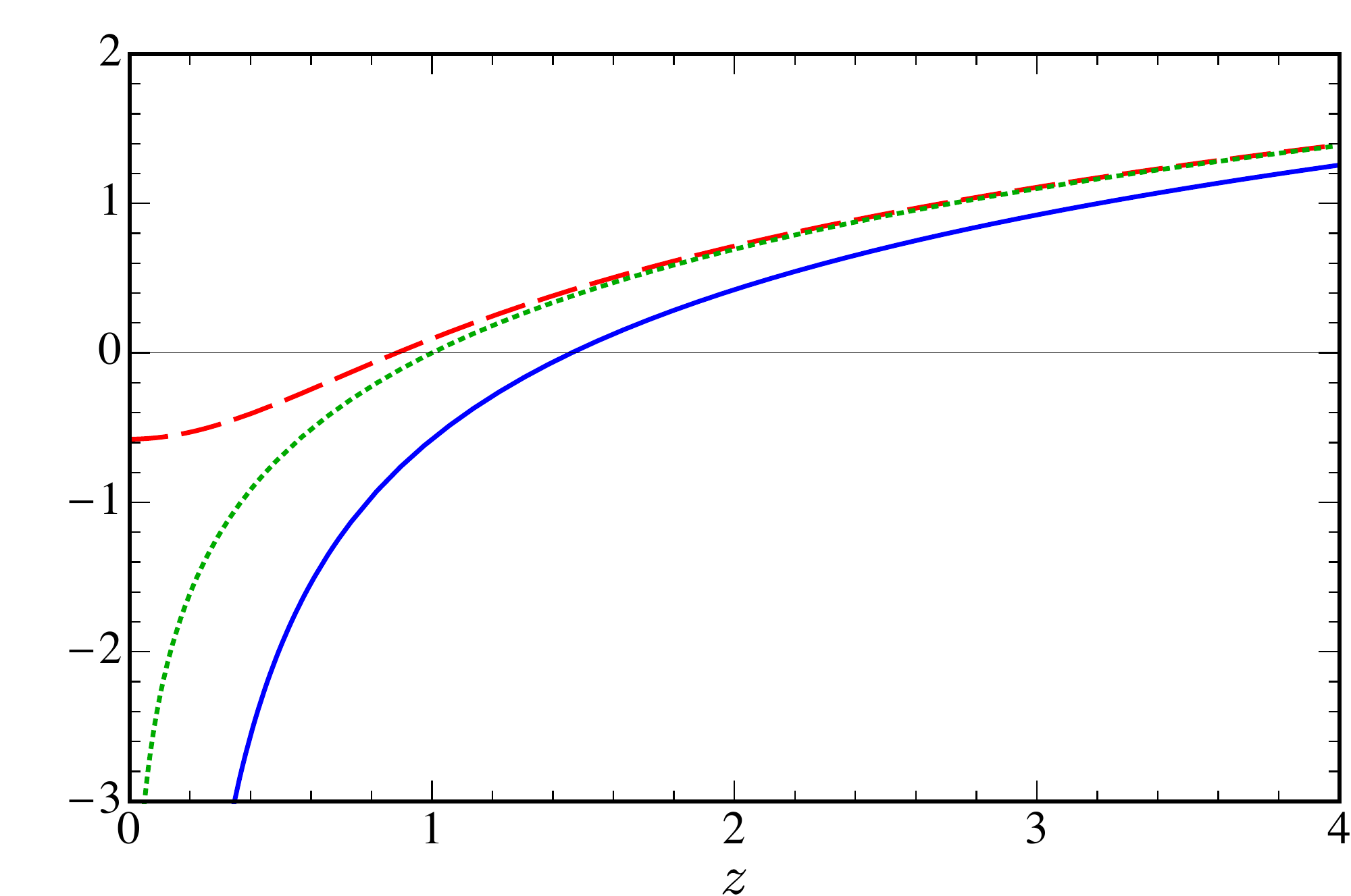}
\caption{(Color online)
Plots of the adimensional functions ${\rm Di}\Gamma(z)$ (continuous) and ${\rm
Re[Di}\Gamma[(iz)]$ (dashed). At large $z$, both functions approach $\log(z)$ (dotted).
} \label{fig:DiGamma}
\end{center}
\end{figure}

The derivation of the anomalous diffusion coefficient $D_{p}$ is more involved. One has
\beq
\label{anomDiffCoeff_1}
D_p(\Omega)
%=-\frac{m\gamma\Lambda^2}{4i\pi}\int_{-\infty}^\infty  \diff \tau\, \int_{-\infty}^\infty  \diff \omega\, \frac{\omega}{\omega^2+\Lambda^2}\coth\left(\frac{\hbar\omega}{2 k_B T}\right){\rm sign}(\tau)e^{i (\omega+\Omega) \tau}\\\nonumber
=-\int_{-\infty}^\infty  \frac{\diff \omega}{2\pi}{\cal P}\left[\frac{m\gamma\Lambda^2}{\omega+\Omega}\frac{\omega}{\omega^2+\Lambda^2}\coth\left(\frac{\hbar\omega}{2 k_B T}\right)\right].
\nonumber
\eeq
To perform the principal part integration with the standard trick
$\int  \diff \omega\, {\cal
P}\left[\frac{f(\omega)}{\omega}\right]=\int \diff \omega\,
\left[\frac{f(\omega)-f(0)}{\omega}\right]$ we need the numerator
to be a polynomial in $\omega$. Inserting the Matsubara representation of the $\coth$ in
\eqref{anomDiffCoeff_1}, one finds
\begin{multline}
\label{anomDiffCoeff}
\frac{\pi(\Omega^2+\Lambda^2)}{m\gamma\Omega \Lambda^2}D_p(\Omega)
%&=-\frac{mk_B T\gamma\Lambda^2}{\pi\hbar}\sum_{n=-\infty}^{\infty}\int_{-\infty}^\infty  \diff \omega\, {\cal P}\left[\frac{1}{\omega+\Omega}\frac{\omega}{\omega^2+\Lambda^2}\frac{\omega}{\omega^2+\nu_n^2}\right]\\\nonumber
%&=-\frac{mk_B T\gamma\Lambda^2}{\pi\hbar}\sum_{n=-\infty}^{\infty}\frac{\Omega}{(\Omega^2+\Lambda^2)(\Omega^2+\nu_n^2)}\int_{-\infty}^\infty  \diff \omega\, \frac{\Omega^2\omega^2-\Lambda^2\nu_n^2}{(\omega^2+\Lambda^2)(\omega^2+\nu_n^2)}\\\nonumber
%&
=-\frac{\pi}{\hbar}\sum_{n=-\infty}^{\infty}\frac{k_B T}{(\Omega^2+\nu_n^2)}\frac{(\Omega^2-\Lambda|\nu_n|)}{\Lambda+|\nu_n|}\\
%=\frac{mk_B T\gamma\Omega\Lambda}{\hbar(\Omega^2+\Lambda^2)}+\frac{m\gamma\Omega\Lambda^2}{\pi(\Omega^2+\Lambda^2)}\left\{{\rm Di}\Gamma\left(\frac{\hbar\Lambda}{2\pi k_BT}\right)-{\rm Re}\left[{\rm Di}\Gamma\left(\frac{i\hbar\Omega}{2\pi k_BT}\right)\right]\right\}.
=\frac{\pi k_B T}{\hbar\Lambda}
+{\rm Di}\Gamma\left(\frac{\hbar\Lambda}{2\pi k_BT}\right)-{\rm Re}\left[{\rm Di}\Gamma\left(\frac{i\hbar\Omega}{2\pi k_BT}\right)\right].
\end{multline}
The function ${\rm Di}\Gamma(z)\equiv\Gamma'(z)/\Gamma(z)$ is the logarithmic derivative of the Gamma function, and it is plotted in Fig.\ \ref{fig:DiGamma} for both real and imaginary arguments.

% #### #### #### #### #### #### #### ####
% #### #### #### #### #### #### #### ####

\begin{figure}[t!]
\begin{center}
\includegraphics[width=\columnwidth]{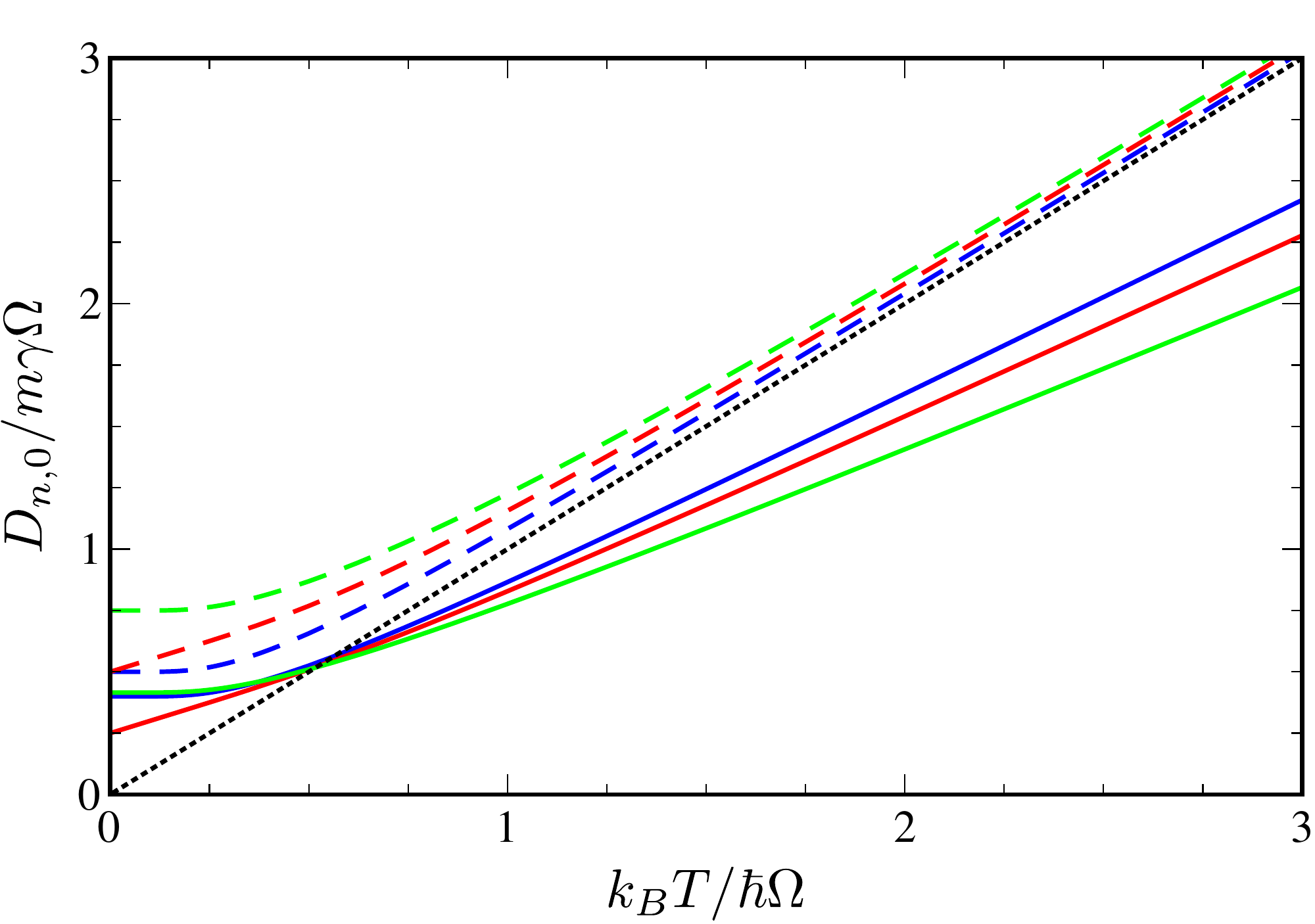}
\caption{(Color online)
Plot of the coefficients $D_{n,0}$, which control the decoherence rate of the off-diagonal elements of the density matrix $\rho(x_1,x_2)$ in the position basis. The lines represent respectively $D_{1,0}=D_x$ (blue), $D_{2,0}=D_{xx}$ (red), and $D_{3,0}$ (green). Continuous lines are for $\Lambda=2\Omega$, dashed lines for $\Lambda=100\Omega$. In the Caldeira-Leggett limit $k_B T/\hbar\gg \Lambda\gg \Omega$,
 we find $D_{n,0}\rightarrow m\gamma k_B T/\hbar$ (dotted line),
  independent of $n$.
} \label{fig:diffRates}
\end{center}
\end{figure}

The $C_x$ term provides a term which strongly renormalizes the harmonic potential frequency. 
The role of the counterterm $V_c$ introduced in the Hamiltonian is exactly to remove this spurious contribution, and from Eq.\ \eqref{master_equation_linear} we see explicitly that a perfect cancellation is obtained by choosing $V_c(x)=-C_x x^2$. 
Regarding the other coefficients, as we will see in the following, $C_p$ provides momentum damping, $D_x$ yields normal momentum diffusion, and $D_p$ contributes to anomalous diffusion. The $D_x$ term may also be seen as the one responsible for decoherence in the position basis \cite{SchlosshauerBook,Zurek2003,Schlosshauer2005}. There, the density matrix may be represented as $\rho(x_1,x_2,t)=\ave{x_1|\rho(t)|x_2}$, and one finds $\partial_t \rho(x_1,x_2,t)=-D_{x}(x_1-x_2)^2\rho(x_1,x_2,t)/\hbar+\ldots$, so that the off-diagonal components of $\rho$ decohere at a rate directly proportional to the square of the distance between them, $\gamma_{x_1,x_2}^{(1)}=D_{x}(x_1-x_2)^2/\hbar$, see Fig.~\ref{fig:diffRates}.

\subsection{Caldeira-Leggett limit (linear case)}

In the high-temperature and large cutoff limits $k_BT/\hbar\gg\Lambda\gg\Omega$, we may use the series expansions 
${\rm Di}\Gamma(z)=-z^{-1}-\tilde\gamma+\pi^{2} z/6+O(z^2)$
 and
${\rm Re}[{\rm Di}\Gamma(iz)]=-\tilde\gamma+O(z^{2})$ (with $\tilde\gamma$ the Euler gamma, and real adimensional argument $z$) to find
\beq
\frac{D_p}{\hbar m\Omega}%\sim
%\frac{\gamma k_BT}{\hbar^2\Lambda}+\frac{\gamma}{\hbar\pi}\left[-\frac{2\pi k_BT}{\hbar\Lambda}-\tilde\gamma+\tilde\gamma+O\left(\frac{\Lambda}{T}\right)\right]=
=-\frac{k_BT\gamma}{\hbar^2\Lambda}+O\left(\frac{\Lambda}{T}\right),
\eeq
  this leading contribution coming from the zero Matsubara frequency term. % in Eq.\ \eqref{anomDiffCoeff}.
Apart from a factor $1/2$, due to a different definition of the damping constant $\gamma$, this expression agrees with Eq. (3.409) of Ref.\ \cite{BreuerBook}, and with Eq.\ (5.54) of Ref.\ \cite{SchlosshauerBook} (mind however that the latter one has a minor typo, i.e., this coefficient appears with the wrong sign). Inserting in the ME, Eq.\
\eqref{master_equation_linear}, at high-$T$ one finds
\begin{multline}
\dot\rho(t)=-\frac{i}{\hbar}\left[H_{\rm sys},\rho(t)\right]
-\frac{i\gamma}{2\hbar} [x,\{p,\rho(t)\}]\\
-\frac{m\gamma k_BT}{\hbar^2}[x,[x,\rho(t)]]
+\frac{\gamma k_BT}{\hbar^2\Lambda}[x,[p,\rho(t)]].
\end{multline}
Since $p$ is of order $m\Omega x$ in an harmonic potential, the last term may be neglected as it scales as $\Omega/\Lambda$, and in this way we recover the usual Caldeira-Leggett ME, Eq. \eqref{CaldeiraLeggettMasterEq}. As such, in the following we will refer to the regime where $k_BT/\hbar\gg\Lambda\gg\Omega$ as {\it the} Caldeira-Leggett limit. Note that in the case of a harmonic potential trapping the Brownian particle, or more generally upon neglecting quantum effects for the general non-harmonic potential, the corresponding time dependent equation for the Wigner function in this regime has a particularly simple interpretation (cf.\ Ref.\ \cite{GardinerBook}): it is a Fokker--Plank equation for the probability distribution in the phase space of a classical Brownian particle undergoing damped motion with a damping constant $\gamma$ under the influence of a Langevin stochastic noise--force $F(t)$. The noise is Gaussian and white, but it fulfills
 the fluctuation--dissipation relation, i.e., the average of the noise correlation satisfies $\langle F(t+\tau)F(t)\rangle= 2\gamma k_BT$. This relation assures that the 
stable stationary state of the dynamics is the classical Gibbs-Boltzmann state. In terms of the coefficients entering the master equation the fluctuation--dissipation relation implies that $D_x/C_p=2 k_B T/\hbar \Omega$.

% #### #### #### #### #### #### #### ####
% #### #### #### #### #### #### #### ####

\subsection{Large cutoff limit (linear case)}
We want to look at the interesting limit $\Lambda\gg\Omega, k_BT/\hbar$, with $\Omega \sim k_BT/\hbar$; in this case we find
\begin{multline}
\dot\rho(t)=-\frac{i}{\hbar}\left[H_{\rm sys},\rho(t)\right]
-\frac{i\gamma}{2\hbar} [x,\{p,\rho(t)\}]\\
-\frac{m\gamma \Omega}{2\hbar}\coth\left(\frac{\hbar\Omega}{2k_B T}\right)[x,[x,\rho(t)]]
-\frac{D_p}{\hbar m\Omega}[x,[p,\rho(t)]].
\label{main-linear}
\end{multline}
For large $z$ we have ${\rm Di}\Gamma(z)\sim\log(z)-1/(2z)+O(z^{-2})$
 and 
 ${\rm Re}[{\rm Di}\Gamma(i z)]\sim\log(z)+1/(12z^2)+O(z^{-3})$,
  and the anomalous diffusion coefficient 
 is proportional to $D_p\sim\frac{m\gamma\Omega}{\pi}\log\left(\frac{\hbar\Lambda}{2\pi k_B T}\right)$.
In this limit, we have moreover $D_x/C_p=\coth\left(\hbar\Omega/2k_B T\right)$. 
Equation \eqref{main-linear}, with the anomalous diffusion coefficient given in Eq.\ \eqref{anomDiffCoeff}, constitutes the main results of this section. 
As we will argue in Section \ref{sec:conclusions} and Appendix \ref{sec:impurityInATrappedGas}, 
in any practical physical application of the present theory the cutoff energy $\hbar\Lambda$ has a very concrete physical meaning: in a trap the bath frequencies are evidently bound by the trap
depth, in an optical lattice by the lowest band's width, and so on. 

\subsection{Ultra-low temperature limit (linear case)}
To conclude the analysis of the linear case, we consider the limit $\Lambda\gg\Omega\gg k_BT/\hbar$. Since both ${\rm Di}\Gamma$ functions in Eq.\ \eqref{anomDiffCoeff} diverge logarithmically, the temperature drops completely out of the QME, which reads now
\begin{multline}
\dot\rho(t)=-\frac{i}{\hbar}\left[H_{\rm sys},\rho(t)\right]
-\frac{i\gamma}{2\hbar} [x,\{p,\rho(t)\}]\\
-\frac{m\gamma \Omega}{2\hbar}[x,[x,\rho(t)]]
-\frac{\gamma}{\hbar \pi}\log\left(\frac \Lambda \Omega \right)[x,[p,\rho(t)]].
\label{main-linear-zeroT}
\end{multline}

\section{BM-QME with quadratic coupling}
\label{sec:QMEquadratic}
Let us now turn to the main subject of this paper: the Born-Markov QME with non-linear coupling in the particle position. We discuss in detail here the case of quadratic coupling, $f(x)=x^2/a$, and leave the presentation of the more involved results for a completely general coupling to the Appendix \ref{sec:genericCoupling}.

  The Heisenberg equation for $x^{2}(\tau)$ yields 
\begin{multline}
 x^{2}(-\tau)=\left(x\cos(\Omega\tau)-\frac{p}{m\Omega}\sin(\Omega\tau)\right)^{2}\\
=x^{2}\cos^{2}(\Omega\tau)-\frac{\{x,p\}}{m\Omega}\sin(\Omega\tau)\cos(\Omega\tau)
+\frac{p^{2}}{m^{2}\Omega^{2}}\sin^{2}(\Omega\tau)
\end{multline}
 so that (using
the linearity of commutators and anti-commutators) one finds
\begin{widetext}
\begin{multline}
\dot\rho(t)=-\frac{i}{\hbar}\left[H_S,\rho(t)\right]
-\frac{iC_{xx}}{\hbar a^2} [x^2,\{x^2,\rho(t)\}]
-\frac{iC_{xp}}{\hbar a^2} \left[x^2,\left\{\frac{\{x,p\}}{m\Omega},\rho(t)\right\}\right]
-\frac{iC_{pp}}{\hbar a^2} \left[x^2,\left\{\frac{p^2}{m^2\Omega^2},\rho(t)\right\}\right]\\
-\frac{D_{xx}}{\hbar a^2}[x^2,[x^2,\rho(t)]]
-\frac{D_{xp}}{\hbar a^2}\left[x^2,\left[\frac{\{x,p\}}{m\Omega},\rho(t)\right]\right]
-\frac{D_{pp}}{\hbar a^2}\left[x^2,\left[\frac{p^2}{m^2\Omega^2},\rho(t)\right]\right],
\label{QMEwithquadraticcoupling}
\end{multline}
\end{widetext}
with the coefficients $C_{\rm \ldots}$
given by
\begin{align}
C_{xx}=&-\int_{0}^\infty  \diff
\tau\,\eta(\tau)\cos^2(\Omega\tau)\nonumber\\
C_{xp}=&\int_{0}^\infty  \diff
\tau\,\eta(\tau)\sin(\Omega\tau)\cos(\Omega\tau)\nonumber\\
C_{pp}=&-\int_{0}^\infty  \diff
\tau\,\eta(\tau)\sin^2(\Omega\tau)\nonumber
\end{align}
 and the $D_{\rm \ldots}$ by
\begin{align}
D_{xx}=&\int_{0}^\infty  \diff
\tau\,\nu(\tau)\cos^2(\Omega\tau)\nonumber\\
D_{xp}=&-\int_{0}^\infty  \diff
\tau\,\nu(\tau)\sin(\Omega\tau)\cos(\Omega\tau)\nonumber\\
D_{pp}=&\int_{0}^\infty  \diff
\tau\,\nu(\tau)\sin^2(\Omega\tau)\nonumber
\end{align}

Using $\sin(x)\cos(x)=\sin(2x)/2$ and introducing the shorthand notation 
\beq
c(\Lambda)=\Lambda^2/(4\Omega^2+\Lambda^2)
\eeq
 for the cutoff function evaluated at frequency $2\Omega$, we may exploit the results for $C_p$ and $D_p$ in the linear case to find
\begin{align}
C_{xp}=&\frac12\int_{0}^\infty  \diff \tau\,\eta(\tau)\sin(2\Omega\tau)=
\frac{C_p(2\Omega)}{2}=\frac{m\gamma\Omega}{2}c(\Lambda)\nonumber\\\nonumber
D_{xp}=&\frac{D_p(2\Omega)}{2}=
\frac{m\gamma\Omega}{\pi}c(\Lambda)
\left\{
 \frac{\pi k_B T}{\hbar\Lambda}+ {\rm Di}\Gamma\left(\frac{\hbar\Lambda/2}{\pi k_BT}\right)
\right.
\\\nonumber
&\left.
-{\rm Re}\left[{\rm Di}\Gamma\left(\frac{i\hbar\Omega}{\pi k_BT}\right)\right]
 \right\}
\end{align}

Similarly, using $\cos^2(x)=[1+\cos(2x)]/2$, $\,
I_\nu\equiv\int_0^\infty  \diff
\tau\,\nu(\tau)=mk_BT\gamma/\hbar$, and $D_x$ for the linear case,
one finds
\begin{align}
D_{xx}=&\frac{I_\nu+D_x(2\Omega)}{2}=\frac{m\gamma\Omega}{2}\left[\frac{k_BT}{\hbar\Omega}+c(\Lambda)\coth\left(\frac{\hbar\Omega}{k_BT}\right)\right] \nonumber\\
D_{pp}=&I_\nu-D_{xx}=\frac{m\gamma\Omega}{2}\left[\frac{k_BT}{\hbar\Omega}-c(\Lambda)\coth\left(\frac{\hbar\Omega}{k_BT}\right)\right]\nonumber
\end{align}

Finally, using $I_\eta\equiv\int_0^\infty  \diff
\tau\,\eta(\tau)=m\gamma\Lambda/2$, and the derivation for
$C_x$ in the linear case, we also find
\begin{align}
C_{xx}=&-\frac{I_\eta}{2}+\frac{C_x(2\Omega)}{2}=-\frac{m\gamma\Lambda (2 \Omega ^2+\Lambda ^2)}{2(4 \Omega ^2+\Lambda ^2)}\nonumber\\
C_{pp}=&-I_\eta-C_{xx}=-\frac{m\gamma  \Omega
^2}{\Lambda}c(\Lambda)\nonumber
\end{align}
In analogy with the linear case, the coefficient
$C_{xx}$ diverges with the cutoff $\Lambda$, but this poses
no problems as $[x^2,\{x^2,\rho\}]=[x^4,\rho]$, so this term may always be canceled exactly by an appropriate counter-term  $V_c(x)=-C_{xx}x^4/a^2$, representing this time a Lamb-shift of the coefficient of the quartic term in the confinement. All
other coefficients remain bounded in the limit of
$\hbar\Lambda/k_BT\rightarrow\infty$, exception made for $D_{xp}$ 
which exhibits a mild logarithmic divergence, in complete analogy with $D_p$ in the linear case.
The generalized QME \eqref{QMEwithquadraticcoupling}, together with the explicit forms of its coefficients, represent a central result of this paper. Here below, we analyze the behavior of the various coefficients in three different limits.

\subsection{Caldeira-Leggett limit (quadratic case)}
 In the usual
high-temperature limit $k_BT/\hbar\gg\Lambda\gg\Omega$, we
have
\begin{align}
D_{xx}\approx&m\gamma k_BT/\hbar\nonumber\\
D_{xp}\approx&-m\gamma (k_BT/\hbar)(\Omega/\Lambda) \longrightarrow 0 \nonumber\\ 
D_{pp}\approx&-m\gamma\hbar\Omega^2/(6k_BT) \longrightarrow 0,  \nonumber\\
\end{align}
and therefore we obtain
\begin{multline}
\dot\rho(t)=-\frac{i}{\hbar}\left[H_{\rm sys},\rho(t)\right] 
-\frac{im\gamma}{2\hbar} \left[\frac{x^2}{a},\left\{\frac{\{x,p\}}{ma},\rho(t)\right\}\right]\\\nonumber
 -\frac{m\gamma k_BT}{\hbar^2}\left[\frac{x^2}{a},\left[\frac{x^2}{a},\rho(t)\right]\right],
\end{multline}
 which agrees with the generalized CLME discussed in the introduction, Eq.\ \eqref{NonLinearCaldeiraLeggettMasterEq}. In this high-temperature limit, it is easy to identify $C_{xp}$ as being proportional to the momentum damping coefficient, and $D_{xx}$ to the normal momentum diffusion coefficient. In analogy with the linear case, this latter term may also be seen as the one responsible for decoherence in the position basis. The off-diagonal components of $\rho$ are in this way found to decohere at a rate $\gamma_{x_1,x_2}^{(2)}=D_{xx}(x_1^2-x_2^2)^2/\hbar a^2$, see Fig.~\ref{fig:diffRates}. This is an important result, providing a typical timescale for decoherence of states entangled in position space in presence of a bath coupling of the form $f(x)\propto x^2$.
In App.\ \ref{sec:genericCoupling} we will provide a general formula which yields the position-space decoherence rate $\gamma_{x_1,x_2}^{(n)}$ associated to a coupling with an arbitrary power of the system's coordinate, $f(x)\propto x^n$. Remarkably, and at odds with what found in Ref.\ \cite{Hu1993}, we find that superposition states which are symmetric around the origin (e.g., sharply localized around both $+x_0$ and $-x_0$) will be protected by decoherence in presence of couplings containing only even powers of $n$.

Note also that in this limit we recover again the classical Gibbs-Boltzmann stationary states, and the dynamics satisfies the fluctuation-dissipation relation.  Namely, in the case of an harmonic potential, or more generally upon neglecting quantum effects induced by an anharmonic potential, the time dependent equation for the Wigner function has the interpretation of a Fokker--Plank equation for the probability distribution in the phase space of a classical Brownian particle undergoing damped motion with an $x$--dependent damping $\gamma (x/a)^2$ under the influence of a multiplicative Langevin stochastic noise--force $F(t)(x(t)/a)$.
The noise is Gaussian and white, and it fulfills the fluctuation--dissipation relation, i.e. the average of the noise correlation yields $\langle F(t+\tau)x(t+\tau)F(t)x(t)\rangle= 2\gamma k_BT\langle x^2\rangle$. This relation assures that the stable stationary state of the dynamics is the classical Gibbs-Boltzmann state.
In terms of the coefficients entering the master equation the fluctuation--dissipation relation implies that $D_{xx}/C_{xp}=2 k_B T /\hbar\Omega$.

\subsection{Large cutoff limit (quadratic case)}
\label{sec:largeCutoffQuadraticCase} 
Taking  the more interesting
limit $\hbar\Lambda\gg\hbar\Omega, k_BT$ limit simply amounts to setting $c(\Lambda)=1$ in the expression for the various coefficients.
In this regime, our QME exhibits several differences in comparison to Eq.\ \eqref{NonLinearCaldeiraLeggettMasterEq}: i) the coefficient $C_{pp}$ (a term contributing to a Lamb-shift of the trap frequency $\Omega$) is suppressed as $\Omega/\Lambda$; 
ii) the normal momentum diffusion (or position-basis decoherence) coefficient $D_{xx}$, which is analogous to the $D_{x}$ of the linear case, develops a non-trivial quantum dependence on $\hbar\Omega/k_BT$;
iii) the coefficient $D_{xp}$ (which contributes to both the Lamb-shift and the anomalous diffusion) becomes log-divergent in $\Lambda$, analogously to $D_{p}$ found in the linear case;
iv) there appears a new coefficient, $D_{pp}$, which
depends on $\hbar\Omega/k_BT$, and vanishes for $k_BT \gg \hbar\Omega$.

We note here that, in this limit, the coefficients of the QME satisfy the generalized fluctuation-dissipation relations $(D_{xx}+D_{pp})/C_{xp}=2k_BT/\hbar\Omega$, and $(D_{xx}-D_{pp})/C_{xp}=2\coth(\hbar\Omega/k_BT)$.
Finally, we note that the usual high temperature limit, \`a la Caldeira-Leggett, $k_B T\gg\hbar\Lambda\gg\hbar\Omega$, should be taken with precaution in the case of non-linear coupling. Indeed, as we will see in the following (cf.\ Fig.\ \ref{fig:quadCase}), for strong damping the system in a purely harmonic trap may become dynamically unstable at sufficiently large temperatures.

\subsection{Ultra-low temperature limit (quadratic case)}
The QME equation for  $k_B T/\hbar\ll\Omega\ll \Lambda$ reads:
\begin{multline}
\dot\rho(t)=-\frac{i}{\hbar}\left[H_{\rm sys},\rho(t)\right]
-\frac{im\gamma}{2\hbar} \left[\frac{x^2}{a},\left\{\frac{\{x,p\}}{ma},\rho(t)\right\}\right]\\
 -\frac{m\gamma\Omega}{2\hbar}\left[\frac{x^2}{a},\left[\frac{x^2}{a}-\frac{p^2}{m^2\Omega^2 a},\rho(t)\right]\right]\\ 
 -\frac{m\gamma}{\hbar \pi}\log\left(\frac{\Lambda}{2\Omega}\right)\left[\frac{x^2}{a},\left[\frac{\{x,p\}}{ma},\rho(t)\right]\right].
\end{multline}
As expected the temperature drops out of the equation, and the $D_{xp}$ term is log-divergent in the cutoff $\Lambda$.

\section{Wigner function approach and stationary solutions}
\label{sec:WignerStationary}
The Quantum Master Equation for the density matrix $\rho$ can be
particularly well analyzed in terms of the Wigner function $W$. To
this aim it is useful to introduce the operators
$x_\pm=x\pm\frac{i\hbar}{2}\frac{\partial}{\partial p}$, and
$p_\pm=p\pm\frac{i\hbar}{2}\frac{\partial}{\partial x}$, which
satisfy the commutation rules
\begin{align}
&[x_+,x_-]=[p_+,p_-]=0\\ \nonumber &[x_+,p_-]=-[x_-,p_+]=i\hbar.
\end{align}
The formal substitutions (see Eqs.(4.5.11) of \cite{GardinerBook})
are of great use in the following:
\begin{align}
\hat x\rho\rightarrow x_+W,\quad \hat p\rho\rightarrow p_-W\\ \nonumber
\rho \hat x\rightarrow x_-W,\quad \rho \hat p\rightarrow p_+W.
\end{align}
We note here that, while in the previous Sections $x$ and $p$ stood for the usual non-commuting operators, from now on the same symbols will be used to represent the commuting variables of the Wigner function $W(x,p)$.
 
\subsection{Linear case}
\label{sec:WignerStationaryLinear}
Let us first analyze the case of linear coupling.
When $f(x)=x$, the QME for general $\Omega$, $\Lambda$ and $T$ in terms of the Wigner function reads\footnote{Note that 
%$[\hat P,\hat \rho]\hat X=[(p_--p_+)\hat \rho] \hat X=x_-(p_--p_+)W$.
$[\hat p,\rho]\hat x=[(p_--p_+)\rho]\hat x=x_-(p_--p_+)W$.
}
\begin{align}
\dot W\!\!=\!\!\left[m\Omega^{2}\partial_p x-\frac{\partial_x p}{m}
+\frac{2C_p}{m\Omega}\partial_p p +\hbar D_x\partial^2_p
-\frac{\hbar D_p}{m\Omega}\partial_x\partial_p\right]\!\!W.
\end{align}
The stationary solution to this equation may be found by inserting a generic quadratic ansatz
\beq
W_{\rm st}\propto\exp\left[-\left(\sigma_p\frac{p^{2}}{2m} +\sigma_x\frac{m\Omega^{2}x^{2}}{2}\right)/(k_B\tilde{T})\right]
\label{gaussianAnsatz}
\eeq
with real parameters $\sigma_p$ and $\sigma_x$, and equating independently the coefficients of $x^{2}$ and $p^{2}$ to zero in the resulting equation.

\begin{figure}\begin{center}
\includegraphics[width=\columnwidth]{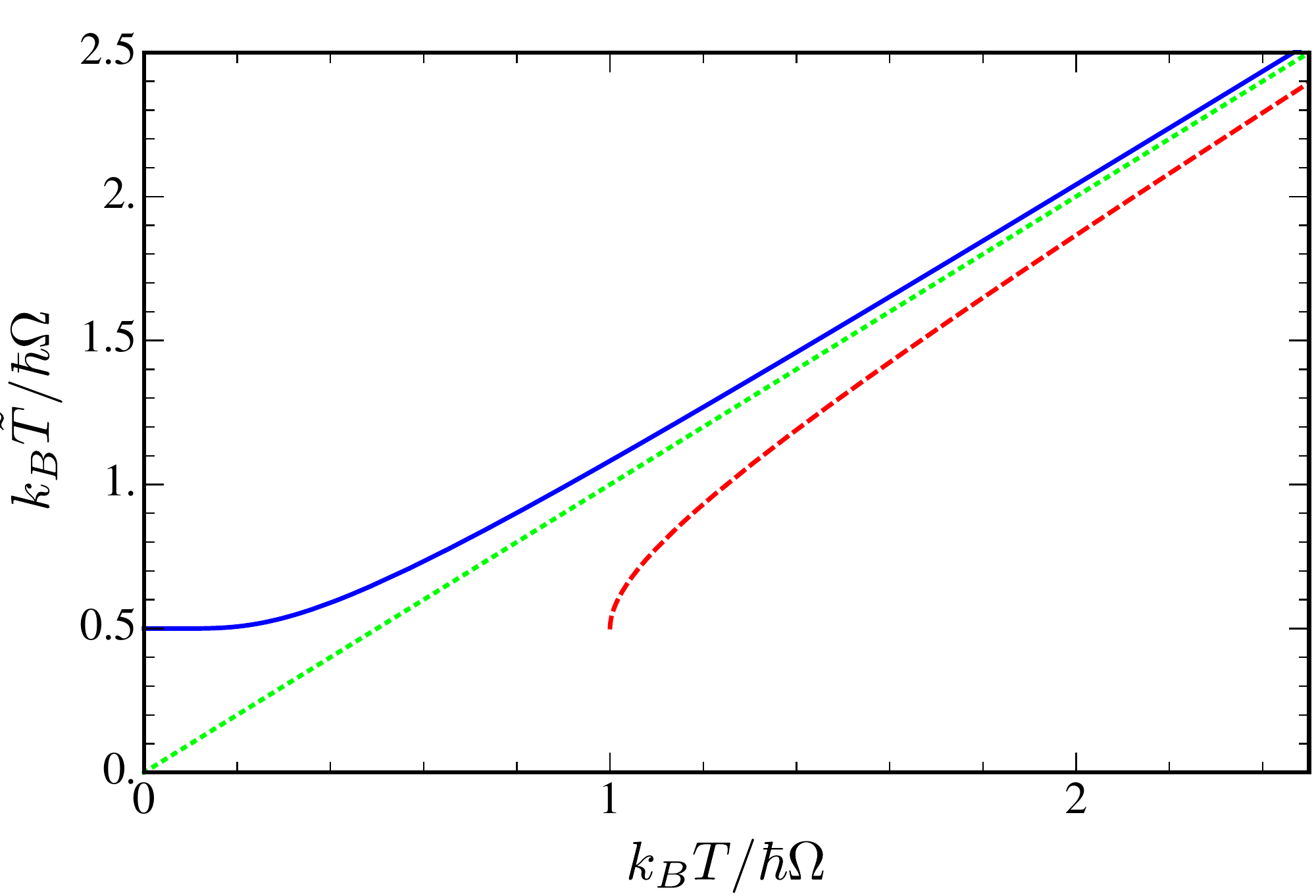}
\end{center}
\caption{(Color online)
Effective temperatures as obtained through the complete quantum treatment, Eq.\
\eqref{effTempRefined} (blue), and by means of an oversimplified approximation discussed in App.\ \ref{sec:highTlimitWithLeadingQuantumCorrections}, Eq.\ \eqref{effTempSimple} (red). The green line is the high-$T$ result, $\tilde T=T$.}
\label{fig:effTemp}
\end{figure}

In the oversimplified high-$T$ limit $k_BT \gg \hbar\Lambda \gg \hbar\Omega$,  \`a la Caldeira-Leggett, one would set $D_x=m\gamma k_BT/\hbar$ and $D_p=0$, and find in this way $\sigma_p=\sigma_x=1$, and $\tilde T=T$. By retaining instead the complete expression of all terms in the equation (and, in particular, a non-zero $D_p$), we find that the stationary Wigner function is obtained by choosing $\sigma_p=1$ and
\beq
\sigma_x
%=\frac{D_x m \Omega ^2}{D_x m \Omega ^2-2 D_p C_p}
=\frac{1}{1-2D_p/(m\Omega^2\coth[\hbar\Omega/2k_BT])},
\label{sigmaXlinear}
\eeq
yielding an effective temperature
\beq
\tilde{T}=\frac{\hbar\Omega}{2k_B}\coth\left(\frac{\hbar\Omega}{2k_BT}\right).
%\approx k_BT(1+ (\hbar\Omega/k_B T)^2+\ldots).
\label{effTempRefined}
\eeq
This results is shown in Fig.\ \ref{fig:effTemp}.
A number of interesting conclusions may now be drawn.
% #### #### #### #### #### #### #### ####
% #### #### #### #### #### #### #### ####

First of all, a careful treatment of the equation at low-$T$ yields an effective temperature which saturates to the zero-point motion energy.
When $\sigma_p=\sigma_x=1$, the Gaussian stationary solution with an effective temperature $\tilde T$ as given by the quantum result \eqref{effTempRefined} corresponds to the exact quantum thermal Gibbs-Boltzmann density matrix of an harmonic oscillator (the system) at the temperature $T$.
In this case, the
contours of the stationary distributions are circles of radius
$\sqrt{2k_B\tilde T/\hbar\Omega}$ for
arbitrary $T$ (i.e., of radius 1 at $T=0$).

\begin{figure}\begin{center}
\includegraphics[width=\columnwidth]{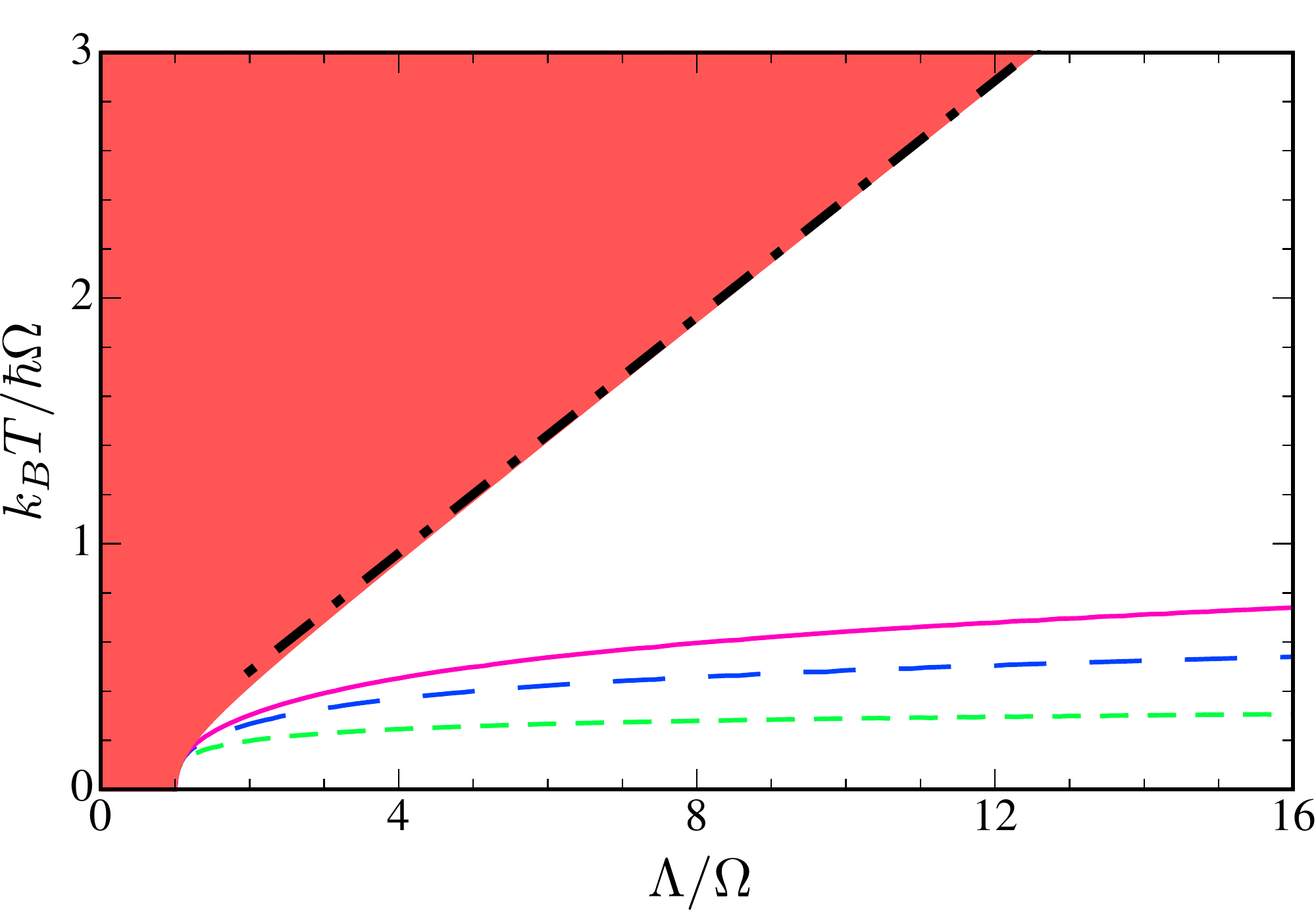}
\end{center}
\caption{(Color online)
Minimal temperature for the fulfillment of the Heisenberg uncertainty principle for an Ohmic spectral function with LD cutoff in the linear case, for $\gamma/\Omega=0.1,\,0.5,\,1$  (from bottom to top). In the red region, the gas displays effective ``heating" and a quenched aspect ratio in $p$ relative to $x$  (i.e., $\delta_x/\delta_p>1$). The black, dot-dashed line is the asymptotic approximation to the boundary of unit aspect ratio, $T=\alpha_{(1)}\Lambda$.}
\label{fig:HeisenbergLimitLinear}
\end{figure}

 More generally, in units of the
normalized standard deviations
\begin{align}
\begin{split}
\delta_x=&2\sqrt{\frac{m\Omega^2\ave{x^2}_{\rm st}}{2\hbar
\Omega}} =\sqrt{\frac{2k_B\tilde T}{\hbar\Omega\sigma_x}}\\
\delta_p=&2\sqrt{\frac{\ave{p^2}_{\rm st}}{2 m \hbar\Omega}}= \sqrt{\frac{2k_B\tilde T}{\hbar\Omega\sigma_p}},
\end{split}
\end{align}
the Heisenberg uncertainty relation requires that
\beq
\delta_x\delta_p\geq 1,
\label{HeisenbergUncertainty}
\eeq
i.e., that
the contour of the distribution encircles an area not smaller than $\pi$.
An important effect of $D_p$ is that it allows for a contraction of the distribution in $x$ vs.\ $p$. The Heisenberg uncertainty principle then puts an important constraint on our theory, forcing us to exclude the region where the inequality is violated. 
In Fig.\ \ref{fig:HeisenbergLimitLinear} we illustrate
this region of validity, as obtained by inserting Eq.\ \eqref{sigmaXlinear} in Eq.\ \eqref{HeisenbergUncertainty}: for any $\Lambda>\Omega$, we find that there exists a critical temperature below which the Heisenberg uncertainty principle is violated. Similar squeezing effects have been discussed \cite{Reibold} in the literature in the context of the so called Ullersma model \cite{Ullersma1,*Ullersma2,*Ullersma3,*Ullersma4}. At $T=0$, the Heisenberg principle requires $\Lambda<\Omega$.

Interestingly, in the linear case there are no log-corrections to $\tilde{T}$
coming from the log-divergent term $D_p$. $D_p$ grows with the cutoff, and at very large values $\sigma_x$ diverges (i.e., $\delta_x^2$ approaches zero) and becomes negative, yielding a non-normalizable solution. However, this bound always lies beyond the one set by the Heisenberg principle, which requires $\delta_x\delta_p\geq 1$.

% #### #### #### #### #### #### #### ####
% #### #### #### #### #### #### #### ####

We may say that the quantum particle immersed in the bath experiences an effective ``heating" if the phase-space area encircled by the normalized standard deviations is larger than the one a quantum Gibbs-Boltzmann (GB) distribution would occupy at the same temperature. Since $\ave{E_k}_{\rm GB}\ave{E_p}_{\rm GB}=(k_B\tilde T/2)^2$, the system is effectively heated if
\beq
\delta_x\delta_p>\coth\left(\frac{\hbar\Omega}{2k_BT}\right),
\label{heatingCondition}
\eeq
or equivalently $\sigma_x\sigma_p< 1$. Since $\sigma_p=1$ in the linear case, this amounts to requiring $D_p<0$, which remarkably does not depend on $\gamma$. Asymptotically, we have $T>\alpha_{(1)}\Lambda+O(\Omega/T)$, with $\alpha_{(1)}\approx0.24$ solution of the implicit equation
 \beq
 \pi\alpha_{(1)}+{\rm Di}\Gamma(1/2\pi\alpha_{(1)})+\tilde\gamma=0.
 \eeq

Finally, we consider the aspect ratio of the phase-space contour described by the standard deviations. Since $\sigma_p$ always equals unity in the linear case, it is easy to see that we have a quenched 
aspect ratio in $x$, relative to $p$ (i.e., $\delta_x/\delta_p<1$) in the ``cooling" region, and the opposite situation ($\delta_x/\delta_p>1$) in the ``heating" region.
In fact the line separating ``heating'' region from the ``cooling'' region corresponds to the regime where $D_p=0$. In this case the Wigner function is exactly given by a Gaussian with effective temperature $\tilde T$, 
and circular shape of the distribution ($\delta_p=\delta_x$); it corresponds precisely to the quantum thermal Gibbs-Boltzmann density matrix.

\pie{It should be noted that, when deriving the stationary solutions from a perturbative treatment of the master equation to order $2n$ in the bath-system coupling constant $\kappa_k$, one gets a reduced equilibrium state which is exact to order $2n-2$, and contains some (but not all)
terms of the order $2n$ solutions. The overall error is therefore of order $(\kappa_k)^{2n}$ itself, 
as pointed out by Fleming and Cummings \cite{FlemPRE2011} (for discussion of the nature of exact reduced equilibrium states see also \cite{Subasi2012}). Indeed, the violation of the Heisenberg uncertainty principle we observe within our BM-QME, which is of second-order in $\kappa_k$, is driven by the unphysical logarithmic divergence of $D_p$, which is itself proportional to $\gamma$, i.e., to $\kappa_k^2$.}
\mac{Obviously, if the exact master equation is used, then Heisenberg uncertainty  violation cannot occur in any parameter regime, ergo this violation is not physical, but is rather  a result of applied approximations. On the other hand, it is to be expected that both 
the degree of cooling and squeezing in  the considered quantum stochastic process should be bounded from below -- and the Heisenberg 
uncertainty violation gives a reasonable  estimate of this bound.}

\subsection{Quadratic case}
We turn now to the most interesting case, the quadratic case with $f(x)=x^2/a$. We consider the complete equation, obtained using the results in Sec.\ (\ref{sec:largeCutoffQuadraticCase}), and as usual we reabsorb the (linearly divergent in $\Lambda$) contribution coming from the $C_{xx}$ term in the Hamiltonian $H_{\rm sys}$, by requiring $V_c(x)=-C_{xx}f(x)^2$. The equation of motion for the Wigner function of a harmonically confined particle reads then 
\begin{widetext}
\begin{align}\label{Wigner-quadratic-with-Cpp}
\dot W=&
-\frac{i}{\hbar}\left[\frac{p_-^2-p_+^2}{2m}+V(x_+)-V(x_-)\right]W
-(x_+^2-x_-^2)\left[
\frac{iC_{xp}\Big(\{x_+,p_-\}+\{x_-,p_+\}\Big)}{\hbar m\Omega a^2}
+\frac{iC_{pp}(p_{-}^2+p_{+}^2)}{\hbar m^2\Omega^2 a^2}\right.\\\nonumber
&\qquad\qquad\qquad\qquad\qquad\qquad\qquad\qquad+\left.\frac{D_{xx}(x_+^2-x_-^2)}{\hbar a^2}
+\frac{D_{xp}\Big(\{x_+,p_-\}-\{x_-,p_+\}\Big)}{\hbar m\Omega a^2}
+\frac{D_{pp}\Big(p_-^2-p_+^2\Big)}{m^2\Omega^2\hbar a^2}\right]W\\\nonumber =&\left[-\frac{\partial_x
p}{m}+m\Omega^{2}\partial_p x
+\frac{8C_{xp}}{m\Omega a^{2}}\left(\partial_p p x^{2} +\frac{\hbar^{2}}{4}\partial_p^{2}(\partial_x x-1)\right)
+\frac{C_{pp}}{(m \Omega a)^{2}}\Big(4\partial_p xp^{2}-\hbar^{2}\partial_p\partial_x^{2}x+2\hbar^{2}\partial_p\partial_x\Big)\right.\\\nonumber
 &\qquad\qquad\qquad\qquad\qquad\qquad\qquad\qquad\left. +\frac{4\hbar D_{xx} \partial_p^2 x^2 }{a^2} 
+\frac{4\hbar D_{xp}(\partial_p^2
x p-\partial_p\partial_x x^2 +\partial_p x)}{m\Omega a^2}
-\frac{4\hbar D_{pp} (\partial_x x -1) \partial_p p}{m^2 \Omega^2 a^2}\right]W
\end{align}
\end{widetext}

Interestingly, the Gaussian ansatz \eqref{gaussianAnsatz} would provide a stationary solution to the above equation if we neglected the terms proportional to $C_{pp}$ and $D_{xp}$. Remembering that $D_{xx}-D_{pp}=2C_{xp}\coth\left(\hbar\Omega/k_BT\right)$, the stationary solution is found when $\sigma_p=\sigma_x=1$ and
\beq \label{standardt}
k_B\tilde{T}\overset{(C_{pp}=D_{xp}=0)}{=}\frac{\hbar\Omega}{2}\coth\left(\frac{\hbar\Omega}{2k_BT}\right),
\eeq
 which coincides with the result found above for the linear case, Eq.\ \eqref{effTempRefined}. Unfortunately however $D_{xp}$ is generally not negligible, as for example it diverges logarithmically with the cut-off $\Lambda$. In order to incorporate the neglected terms, one may try to generalize the ansatz by including in the exponent terms
proportional to higher polynomials in $x^2$ and $p^2$ (i.e., terms such as $x^4$, $x^2p^2$, or $p^{4}$), but no closed solution can be be found in this way, as moments of a given order always couple with higher ones.

The contributions higher than quadratic can, however, be reasonably taken into account by means of the so-called {\it self-consistent Gaussian (or pairing) approximation} \cite{Gardiner-cl,Risken}. The $D_{xp}$ term is proportional to
\begin{multline}
\partial_p^2 x p-\partial_p\partial_x x^2 + \partial_p x \simeq 
 \partial_p^2 \ave{x p}_{\rm st} -\partial_p\partial_x \ave{x^2}_{\rm st} + \partial_p x\\
= -\partial_p\partial_x \frac{k_B\tilde T}{\sigma_x m\Omega^2} + \partial_p x.
\end{multline}
As a general rule, averages of odd functions or partial derivatives vanish when performed with respect  to the Gaussian distribution \eqref{gaussianAnsatz}.
Similarly, the $C_{pp}$ term contributes 
\begin{equation}
4\partial_p xp^{2}-\hbar^{2}\partial_p\partial_x^{2}x+2\hbar^{2}\partial_p\partial_x\approx
\frac{4mk_B\tilde T}{\sigma_p}\partial_p x+2\hbar^{2}\partial_p\partial_x,
\end{equation}
as (mixed) derivatives of order higher than two vanish in this approximation. In this way, we get the two equations
\begin{align}
%\begin{split}
\label{deltax2overdeltap2}
\delta_p^2&=\frac{\delta_x^2}{\zeta}+\Gamma c_{pp}\left(\frac{\delta_x^2\delta_p^2}{2}-1\right)\\
\label{deltax2timesdeltap2}
\delta_x^2\delta_p^2&=\frac{\delta_x^2d_{xx}-\delta_p^2d_{pp}}{c_{xp}}-1.
%\end{split}
\end{align}
To simplify notation, we have introduced the normalized damping $\Gamma\equiv 2\hbar\gamma/(m\Omega^2a^2)$, the adimensional variables $c_{xp}=2C_{xp}/(m\gamma\Omega)$ (and similarly for $c_{pp},\,d_{xp},\,\ldots$), and the quantity $\zeta=1/(1+2\Gamma d_{xp})$.

% #### #### #### #### #### #### #### ####
\begin{figure*}
\subfloat{ \label{fig:HeisenbergLimitQuadraticWithCpp} \includegraphics[width=\linewidth]{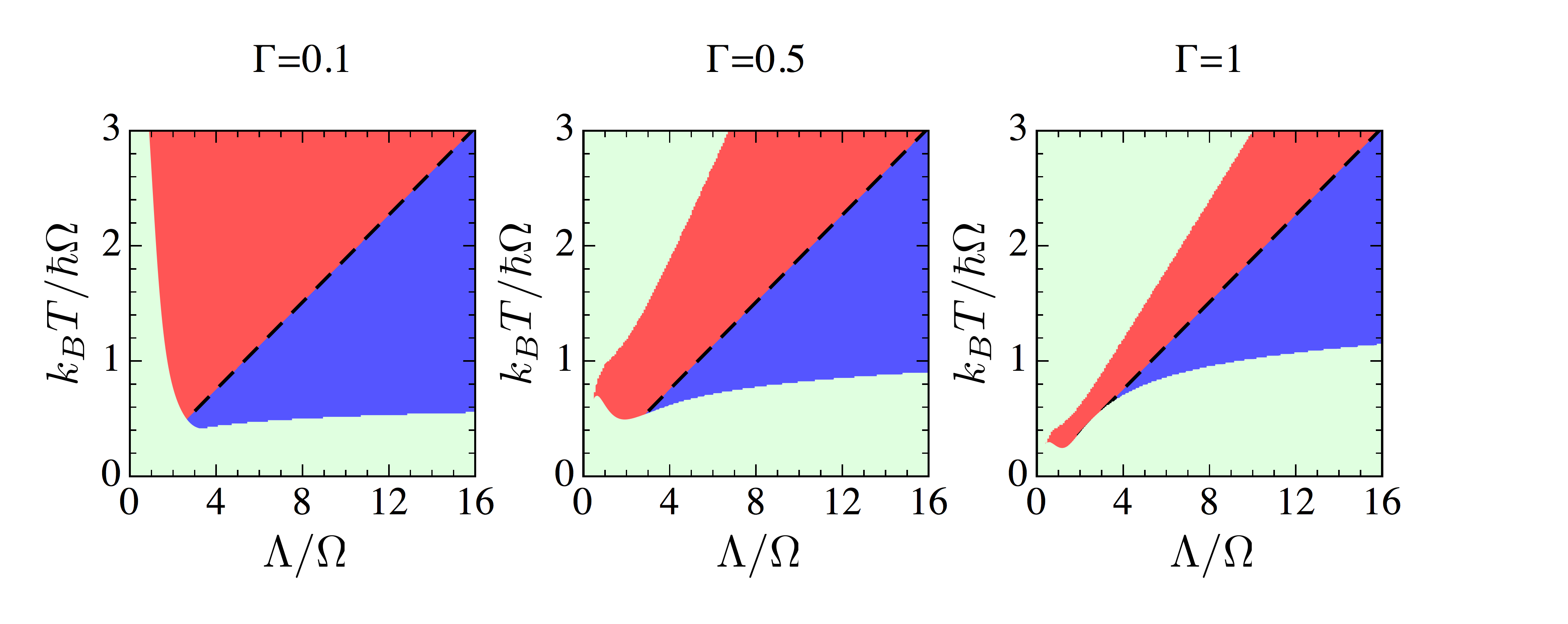}}\\
\subfloat{\label{fig:SqueezingQuadraticWithCpp} \includegraphics[width=\linewidth]{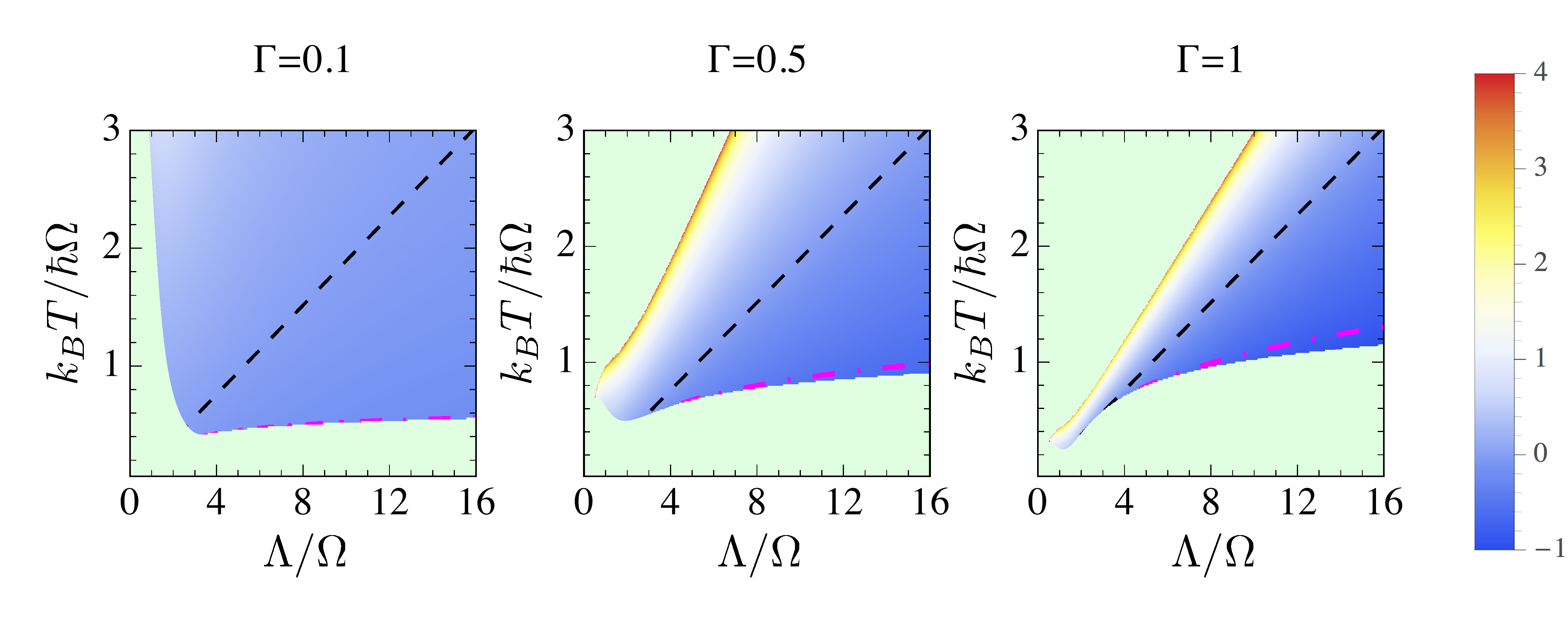}}\\
\subfloat{\label{fig:quadEigs}\includegraphics[width=\linewidth]{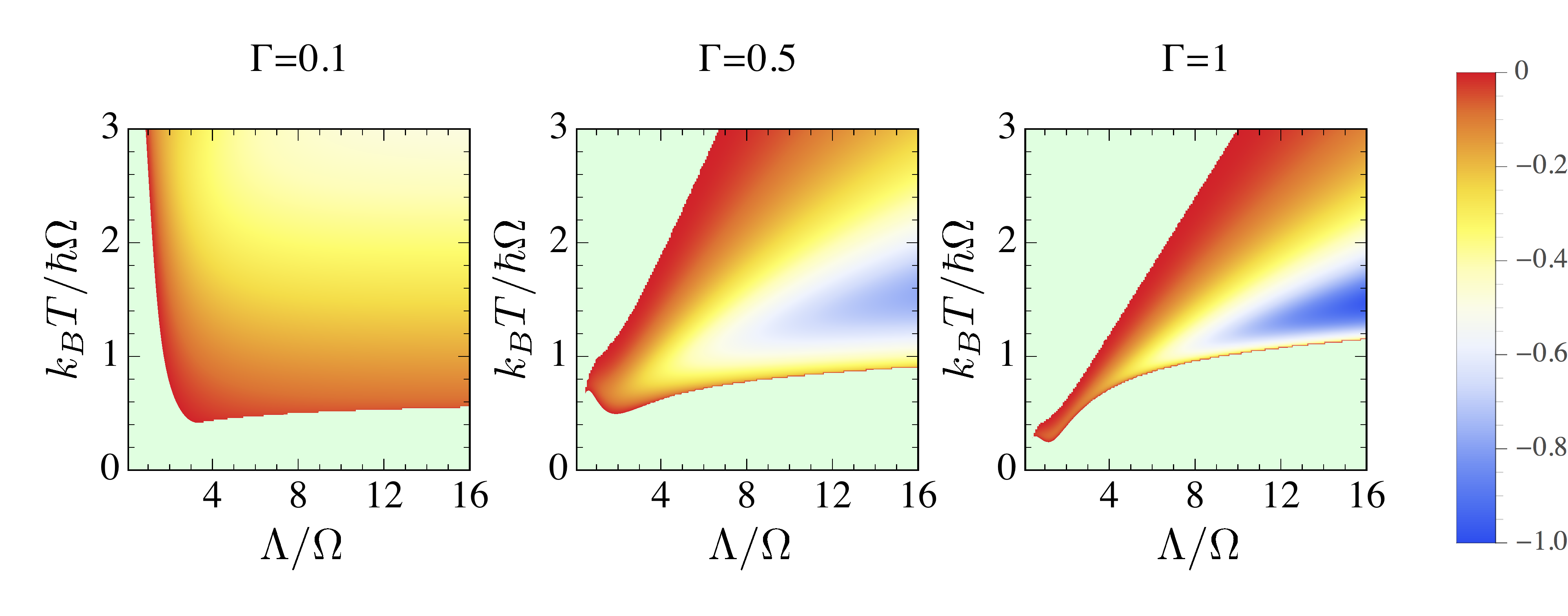}}
\caption{
\label{fig:quadCase}
(Color online)
Phase diagram of our equation for a quadratic coupling, under the self-consistent Gaussian approximation. From left to right, plots are for $\Gamma=0.1,\,0.5,\,1$. 
Top {\bf (a)}: the gas experiences an effective ``cooling" in the blue regions, and an effective ``heating" in the red regions. 
Center {\bf(b)}: density plot of the logarithm of the aspect ratio $\log(\delta_x^2/\delta_p^2)$.
Bottom {\bf(c)}: maximum of the real part of the eigenvalues of the matrix of coefficients of the linear system defined in Eq.\ \eqref{dynamics-linearized}.
In the green regions, one of the validity conditions is violated, i.e., either the Heisenberg principle is not satisfied, or one of the eigenvalues of the stability equations becomes positive, or fluctuations $\delta_x^2$ and $\delta_p^2$ are complex numbers.
The black dashed lines are the boundaries of unity aspect ratio, where $\delta_x^2=\delta_p^2$.
In this way, we see we have ``cooling" for  $\delta_x^2/\delta_p^2<1$, and ``heating" for $\delta_x^2/\delta_p^2>1$.
We have quantum squeezing with $\delta_x^2<1$ below the magenta dot-dashed lines, while $\delta_p^2$ is never smaller than 1 in the allowed region.  }
\end{figure*}
% #### #### #### #### #### #### #### ####

The two coupled equations \eqref{deltax2overdeltap2} and \eqref{deltax2timesdeltap2} may be combined to obtain a single quadratic equation determining, e.g., $\delta_x^2$, from which we may then extract $\delta_p^2$. The quadratic equation has two solutions, and the correct one may selected by looking at its behaviour in the regime $\Omega\ll k_B T/\hbar\ll \Lambda$. The (-) solution unphysically tends towards zero there. On the other hand, the (+) solution correctly yields $\delta_x^2\sim 2k_BT/\hbar\Omega$, i.e., an effective temperature $\tilde T\sim T$. At odds with the linear case, however, $\tilde T$ strongly deviates from $T$ when $T\sim O(\Lambda/\Omega)$.

A detailed phase diagram for the present case of quadratic coupling is presented in Fig.\ \ref{fig:quadCase}. 
The Heisenberg principle requires $\delta_x \delta_p\geq 1$, a condition which gives rise to a minimal acceptable temperature which grows as $T_{min}\sim\log(\Lambda)$ for large $\Lambda/\Omega$, in close analogy to the linear case. The Heisenberg bound is shown in Fig.\ \ref{fig:HeisenbergLimitQuadraticWithCpp}, together with the region where the gas experiences an effective heating, or cooling, with respect to its Gibbs-Boltzmann counterpart.

 The corresponding degree of deformation of the phase-space distribution, as measured by the logarithm of the aspect ratio $\log(\delta_x^2/\delta_p^2)=\log(\sigma_p/\sigma_x)$, is shown in Fig.\ \ref{fig:SqueezingQuadraticWithCpp}. At small temperatures, we observe the emergence of a region (below the magenta, dot-dashed lines) where $\delta_x^2<1$, i.e., of {\it genuine quantum squeezing}.
 Notice that, for damping $\Gamma\gtrsim 0.1$, at large temperatures the aspect ratio of the distribution displays a very sharp increase; beyond a certain point, the solution of Eqs.\ \eqref{deltax2overdeltap2} and \eqref{deltax2timesdeltap2} yields a value for the fluctuations $\delta_x^2$ which diverges and turns negative, a clearly unphysical feature signaling the breakdown of the Gaussian Ansatz in that region.

It may be noticed by comparing Figs.\ \ref{fig:HeisenbergLimitQuadraticWithCpp} and \ref{fig:SqueezingQuadraticWithCpp}  that, as in the linear case, the Gibbs-Boltzmann boundary coincides with the one of unit aspect ratio, a condition which again is independent of $\Gamma$. This may be explicitly checked by employing the trial GB solution $\delta_x^2=\delta_p^2=\coth(\hbar\Omega/2k_B T)$, which is an identical solution of Eq.\ \eqref{deltax2timesdeltap2} for every $\{\Lambda,\Omega,T\}$, and a solution of Eq.\ \eqref{deltax2overdeltap2} for every $\Gamma$ provided that $T=\alpha_{(2)}\Lambda+O(\Omega/T)$, with $\alpha_{(2)}\approx 0.189$ satisfying the implicit equation
\beq
\pi\alpha_{(2)}+2[{\rm Di}\Gamma(1/2\pi\alpha_{(2)})+\tilde\gamma]=0.
\eeq

At odds with the linear case seen above, the equations for a quadratic coupling determine the two ratios $\delta_x^2\propto\tilde T/\sigma_x$ and $\delta_p^2\propto\tilde T/\sigma_p$, but do not provide an explicit expression for $\tilde T$, $\sigma_x$ and $\sigma_p$ separately, leaving therefore open various possible applications of this theory.

 As an example, we may fix $\tilde T$ in accordance to the standard formula for the
quantum mechanical harmonic oscillator, Eq.\ \eqref{effTempRefined}, and then interpret $\sigma_p$ and $\sigma_x$ as quantum corrections to the inverse mass $1/m$ and
the spring constant $m\Omega^2$. Such ``renormalization" should
be used if we considered the starting model as a fundamental
quantum field theoretic construct. 

 Alternatively, one may set, say, $\sigma_p=1$, and consider quantum
modification of the effective temperature,  and the spring
constant. From Eq.\ \eqref{deltax2overdeltap2} one finds in this way
\beq
k_B\tilde T=\frac{\hbar\Omega}{2}\frac{\delta_x^2/\zeta-\Gamma c_{pp}}{1-\Gamma c_{pp}\delta_x^2/2}.
\label{eq:effectiveTemperatureQuadratic}
\eeq

\mac{Similarly as in the case of the {\it linear coupling}, one needs to examine the nature of  Heisenberg uncertainty pathologies in the present quadratic case.  Obviously, the exact stationary state should not violate the Heisenberg uncertainty inequality. In the quadratic case, however, the exact solution is not known, and the results of Ref.~\cite{FlemPRE2011}  cannot be applied directly. 
The pathologies  may  result from solutions being of mixed order as in  Ref.~\cite{FlemPRE2011}, or from the non-Gaussian form of the unknown exact solution.  In any case the pathologies signal the invalidity  of applied approximations and offer 
a reasonable bound for 
the degree of cooling and squeezing in  the considered quantum stochastic process.}

\section{Near-equilibrium dynamics in self-consistent Gaussian approximation}
\label{sec:nearEqDyn}

In the last Section before Conclusions, we investigate the
near-equilibrium dynamics and stability of stationary solutions
found in the previous Section. We use the self-consistent Gaussian
approximation, which actually is exact in the case of linear
coupling provided the initial state was Gaussian.

\subsection{Linear case}

It is elementary to derive the equations for the first and second
moments of the Wigner distribution -- these moments characterize
the Gaussian state fully, and in the linear case form two closed
systems of linear equations:
\begin{align}
\dot{\ave{x}}=& \ave{p}/m, \\
\nonumber \dot{\ave{p}} = &-m\Omega^2 \ave{x}  -\frac{2C_p}{m\Omega}\ave{p},
\end{align}
and
\begin{align}
\dot{\ave{x^2}}=& 2\ave{xp}/m, \\\nonumber
\dot{\ave{xp}} = &\frac{\ave{p^2}}{m} -m\Omega^2 \ave{x^2} -\frac{2C_p}{m\Omega} \ave{xp} - \frac{\hbar D_p}{m\Omega},\\\nonumber
\dot{\langle p^2 \rangle}=&  -2m\Omega^2 \ave{xp} -\frac{4C_p}{m\Omega} \langle p^2 \rangle + 2\hbar
D_x.
\end{align}
Clearly, the solutions tend to their stable stationary values,
$\ave{x}_{\rm st}=\ave{p}_{\rm st}=\ave{xp}_{\rm st}=0$, $\ave{p^2}_{\rm st} =\hbar m\Omega D_x/2C_p$, and $(m^2\Omega^2)\ave{x^2}_{\rm st}=\hbar(m\Omega D_x/2C_p -D_p/\Omega)$. The only constraint is imposed by the Heisenberg principle
\beq
\frac{m\Omega^2\ave{x^2}}{2} \frac{\langle p^2 \rangle}{2m} \ge \left(\frac{\hbar \Omega}{4}\right)^2.
\eeq
The equations for $\ave{x^2}_{\rm st}$ and $\ave{p^2}_{\rm st}$ and the resulting Heisenberg bound
coincides with the one found for $\sigma_x,\,\sigma_p,\,\textrm{and }\delta_x\delta_p$ in Sec.\ \ref{sec:WignerStationaryLinear}, a fact which should not surprise, as we have seen that a Gaussian Ansatz was providing an exact solution of the problem.

\subsection{Quadratic case}

 In this case, the Gaussian Ansatz provides only an approximate solution. Again, the first and second
moments of the Wigner distribution characterize the Gaussian state
fully, but this time they couple to higher moments, so that
Wick (Gaussian) de-correlation techniques have to be used. We
obtain for the first moments
\begin{align}
\dot{\ave{x}}=&\ave{p}/m, \\\nonumber
 \dot{\ave{p}} = &  -m\Omega^2 \ave{x} -\frac{8C_{xp}}{m\Omega a^2} \ave{x^2p}
 -\frac{4C_{pp}}{(m\Omega a)^2} \ave{xp^2} \\\nonumber
 &- \frac{4\hbar D_{xp}}{m\Omega a^2}\ave{x} - \frac{4\hbar D_{pp}}{m^2\Omega^2a^2}\ave{p}.
\end{align}
The Wick's theorem allows to replace $\ave{x^2 p}=\ave{x}^2\ave{p} + 2\langle \Delta_x\Delta_p \rangle
\ave{x} + \langle \Delta_x^2 \rangle \ave{p}$, and similarly for $\ave{xp^2}$,
where we represent the Gaussian random variables $x =\langle x\rangle +\Delta_x$, $p=\langle p\rangle+\Delta_p$. We obtain thus
\pie{
\begin{align}
 \dot{\ave{p}} = &  -m\Omega^2\ave{x} 
 -\frac{8C_{xp}(\langle x \rangle^2+\langle \Delta_x^2 \rangle)}{m\Omega a^2} \ave{p}
 \nonumber\\
 &-\frac{4C_{pp}(\ave{p}^2+\ave{\Delta_p^2})}{m^2\Omega^2 a^2} \ave{x}
 - \frac{4\hbar D_{xp}}{m\Omega a^2}\ave{x}
 \nonumber\\
&
 - \frac{4\hbar D_{pp}}{m^2\Omega^2a^2}\ave{p}
 -\frac{8\ave{\Delta_x\Delta_p}}{m^2 \Omega^2 a^2}(C_{pp}\ave{p}+2m\Omega C_{xp}\ave{x}).
\end{align}
}
These equations have a stable stationary solution $\ave{x}_{\rm st}=\ave{p}_{\rm st}=0$, provided that they describe a damped harmonic oscillator. If such a solution exists, in its vicinity we may identify $\ave{\Delta_x^2}_{\rm st}=\ave{x^2}_{\rm st}=\delta_x^2\hbar/(2m\Omega)$ and $\ave{\Delta_p^2}_{\rm st}=\ave{p^2}_{\rm st}=\hbar m \Omega\delta_p^2/2$ (since by hypothesis the first moments are zero), and we may neglect the quadratic terms $\ave{x}^2$ and $\ave{p}^2$ and the crossed fluctuation term $\ave{\Delta_x\Delta_p}$, to obtain the two simultaneous conditions
%\begin{align}
\pie{
\beq
1+\Gamma d_{xp}+\Gamma c_{pp}\delta_p^2 /2\ge 0,
\qquad
c_{xp}\delta_x^2 +d_{pp}\ge 0
\label{stability-of-first-moments}
\eeq
}
These, in turn, depend self-consistently on the equations for the second moments,
\begin{align}
\label{dynamics-nl}
\dot{\ave{x^2}}=& \frac{2}{m}\ave{xp}, \\\nonumber
\dot{\ave{xp}} = & 
 \frac{\ave{p^2}}{m} -m\Omega^2 \ave{x^2} 
-\frac{8}{m\Omega a^2}\left[C_{xp}\ave{x^3p}+\hbar D_{xp}\ave{x^2}\right]
%-\frac{8}{m\Omega a^2}
\\\nonumber
&-\frac{1}{m^2\Omega^2 a^2}\Big[C_{pp}\Big(4\ave{x^2p^2}-2\hbar^2\Big)+8\hbar D_{pp}\ave{xp}\Big]
%-\frac{8\hbar D_{pp}}{m^2\Omega^2 a^2}
,\\\nonumber
\dot{\ave{p^2}}=&  -2m\Omega^2 \ave{xp} 
-\frac{4C_{xp}}{m\Omega a^{2}} \Big(4\ave{x^2p^2} +\hbar^2\Big)\\\nonumber
&-\frac{8C_{pp}}{m\Omega a^{2}}\ave{xp^3}
+\frac{8\hbar D_{xx}}{a^2}\ave{x^2} 
-\frac{8\hbar D_{pp}}{m^2\Omega^2 a^2}\ave{p^2}.
\end{align}
From the first equation, we see that if a stable stationary solution exists then $\ave{xp}_{\rm st}=0$. 
 The quartic terms may be decomposed as above, using the Wick's method, and in this way one may compute
the stationary solution. A straightforward calculation then shows that in the stationary state $\ave{x^2}_{\rm st}$ and $\ave{p^2}_{\rm st}$ satisfy the same two equations found in the preceding Section, Eqs.\ \eqref{deltax2overdeltap2} and \eqref{deltax2timesdeltap2}. To check the stability of the  steady-state, we write $\langle
x^2 \rangle= \ave{x^2}_{\rm st} + \Delta_{x^2}$, $\langle
p^2 \rangle= \langle p^2 \rangle_{\rm st} + \Delta_{p^2}$, $\langle
xp \rangle= \Delta_{xp}$, and perform linear stability analysis
in $\Delta$'s,
\begin{align} \label{dynamics-linearized}
\partial_t(\Delta_{x^2})=& \frac{2}{m} \Delta_{xp} \\\nonumber
\partial_t(\Delta_{xp}) = & \frac{\Delta_{p^{2}}}{m}  -m\Omega^2 \Delta_{x^2}  
-\frac{24C_{xp}\ave{x^2}_{\rm st} \Delta_{xp} + 8 \hbar D_{xp} \Delta_{x^2}}{m\Omega a^2} \\\nonumber
&-\frac{4C_{pp}(\ave{x^2}_{\rm st} \Delta_{p^2}+\ave{p^2}_{\rm st} \Delta_{x^2})+8\hbar D_{pp}\Delta_{xp}}{m^2\Omega^2 a^2} 
\\\nonumber
\partial_t(\Delta_{p^2})=&  -2m\Omega^2 \Delta_{xp} 
-\frac{16C_{xp}}{m\Omega a^2} [\langle p^2 \rangle_{\rm st}\Delta_{x^2} +\ave{x^2}_{\rm st}\Delta_{p^2} 
]\\\nonumber
&-\frac{24C_{pp}}{m^2\Omega^2 a^2}\ave{p^2}_{\rm st}\Delta_{xp}
+\frac{8\hbar D_{xx}}{a^2}\Delta_{x^2}  
-\frac{8\hbar D_{pp}}{m^2\Omega^2 a^2}\Delta_{p^2}.
\end{align}
The stability requires that the real parts of all eigenvalues of
the matrix governing the above linear evolution have to be
negative, i.e., have to describe damping. 
Numerical analysis of the eigenvalues of this matrix is presented in Fig. \ref{fig:quadEigs}. The plot indicates that all eigenvalues are negative in most of the region of
existence of the physically sound Gaussian stationary solution, but at the same time that the region of validity rapidly shrinks with increasing damping $\Gamma$.
To resume, regions colored in green are not accessible by the system because either the normalized standard deviations $\delta_x^2$ and $\delta_p^2$ have an unphysical imaginary part, or they do not satisfy the Heisenberg bound $\delta_x^2\delta_p^2\geq 1$, or the equations for the first moments do not describe a damped harmonic oscillator (i.e., inequalities in \eqref{stability-of-first-moments} are not satisfied), or at least one of the eigenvalues of the linear stability matrix of the second moments \eqref{dynamics-linearized} becomes positive.

Note that besides the stability question, Eqs. \eqref{dynamics-nl}
and \eqref{dynamics-linearized} incorporate quantum dynamical
effects: they describe dynamics clearly different from their high
$T$ classical analogues, due to the quantum form/origin of the
diffusion coefficients $D_{xx}$, $D_{xp}$ and $D_{pp}$.

Finally, let us comment about the  large prohibited region we find in the quadratic
case at large $T$. This region is generally dynamically unstable, and  arises because of the diverging fluctuations in $x$ caused by a large Lamb-shift of the effective trap frequency, which turns the attractive harmonic potential into an effectively repulsive one.  
It is reasonable to expect that this region would become allowed if we added a quartic term to the confinement, on top of the usual quadratic one. Indeed, Hu, Paz and Zhang considered only this case, for non-linear couplings \cite{Hu1993}. However, traps for ultracold atoms are generally to a very high approximation purely quadratic in the region where the atoms are confined, so that the presence of a quartic component may be unjustified in a real experiment.

\section{Conclusions}
\label{sec:conclusions}

We have presented in this paper a careful discussion of quantum Brownian motion in the case when the reservoir exhibits an energy cutoff $\hbar\Lambda$ much larger than other energy scales. We considered a Brownian particle in a harmonic trap, and  derived and discussed validity of QME in this limit for the case of linear and various forms of nonlinear couplings to the bath. We have pointed out that stationary distributions exhibit elliptical deformations, and in the case of non-linear coupling even genuine quantum squeezing along $x$.

An ideal application of this theory would be the study of the properties of impurity atoms embedded into a Bose-Einstein condensate or an ultracold Fermi gas. A possible detection of predicted effects would require to: i) embed a dilute and weakly-interacting gas of impurities in a degenerate ultracold gas; ii) monitor the stationary distribution of impurities; iii) eventually, monitor their approach toward equilibrium. The application of our theory to such situations may be implemented along the lines sketched in Appendix \ref{sec:impurityInATrappedGas}.

Another interesting question concerns the Smoluchowski-Kramers limit \cite{Smoluchowski,Kramers}, which can be considered as a regime of over-damped quantum Brownian motion, or the case where the mass $m$ of the Brownian particle tends to zero.
This limit is already highly non-trivial at the classical level, in the presence of the inhomogeneous damping and diffusion, and it requires a careful application of homogenization theory (cf.\ \cite{Hottovy2012,Hottovy2014,papa-book,Pavliotis}). 
Of course, the theoretical approach here is based on the separation of time scales, and has been in other contexts studied in the theory of classical and quantum stochastic process \cite{Gardiner-cl,Risken}.  In particular, the theory of adiabatic elimination has been developed to include the short time non-Markovian ``initial slip" effects and the effective long time dynamics of the systems and the bath (``adiabatic drag") (cf.\ \cite{Haake-ZfP,MLFH,Reibold1} and references therein).

The Smoluchowski-Kramers (SK) limit was also intensively studied in the contexts of Caldeira-Leggett model and quantum Brownian motion (cf.\ \cite{Maier2010,Ankerhold1,*Ankerhold2} and references therein). The problem with this limit is that it corresponds to strong damping, and evidently cannot be described using weak coupling approach that is normally used to derive the QME from the microscopic model in the Born-Markov approximation. We envisage here two possible and legitimate lines of investigation.

One can forget about the microscopic derivation, and take the Born-Markov QME as a starting point. The SK limit corresponds then to setting the spring constant $m\Omega^2$ and friction $\eta$ to constants, and letting the mass $m\rightarrow  0$, so that $\gamma \rightarrow \infty$ as $1/m$ and $\Omega \rightarrow \infty$ as $1/\sqrt{m}$. The aim is to eliminate the fast variable (the momentum) and to obtain the resulting equation for the position of the Brownian particle; again, the Wigner function formalism is particularly suited for such a task.

More ambitious and physically more sound is the approach in which the microscopic model is treated seriously, and appropriate scalings are introduced at the microscopic level. One can then start, for instance, from the formally exact path integral expression for the reduced dynamics, as pursued by Ankerhold and collaborators
\cite{Maier2010,Ankerhold1,Ankerhold2}. The other possibility is to use a restricted version of the weak coupling assumption, only demanding that the system does not influence the bath, and use Eq.\ \eqref{preRedfield} combined with Laplace transform techniques and Zwanzig's approach \cite{Zwanzig}.

To our knowledge, neither of the two above proposed research tasks has been so far realized for the case of inhomogeneous damping and diffusion.

Last, but not least we must bear in mind that the QMEs derived and discussed in this work suffer from the fact that they do not, in general, have the Lindblad form, and thus their solutions are not guaranteed to correspond to physically sound, non-negatively defined density matrices. One should stress that, similarly as in the case of the (in)famous sign problem in the Monte Carlo studies of many-fermion systems, these solutions still may serve very well as generators of averages and moments, as long as the negative part of the density matrix is relatively  small with respect to the positive part (in any "reasonable" matrix norm). If this is not the case,
or just for formal reasons, one may add artificially ``minimal" terms that assure the Lindblad form of the master equation \cite{BreuerBook,SchlosshauerBook,Gao1997,Diosi1993}. It would eventually be very interesting to generalize these methods to the QMEs describing inhomogeneous damping and diffusion, and to see how these terms affect the stationary solutions and dynamics discussed in this paper.

\acknowledgments Insightful discussion with Alessio Celi, Eugene Demler, Maria Garc\'ia Parajo, Crispin Gardiner, John Lapeyre, Carlo Manzo, Piotr Sza\'nkowski,  Marek
Trippenbach and Giovanni Volpe are gratefully acknowledged. This
research has been supported through ERC Advanced Grant OSYRIS (led by
M.L.),  EU IP SIQS, EU STREP EQuaM, John Templeton Foundation, and
Spanish Ministry Project FOQUS. P.M.\ acknowledges funding from a Spanish MINECO ``Ram\'on y Cajal" fellowship.

\appendix

\section{Markovian QME for generic coupling}
\label{sec:genericCoupling}

We consider here an interaction term with a completely general coupling in the position of the particle:
\pie{
\begin{equation}
H_{int}=\sum_{k}\kappa_{k}\sqrt{\frac{\hbar}{2 m_k \omega_{k}}}f(x)\left(g^{\dagger}_{k}+g_{k}\right).
\end{equation}
}
If $f\in\mathcal{C}^{\infty}(I)$ and thus may be expanded in
Taylor series, the master equation can be written in the form:
\begin{widetext}
\beq
\dot{\rho}=-i\comm{H_{S}}{\rho}-\sum^{\infty}_{j,n=0}\sum^{n}_{k=0}\frac{\tilde{f}^{(j)}\tilde{f}^{(n)}}{a^{j+n-2}j!n!(m\Omega)^{k}}\left[x^{j},\frac{i C_{n,k}}\hbar\{\sigma(x^{n-k}p^{k}),\rho\}+\frac{D_{n,k}}{\hbar}\left[\sigma(x^{n-k}p^{k}),\rho\right]\right],
\label{QME_generic_coupling}
\eeq
\end{widetext}
 where $\sigma(x^{m}p^{k})$ is the sum of the $\frac{(m+k)!}{m!k!}$ distinguishable permutations of the $m+k$ operators in the polynomial $x^{m}p^{k}$ [e.g.,
$\sigma(x^{2}p)=x^{2}p+xpx+px^{2}$]. In analogy with the preceding
sections, we have introduced here
\begin{align}
C_{n,k}(\Omega)=&(-1)^{k+1}\int_0^\infty{\rm
d}\tau\,\eta(\tau)\cos^{n-k}(\xi)\sin^{k}(\xi)\\\nonumber
D_{n,k}(\Omega)=&(-1)^{k}\int_0^\infty{\rm
d}\tau\,\nu(\tau)\cos^{n-k}(\xi)\sin^{k}(\xi)
\end{align}
where $\xi=\Omega\tau$. These integrals may be calculated by Laplace transformation, as detailed in Appendix \ref{sec:LaplaceTransforms}. Alternatively, we will outline in Appendix \ref{sec:trigIdentities} a simpler method which employs standard trigonometric identities
to straightforwardly reduce every $C_{n,k}$ to a linear combination of $C_x$ and $C_p$ (the ones computed in the linear case), and similarly every $D_{n,k}$ in terms of $D_x$ and $D_p$. As an example, since $\cos^3(\xi)\sin(\xi)=[2\sin(2\xi)+\sin(4\xi)]/8$, it is obvious that
$D_{4,1}(\Omega)=[2D_p(2\Omega)+D_p(4\Omega)]/8$.

In complete analogy with the linear and quadratic cases, for a power law coupling with $f(x)=a(x/a)^n$ the coefficient $D_{n,0}$ determines the decoherence in the position basis, which for a quantum superposition of two states centered respectively at $x$ and $x'$ happens with a characteristic rate $\gamma_{x_1,x_2}^{(n)}=D_{n,0}(x_1^n-x_2^n)^2/\hbar a^{2n-2}$. As a consequence, for an even more general coupling containing various powers of $(x/a)$, the total decay rate in position space reads
\beq
\gamma_{x_1,x_2}=\sum^{\infty}_{j,n=0}\frac{\tilde{f}^{(j)}\tilde{f}^{(n)}D_{n,0}(x_1^n-x_2^n)^2}{\hbar j!n!a^{j+n-2}}.
\eeq
In contrast with Ref.\ \cite{Hu1993}, we find here that quantum superpositions which are sharply localized at positions symmetric with respect to the origin (e.g., in the vicinity of, say, $x_0$ and $-x_0$) will be characterized by a vanishing decoherence rate (i.e., a diverging lifetime) in presence of couplings which contain only even powers of $n$.

\subsubsection{Large cut-off limit (general case)}
In the limit $\Lambda\gg T,\Omega$, we find:
\begin{itemize}
\item $C_{n,k}\propto\Lambda^{1-k}$, such that at every order $n$ the only divergent term is linear, and it is the one which may be re-absorbed in the Hamiltonian; indeed, $C_{n,0}$ is the coefficient in front of the term $i[x^n,\{x^n,\rho\}]=i[x^{2n},\rho]$, so that the divergent term is cancelled by taking $H_{\rm sys}=H_S-C_{n,0}f(x)^2$.  Moreover, for every $n$ we have $C_{n,1}=m\gamma\Omega/2$.
 \item between the coefficients $D_{n,k}$, only the term with $k=1$ diverges, logarithmically as  $D_{n,1}\sim \frac{m\gamma\Omega}{\pi}\log\left(\frac{\hbar\Lambda}{2\pi k_BT}\right)+\ldots$. All terms with $k\neq 1$ are instead finite.
 \end{itemize}

\subsubsection{High-temperature limit (general case)}
In the high-temperature limit $k_BT\gg\Lambda\gg\Omega$, the
coefficients $C$ are as in the large-cutoff limit, as they do
not depend on $T$. In the set of $D$ coefficients, only
$D_{n,0}\sim m\gamma k_BT/\hbar$ remains finite, while all others
go to zero.
Using the identity $\sigma(x^{n-1}p)=n\{x^{n-1},p\}/2$, it is easy to show that the master equation \eqref{QME_generic_coupling} reduces to \eqref{NonLinearCaldeiraLeggettMasterEq} at high temperatures. In this {\it classical} limit, we see that in presence of a non-linear coupling the coefficients of the QME satisfy a generalized fluctuation-dissipation theorem, since for any $n$ we have $D_{n,0}/C_{n,1}\approx2k_BT/\hbar\Omega$.

\section{Laplace transforms}
\label{sec:LaplaceTransforms}
Here we show how to compute the coefficients of the QME with a generic coupling  by direct Laplace transform. We have 
\begin{align}
C_{n,k}&(\Omega)=(-1)^{k+1}\frac{m\gamma\Lambda^2}{2}\mathcal{L}[\cos^{n-k}(\xi)\sin^{k}(\xi)]_{\Lambda}\\\nonumber
D_{n,k}&(\Omega)=\frac{mk_BT\gamma\Lambda^2}{\hbar}\sum^{+\infty}_{p=-\infty}\frac{1}{\Lambda^2-\nu_{p}^2}\\
&\left(\Lambda\mathcal{L}[\cos^{n-k}(\xi)\sin^{k}(\xi)]_{\Lambda}-|\nu_{p}|\mathcal{L}[\cos^{n-k}(\xi)\sin^{k}(\xi)]_{|\nu_{p}|}\right),
\end{align}
where $\mathcal{L}[a(\xi)]_s=\int_0^\infty {\rm
d}\xi\,a(\xi)e^{-s\xi}$ stands for the Laplace transform of
$a(\xi)$ with respect to the variable $s$. Using the following
identity, valid for $s> 0$,
\begin{multline}
\mathcal{L}\left[\cos^{(n-k)}(\xi)\sin^{(k)}(\xi)\right]_s=
\nonumber\\
=\sum^{n-k}_{l=0}\sum^{k}_{j=0}(-1)^{j+k}\frac{i^{k}}{2^{n}}\binom{n-k}{l}\binom{k}{j}\mathcal{L}\left[e^{i[n-2(j+l)]\xi}\right]_s=\nonumber\\
=\sum^{n-k}_{l=0}\sum^{k}_{j=0}(-1)^{j+k}\frac{i^{k}}{2^{n}}F_{njl}(s)\label{Laplace},
\end{multline}
with 
\beq
\nonumber
F_{njl}(s)\equiv\binom{n-k}{l}\binom{k}{j}\frac{1}{s-i[n-2(j+l)]\Omega},
\eeq
one readily finds 
\beq
C_{n,k}=\frac{m\gamma\Lambda^2}{2}\sum^{n-k}_{l=0}\sum^{k}_{j=0}(-1)^{j+1}\frac{i^{k}}{2^{n}}
F_{njl}(\Lambda).
\eeq
 In the expression for $D_{n,k}$, the zero Matsubara-frequency term
should must be treated separately, so that one obtains:
\begin{multline}
D_{n,k}=\frac{i^{k}}{2^{n}}\frac{mk_BT\gamma}{\hbar}\sum^{n-k}_{l=0}\sum^{k}_{j=0}(-1)^{j}\Bigg\{\Lambda F_{njl}(\Lambda)\\
+2\sum^{+\infty}_{p=1}\frac{\Lambda^2}{\Lambda^2-\nu_{p}^2}\left[\Lambda F_{njl}(\Lambda)-\nu_p F_{njl}(\nu_p)\right]\Bigg\}
\end{multline}

\section{Trigonometric identities}
\label{sec:trigIdentities}
  The identities presented here provide a very simple method (alternative to the one described in App.\ \ref{sec:LaplaceTransforms}) to compute the $2n+2$ coefficients needed to describe the QME for an arbitrary coupling $f(x)\propto x^n$ in terms of just the two integrals $I_\nu\equiv\int_0^\infty  \diff
\tau\,\nu(\tau)$ and $I_\eta\equiv\int_0^\infty  \diff
\tau\,\eta(\tau)$, and of the four coefficients $\{C_x,C_p,D_x,D_p\}$ we derived for a linear coupling.
 Take $p+q=n$.

Whenever $p$ is even (or zero), we have
 \begin{multline}
 \sin^{p}(x)\cos^{q}(x)=[1-\cos^{2}(x)]^{p/2}\cos^{q}(x)\\
 =c_0+\sum_{k=0}^{{\mathcal F}[(n-1)/2]}\alpha_k\cos[(n-2k)x],
 \end{multline}
 where ${\mathcal F}(x)$ is the "floor" function (giving the greatest integer less than or equal to x), and $c_0$ and $\{\alpha_k\}$ are constants which may be determined using the power reduction trigonometric formulas \cite{Gradshteyn2000}. As an example, we find
 \begin{multline}
 \sin^{2}(x)\cos^{3}(x)=\frac{3\cos(x)+\cos(3x)}{4}\\
 -\frac{10\cos(x)+5\cos(3x)+\cos(5x)}{16}
 \end{multline}
 This formula reduces high powers of the trigonometric quantity to a sum of cosine-functions of multiples of its argument, thereby reducing the desired integrals to known ones.

 Similarly, whenever $q$ is even (or zero), we have
\begin{multline}
\sin^{p}(x)\cos^{q}(x)=\sin^{p}(x)[1-\sin^{2}(x)]^{q/2}\\
 =c_0+\sum_{k=0}^{{\mathcal F}[(n-1)/2]}\alpha_k\sin[(n-2k)x].
\end{multline}

 In the case where both $p$ and $q$ are odd integers, we may write
 \begin{multline}
 \sin^{p}(x)\cos^{q}(x)=\sin(x)\cos(x)[1-\cos^{2}(x)]^{\frac{p-1}{2}}\cos^{q-1}(x)\\
 =\frac{\sin(2x)}{2}\left[c_0+\sum_{k=0}^{{\mathcal F}[(n-3)/2]}\alpha_k\cos[(n-2k)x]\right],
 \end{multline}
and the resulting integrals may be computed using the simple identity, valid for $n>0$,
 \beq
 \sin(2x)\cos(2nx)=\frac{\sin[(2n+2)x]-\sin[(2n-2)x]}{2}.
 \eeq
 
 \section{Asymptotic values of the QME coefficients for the linear and quadratic cases}
 \label{sec:asymptoticTable}
 We provide here below a table summarizing the asymptotic values of the coefficients of the QME for an Ohmic spectral with a Lorentz-Drude cutoff, in presence of linear and quadratic couplings, and in various interesting limits. For simplicity of notation, we give here the values of the dimensionless quantities $c_{\ldots}\equiv 2C_{\ldots}/(m\gamma\Omega)$ (and similarly for $d_{\ldots}$). In the central column, $\hbar\Omega$ and $k_B T$ are assumed to be of the same order of magnitude, and both much smaller than $\hbar\Lambda$.
 
 The coefficients for a linear coupling read:
 \setlength{\tabcolsep}{4pt}
 \begin{equation}\nonumber
 \begin{tabular}[c]{l|c|c|c|}
 %\hline
 &$\frac{k_BT}{\hbar}\gg\Lambda\gg\Omega$ & $\Lambda\gg %\Omega,\frac{k_BT}{\hbar}$;$
 \Omega\sim\frac{k_BT}{\hbar}$ & $\Lambda\gg\Omega\gg\frac{k_BT}{\hbar}$\\
 \hline
 $c_{x}$ & $-\Lambda/\Omega$   &  $-\Lambda/\Omega$ &  $-\Lambda/\Omega$\\
 \hline
 $c_{p}$ & 1  &  1 & 1\\
 \hline
 $d_{x}$ & $\frac{2 k_BT}{\hbar\Omega} $  & $\coth\left(\frac{\hbar\Omega}{2k_BT}\right)$ & $1$\\
 \hline
 $d_{p}$ & $-\frac{2 k_BT}{\hbar\Lambda}$  & $\frac{2}{\pi}\log\left(\frac{\hbar\Lambda}{2\pi k_BT}\right)$ & $\frac{2}{\pi}\log\left(\frac{\Lambda}{\Omega}\right)$\\
 \hline
 \end{tabular}
 \end{equation}

 The coefficients for a quadratic coupling instead read:
 \begin{equation}\nonumber
 \begin{tabular}{l|c|c|c|}
 %\hline
 &$\frac{k_BT}{\hbar}\gg\Lambda\gg\Omega$ & $\Lambda\gg %\Omega,\frac{k_BT}{\hbar}$;$
 \Omega\sim\frac{k_BT}{\hbar}$ & $\Lambda\gg\Omega\gg\frac{k_BT}{\hbar}$\\
 \hline
 $c_{xx}$ & $-\Lambda/\Omega$   & $-\Lambda/\Omega$ & $-\Lambda/\Omega$\\
 \hline
 $c_{xp}$ & 1  &  1 & 1\\
 \hline
 $c_{pp}$ & $-2\Omega/\Lambda$  & $-2\Omega/\Lambda$ & $-2\Omega/\Lambda$\\
 \hline
 $d_{xx}$ & $\frac{2 k_BT}{\hbar\Omega} $  & $\frac{k_BT}{\hbar\Omega}+\coth\left(\frac{\hbar\Omega}{k_BT}\right)$ & 1\\
 \hline
 $d_{xp}$ & $-\frac{2 k_BT}{\hbar\Lambda}$  & $\frac{2}{\pi}\log\left(\frac{\hbar\Lambda}{2\pi k_BT}\right)$ & $\frac{2}{\pi}\log\left(\frac{\Lambda}{2\Omega}\right)$\\
 \hline
 $d_{pp}$ & $-\frac{\hbar\Omega}{3k_BT}$  & $\frac{k_BT}{\hbar\Omega}-\coth\left(\frac{\hbar\Omega}{k_BT}\right)$ & $-1$\\
 \hline
 
 \end{tabular}
 \end{equation}

\section{High-$T$ limit with leading quantum corrections}
\label{sec:highTlimitWithLeadingQuantumCorrections}

Let us now apply the Wigner function formalism  to the generalized
ME, Eq.\ \eqref{NonLinearCaldeiraLeggettMasterEq} valid in the
oversimplified high-$T$ limit,  and obtain\footnote{Note that
$\{\dot f(\hat x),\rho\} f(\hat x)=
\frac{\{p_-,f'(x_+)\}+\{p_+,f'(x_-)\}}{2m}\rho f(\hat x)=
f(x_-)\frac{\{p_-,f'(x_+)\}+\{p_+,f'(x_-)\}}{2m}W $}
\begin{multline}
\dot{W}=-\frac{i}{\hbar}\left[\frac{p_-^2-p_+^2}{2m}+V(x_+)-V(x_-)\right]W\\
-\frac{i\gamma}{4\hbar}[f(x_+)-f(x_-)]\Big(\{p_-,f'(x_+)\}+\{p_+,f'(x_-)\}\Big)W\\
-\frac{\gamma m k
T}{\hbar^2}[f^2(x_+)+f^2(x_-)-2f(x_+)f(x_-)]W
\end{multline}
In the case, when the potential $V(x)$ is non-harmonic and/or
$f(x)$ is not a linear or quadratic function of $x$, to proceed
further we perform a Taylor expansion in $\hbar$, and keep the
leading terms only. In other words we attempt to include the
leading quantum corrections. One finds then
\begin{multline} 
\label{WignerME} \dot{W}= \left[-\partial_x\frac{p}{m}+\partial_p
V'(x)-\frac{\hbar^2}{24}\partial_p^3V'''(x)+\ldots\right]W\\
+\gamma\bigg[\partial_p p [f'(x)]^{2} +\frac{\hbar^{2}\partial_p^2}{8}\bigg(2\partial_x f'(x) f''(x)\\
- 2[f''(x)]^2 -\frac{4}{3}\partial_p p f'(x) f'''(x) \bigg)+\ldots\bigg]W\\
+m\gamma k_BT\bigg[\partial_p^{2}[f'(x)]^2-\frac{\hbar^{2}}{12}\partial_p^{4}f'(x)f'''(x)+\ldots\bigg]W
\end{multline}
The above equation is the main result of this subsection -- it
combines the (oversimplified) high-$T$ limit with the leading
quantum corrections.
 To zeroth order in $\hbar$, the ME for the
Wigner matrix reads 
\begin{multline}
\dot{W}=\left[-\frac{p}{m}\partial_x+V'(x)\partial_p+\gamma[f'(x)]^{2}\partial_p p\right.\\
\left. +m\gamma k_BT[f'(x)]^2\partial_p^{2}\right]W 
\end{multline}

\subsection{Quadratic case -- high-$T$ solution}
As an example we  consider the simplest non-linear coupling to the
bath, a quadratic one, which we write in the form $f(x)=x^{2}/a$.
We also take the potential to be quadratic,
$V(x)=m\Omega^{2}x^{2}/2$. Since $f'''(x)=0$, from
Eq.~\eqref{WignerME} truncated to $O(\hbar^{2})$ we have 
\begin{multline}
\dot{W}=\bigg[-\frac{p}{m}\partial_x+m\Omega^{2}x\partial_p\\
+\frac{4 \gamma x^{2}}{a^2}\left(\partial_p p
+m k_B T\partial_p^{2}+\frac{\hbar^{2}}{4x}\partial_p^{2}\partial_x\right)
\bigg]W 
\end{multline}
 A stationary solution of this equation is in the
form of Eq.\ \eqref{gaussianAnsatz} with $\sigma_p=\sigma_x=1$ and
\beq
\tilde{T}=\frac{T}{2}\left[1\pm\sqrt{1-\left(\frac{\hbar\Omega}{k_BT}\right)^{2}}\right].
\label{effTempSimple}
\eeq
Only the $+$ solution is physically
acceptable, as can be seen by looking at large temperature
$k_BT\gg\hbar\Omega$, where the $+$ solution becomes \beq
\tilde{T}=T\left[1-\left(\frac{\hbar\Omega}{2kT}\right)^{2}\right]
\eeq This result is plotted as a red curve in Fig. \ref{fig:effTemp},
and may be interpreted as an effective cooling, since $\tilde{T}<T$, or as a breakdown of the dissipation-fluctuation relation, or as quantum localization in phase space.
However, as we have seen, this result is incorrect.
Obviously, it cannot be correct when $k_BT\simeq \hbar\Omega$, but
it loses validity already at larger temperatures, when $k_BT\lesssim
\hbar\Lambda$, since then neither  $D_{xp}$ nor $D_{pp}$ terms
can be neglected. Looking from another angle,  this result contains a quantum correction of order
$\hbar\Omega/k_BT$, which is simply non-systematic, and moreover it depends
on the order of limits: high temperature  $T\rightarrow \infty$, 
and stationarity, long time limit $t\rightarrow \infty$.

\section{Harmonically trapped particle  inside a Bose-Einstein condensate}
\label{sec:impurityInATrappedGas}
The problem of dilute impurities in an ultracold gas can be studied from
various points of view: as a polaron problem in a Fermi (cf.
\cite{Schirotzek2009,Kohstall2012,Koschorreck2012,MassignanPolRev2014,Lan2014,Levinsen2014}) or Bose (cf. \cite{Cote2002,Massignan2005,Cucchietti2006,Palzer2009,Catani2012,Rath2013,Fukuhara2013,Shashi2014,Grusdt2014a,Grusdt2014b}) gas, or as problem
of orthogonality catastrophe in a Fermi gas (cf.
\cite{Demler-cata,Kim2012}), or with established techniques for studying polarons in condensed matter systems \cite{Devreese2009}. We propose yet another point of view.
We consider a condensate of $N\gg1$ identical bosonic atoms of
mass $M$ inside an harmonic trap of frequency $\omega$,
interacting with scattering length $a_{s}$. Denoting by
$\psi^\dagger(\br)$ and $\hat\psi(\br)$ atomic creation and
annihilation  operators, the Hamiltonian of the Bose-Einstein condensate (BEC) is given by
\begin{multline}
 H_{\rm BEC}=\int{\rm
d}^3\br\,\hat\psi^\dagger(\br)\left[-\frac{\hbar^2\nabla^2}{2M}+\frac{M\omega^2\br^2}{2}\right.\\
\left.+\frac{4\pi\hbar^2a_s}{M}\hat\psi^\dagger(\br)\hat\psi(\br)\right]
\hat\psi(\br).
\end{multline}
We consider a single impurity trapped inside the BEC. The impurity is described as an harmonic oscillator of mass $m$ and frequency $\Omega$, interacting with the BEC atoms through a short-range (contact) potential characterized by a scattering length $a_s$. Its mean-field
Hamiltonian is 
\beq H_{\rm imp}=-\frac{\hbar^2\nabla^2}{2m}+\frac{m\Omega^2\br^2}{2}+\frac{2\pi\hbar^2a_s}{\mu}n(\br),
\eeq
where $\mu=mM/(m+M)$ is the reduced mass, and $n(\br)=\hat\psi^\dagger(\br)\hat\psi(\br)$ is the BEC density. We
follow the Bogolyubov-de Gennes (BdG) formalism \cite{Pita-String}, and write $\hat\psi(\br)=\sqrt{N}\varphi(\br)+\delta\hat\psi(\br)$ with $\varphi$ real, $\int{\rm d}^3\br\, \varphi^2(\br)=1$, and 
\begin{align}
\nonumber \delta\hat\psi(\br)=&\sum_\bk \hat g_\bk u_\bk(\br)+\hat
g^\dagger_{-\bk} v^*_\bk(\br)\\ \nonumber
\delta\hat\psi^\dagger(\br)=&\sum_\bk \hat g^\dagger_\bk
u^*_\bk(\br)+\hat g_{-\bk} v_\bk(\br),
\end{align}
where $\hat g^\dagger_\bk$ and $\hat g_\bk$ are the Bogolyubov quasi-particles' creation and annihilation operators, while $v_{-\bk}^*(\br)$ and $u_\bk(\br)$ are the corresponding mode functions.  We approximate 
\begin{multline}
\hat\psi^\dagger(\br)\hat\psi(\br)\simeq
N\varphi^2(\br)+\sqrt{N}\varphi(\br)\left[\delta\hat\psi(\br)
+\delta\hat\psi^\dagger(\br)\right]=\\
n(\br)+\sqrt{n(\br)}\left[\sum_\bk \hat g_\bk f_\bk(\br) +  \hat
g^\dagger_\bk f^*_\bk(\br) \right] 
\end{multline}
 with
$f_\bk(\br)=u_\bk(\br)+v_{-\bk}(\br)$. As the phases of $u$ and $v$ are arbitrary, we may choose them real, such that $f_\bk(\br)=f^{*}_\bk(\br)$. The BdG Hamiltonian for the impurity + BEC becomes then 
\begin{multline}
 H_{\rm
BdG}=-\frac{\hbar^2\nabla^2}{2m}+\frac{m\Omega^2\br^2}{2} + \sum_\bk \hbar \omega_\bk \hat g^\dagger_\bk \hat g_\bk\\
+\frac{2\pi\hbar^2a_s}{\mu}\left[n(\br)+\sqrt{n(\br)}\sum_\bk f_\bk( \br)(\hat g_\bk+\hat g^\dagger_\bk )\right].
\label{BdG-model}
\end{multline}
There are several important differences
  between the BdG model \eqref{BdG-model} and the
  Caldeira-Leggett model:
  \begin{itemize}
\item In the CLM the interaction Hamiltonian has a simple
separable form, $H_I=-\hat B f(\hat x)$, where $\hat B$ and $f(\hat
x)$ are bath and system operators, respectively. This is not the
case in the BdG model: different Bogolyubov modes couple
differently to the system via different mode functions.
\item The spectral density for a BEC is not necessarily Ohmic. It depends on the dimension, and the dispersion
relation of the Bogolyubov modes, $\hbar \omega_\bk$; this
relation generally interpolates between a low-energy phonon-like ($\hbar
\omega_\bk\propto |\bk|$) and a high-energy free-particle-like ($\hbar
\omega_\bk\propto \bk^2$) behaviors (cf. \cite{Pita-String}), and it may even exhibit a roton minimum at intermediate energies (cf. \cite{Lahaye2009}).
\item In any practical physical application of the present theory the cutoff energy $\hbar\Lambda$ has a very concrete physical sense: in a trap the bath frequencies are evidently bound by the trap
depth, in an optical lattice by the lowest band's width, and so on. Even more seriously: in any tight trap the high energy excitation modes will be concentrated at the semi-classical edges, as determined by the trap potential at a given energy; their overlap with the condensate, which has a size limited, say, by the Thomas--Fermi radius, will then be very small, and will decrease rapidly with the energy of excitations.
\end{itemize}

Radically different is the case of a Fermi bath. In this case there is no condensate, so the density fluctuations are from  the very beginning quadratic functions of the fermionic creation and annihilation operators. Still, a theory similar to the one presented here may be used in situations where bosonization theory works \cite{GiamarchiBook,GogolinBook}, i.e., typically in specific 1D systems. There are rare examples of Fermi surfaces for which bosonization, or in this case better to say Luttinger-Tomonaga theory, works \cite{MahanBook}. If we cannot use bosonization theory, the Fermi bath has to be treated according to its fermionic identity. These problems lead, however, far beyond the scope of the present paper.

\bibliography{quantumBrownianMotion}

\begin{thebibliography}{109}%
\makeatletter
\providecommand \@ifxundefined [1]{%
 \@ifx{#1\undefined}
}%
\providecommand \@ifnum [1]{%
 \ifnum #1\expandafter \@firstoftwo
 \else \expandafter \@secondoftwo
 \fi
}%
\providecommand \@ifx [1]{%
 \ifx #1\expandafter \@firstoftwo
 \else \expandafter \@secondoftwo
 \fi
}%
\providecommand \natexlab [1]{#1}%
\providecommand \enquote  [1]{``#1''}%
\providecommand \bibnamefont  [1]{#1}%
\providecommand \bibfnamefont [1]{#1}%
\providecommand \citenamefont [1]{#1}%
\providecommand \href@noop [0]{\@secondoftwo}%
\providecommand \href [0]{\begingroup \@sanitize@url \@href}%
\providecommand \@href[1]{\@@startlink{#1}\@@href}%
\providecommand \@@href[1]{\endgroup#1\@@endlink}%
\providecommand \@sanitize@url [0]{\catcode `\\12\catcode `\$12\catcode
  `\&12\catcode `\#12\catcode `\^12\catcode `\_12\catcode `\%12\relax}%
\providecommand \@@startlink[1]{}%
\providecommand \@@endlink[0]{}%
\providecommand \url  [0]{\begingroup\@sanitize@url \@url }%
\providecommand \@url [1]{\endgroup\@href {#1}{\urlprefix }}%
\providecommand \urlprefix  [0]{URL }%
\providecommand \Eprint [0]{\href }%
\providecommand \doibase [0]{http://dx.doi.org/}%
\providecommand \selectlanguage [0]{\@gobble}%
\providecommand \bibinfo  [0]{\@secondoftwo}%
\providecommand \bibfield  [0]{\@secondoftwo}%
\providecommand \translation [1]{[#1]}%
\providecommand \BibitemOpen [0]{}%
\providecommand \bibitemStop [0]{}%
\providecommand \bibitemNoStop [0]{.\EOS\space}%
\providecommand \EOS [0]{\spacefactor3000\relax}%
\providecommand \BibitemShut  [1]{\csname bibitem#1\endcsname}%
\let\auto@bib@innerbib\@empty
%</preamble>
\bibitem [{\citenamefont {Gardiner}\ and\ \citenamefont
  {Zoller}(2004)}]{GardinerBook}%
  \BibitemOpen
  \bibfield  {author} {\bibinfo {author} {\bibfnamefont {C.}~\bibnamefont
  {Gardiner}}\ and\ \bibinfo {author} {\bibfnamefont {P.}~\bibnamefont
  {Zoller}},\ }\href@noop {} {\emph {\bibinfo {title} {Quantum Noise: A
  Handbook of Markovian and Non-Markovian Quantum Stochastic Methods with
  Applications to Quantum Optics}}},\ Springer Series in Synergetics\ (\bibinfo
   {publisher} {Springer},\ \bibinfo {address} {Berlin},\ \bibinfo {year}
  {2004})\BibitemShut {NoStop}%
\bibitem [{\citenamefont {Breuer}\ and\ \citenamefont
  {Petruccione}(2007)}]{BreuerBook}%
  \BibitemOpen
  \bibfield  {author} {\bibinfo {author} {\bibfnamefont {H.}~\bibnamefont
  {Breuer}}\ and\ \bibinfo {author} {\bibfnamefont {F.}~\bibnamefont
  {Petruccione}},\ }\href {http://books.google.es/books?id=DkcJPwAACAAJ} {\emph
  {\bibinfo {title} {The Theory of Open Quantum Systems}}}\ (\bibinfo
  {publisher} {OUP},\ \bibinfo {address} {Oxford},\ \bibinfo {year}
  {2007})\BibitemShut {NoStop}%
\bibitem [{\citenamefont {Schlosshauer}(2007)}]{SchlosshauerBook}%
  \BibitemOpen
  \bibfield  {author} {\bibinfo {author} {\bibfnamefont {M.}~\bibnamefont
  {Schlosshauer}},\ }\href
  {http://www.springer.com/physics/quantum+physics/book/978-3-540-35773-5}
  {\emph {\bibinfo {title} {Decoherence and the Quantum-To-Classical
  Transition}}},\ The Frontiers Collection\ (\bibinfo  {publisher} {Springer},\
  \bibinfo {year} {2007})\BibitemShut {NoStop}%
\bibitem [{\citenamefont {Waldenfels}(2014)}]{WaldenfelsBook}%
  \BibitemOpen
  \bibfield  {author} {\bibinfo {author} {\bibfnamefont {W.~v.}\ \bibnamefont
  {Waldenfels}},\ }\href@noop {} {\emph {\bibinfo {title} {A Measure
  Theoretical Approach to Quantum Stochastic Processes}}},\ Lecture Notes in
  Physics\ (\bibinfo  {publisher} {Springer},\ \bibinfo {address}
  {Heidelberg},\ \bibinfo {year} {2014})\BibitemShut {NoStop}%
\bibitem [{\citenamefont {Weiss}(2008)}]{Weiss}%
  \BibitemOpen
  \bibfield  {author} {\bibinfo {author} {\bibfnamefont {U.}~\bibnamefont
  {Weiss}},\ }\href@noop {} {\emph {\bibinfo {title} {Quantum Dissipative
  Systems}}}\ (\bibinfo  {publisher} {World Scientific},\ \bibinfo {address}
  {Singapore},\ \bibinfo {year} {2008})\BibitemShut {NoStop}%
\bibitem [{\citenamefont {Landauer}(1957)}]{Landauer1957}%
  \BibitemOpen
  \bibfield  {author} {\bibinfo {author} {\bibfnamefont {R.}~\bibnamefont
  {Landauer}},\ }\bibfield  {title} {\bibinfo {title} {\emph {Spatial Variation
  of Currents and Fields Due to Localized Scatterers in Metallic Conduction}},\
  }\href@noop {} {\bibfield  {journal} {\bibinfo  {journal} {I.B.M. J. Res.
  Develop.}\ }\textbf {\bibinfo {volume} {149}},\ \bibinfo {pages} {223}
  (\bibinfo {year} {1957})}\BibitemShut {NoStop}%
\bibitem [{\citenamefont {Dykman}\ and\ \citenamefont
  {Krivoglaz}(1975)}]{Dykman1975}%
  \BibitemOpen
  \bibfield  {author} {\bibinfo {author} {\bibfnamefont {M.~I.}\ \bibnamefont
  {Dykman}}\ and\ \bibinfo {author} {\bibfnamefont {M.~A.}\ \bibnamefont
  {Krivoglaz}},\ }\bibfield  {title} {\bibinfo {title} {\emph {Spectral
  distribution of nonlinear oscillators with nonlinear friction due to a
  medium}},\ }\href {\doibase 10.1002/pssb.2220680109} {\bibfield  {journal}
  {\bibinfo  {journal} {Physica Status Solidi (B)}\ }\textbf {\bibinfo {volume}
  {68}},\ \bibinfo {pages} {111} (\bibinfo {year} {1975})}\BibitemShut
  {NoStop}%
\bibitem [{\citenamefont {Hu}\ \emph {et~al.}(1993)\citenamefont {Hu},
  \citenamefont {Paz},\ and\ \citenamefont {Zhang}}]{Hu1993}%
  \BibitemOpen
  \bibfield  {author} {\bibinfo {author} {\bibfnamefont {B.~L.}\ \bibnamefont
  {Hu}}, \bibinfo {author} {\bibfnamefont {J.~P.}\ \bibnamefont {Paz}}, \ and\
  \bibinfo {author} {\bibfnamefont {Y.}~\bibnamefont {Zhang}},\ }\bibfield
  {title} {\bibinfo {title} {\emph {Quantum Brownian motion in a general
  environment. II. Nonlinear coupling and perturbative approach}},\ }\href
  {\doibase 10.1103/PhysRevD.47.1576} {\bibfield  {journal} {\bibinfo
  {journal} {Phys. Rev. D}\ }\textbf {\bibinfo {volume} {47}},\ \bibinfo
  {pages} {1576} (\bibinfo {year} {1993})}\BibitemShut {NoStop}%
\bibitem [{\citenamefont {Brun}(1993)}]{Brun1993}%
  \BibitemOpen
  \bibfield  {author} {\bibinfo {author} {\bibfnamefont {T.~A.}\ \bibnamefont
  {Brun}},\ }\bibfield  {title} {\bibinfo {title} {\emph {Quasiclassical
  equations of motion for nonlinear Brownian systems}},\ }\href {\doibase
  10.1103/PhysRevD.47.3383} {\bibfield  {journal} {\bibinfo  {journal} {Phys.
  Rev. D}\ }\textbf {\bibinfo {volume} {47}},\ \bibinfo {pages} {3383}
  (\bibinfo {year} {1993})}\BibitemShut {NoStop}%
\bibitem [{\citenamefont {Banerjee}\ and\ \citenamefont
  {Ghosh}(2003)}]{Banerjee2003}%
  \BibitemOpen
  \bibfield  {author} {\bibinfo {author} {\bibfnamefont {S.}~\bibnamefont
  {Banerjee}}\ and\ \bibinfo {author} {\bibfnamefont {R.}~\bibnamefont
  {Ghosh}},\ }\bibfield  {title} {\bibinfo {title} {\emph {General quantum
  Brownian motion with initially correlated and nonlinearly coupled
  environment}},\ }\href {\doibase 10.1103/PhysRevE.67.056120} {\bibfield
  {journal} {\bibinfo  {journal} {Phys. Rev. E}\ }\textbf {\bibinfo {volume}
  {67}},\ \bibinfo {pages} {056120} (\bibinfo {year} {2003})}\BibitemShut
  {NoStop}%
\bibitem [{\citenamefont {{Smoluchowski}}(1916)}]{Smoluchowski}%
  \BibitemOpen
  \bibfield  {author} {\bibinfo {author} {\bibfnamefont {M.~V.}\ \bibnamefont
  {{Smoluchowski}}},\ }\bibfield  {title} {\bibinfo {title} {\emph {{Drei
  Vortr\"age \"uber Diffusion, Brownsche Bewegung und Koagulation von
  Kolloidteilchen}}},\ }\href@noop {} {\bibfield  {journal} {\bibinfo
  {journal} {Zeitschrift f\"ur Physik}\ }\textbf {\bibinfo {volume} {17}},\
  \bibinfo {pages} {557} (\bibinfo {year} {1916})}\BibitemShut {NoStop}%
\bibitem [{\citenamefont {Kramers}(1940)}]{Kramers}%
  \BibitemOpen
  \bibfield  {author} {\bibinfo {author} {\bibfnamefont {H.}~\bibnamefont
  {Kramers}},\ }\bibfield  {title} {\bibinfo {title} {\emph {Brownian motion in
  a field of force and the diffusion model of chemical reactions}},\ }\href
  {\doibase http://dx.doi.org/10.1016/S0031-8914(40)90098-2} {\bibfield
  {journal} {\bibinfo  {journal} {Physica}\ }\textbf {\bibinfo {volume} {7}},\
  \bibinfo {pages} {284 } (\bibinfo {year} {1940})}\BibitemShut {NoStop}%
\bibitem [{\citenamefont {Hottovy}\ \emph
  {et~al.}(2012{\natexlab{a}})\citenamefont {Hottovy}, \citenamefont {Volpe},\
  and\ \citenamefont {Wehr}}]{Hottovy2012}%
  \BibitemOpen
  \bibfield  {author} {\bibinfo {author} {\bibfnamefont {S.}~\bibnamefont
  {Hottovy}}, \bibinfo {author} {\bibfnamefont {G.}~\bibnamefont {Volpe}}, \
  and\ \bibinfo {author} {\bibfnamefont {J.}~\bibnamefont {Wehr}},\ }\bibfield
  {title} {\bibinfo {title} {\emph {Noise-induced drift in stochastic
  differential equations with arbitrary friction and diffusion in the
  Smoluchowski-Kramers limit}},\ }\href@noop {} {\bibfield  {journal} {\bibinfo
   {journal} {J. Stat. Phys.}\ }\textbf {\bibinfo {volume} {146}},\ \bibinfo
  {pages} {762} (\bibinfo {year} {2012}{\natexlab{a}})}\BibitemShut {NoStop}%
\bibitem [{\citenamefont {Hottovy}\ \emph {et~al.}(2014)\citenamefont
  {Hottovy}, \citenamefont {McDaniel}, \citenamefont {Volpe},\ and\
  \citenamefont {Wehr}}]{Hottovy2014}%
  \BibitemOpen
  \bibfield  {author} {\bibinfo {author} {\bibfnamefont {S.}~\bibnamefont
  {Hottovy}}, \bibinfo {author} {\bibfnamefont {A.}~\bibnamefont {McDaniel}},
  \bibinfo {author} {\bibfnamefont {G.}~\bibnamefont {Volpe}}, \ and\ \bibinfo
  {author} {\bibfnamefont {J.}~\bibnamefont {Wehr}},\ }\bibfield  {title}
  {\bibinfo {title} {\emph {The Smoluchowski-Kramers Limit of Stochastic
  Differential Equations with Arbitrary State-Dependent Friction}},\ }\href
  {\doibase 10.1007/s00220-014-2233-4} {\bibfield  {journal} {\bibinfo
  {journal} {Communications in Mathematical Physics}\ ,\ \bibinfo {pages} {1}}
  (\bibinfo {year} {2014})}\BibitemShut {NoStop}%
\bibitem [{\citenamefont {{McDaniel}}\ \emph {et~al.}(2014)\citenamefont
  {{McDaniel}}, \citenamefont {{Duman}}, \citenamefont {{Volpe}},\ and\
  \citenamefont {{Wehr}}}]{McDaniel2014}%
  \BibitemOpen
  \bibfield  {author} {\bibinfo {author} {\bibfnamefont {A.}~\bibnamefont
  {{McDaniel}}}, \bibinfo {author} {\bibfnamefont {O.}~\bibnamefont {{Duman}}},
  \bibinfo {author} {\bibfnamefont {G.}~\bibnamefont {{Volpe}}}, \ and\
  \bibinfo {author} {\bibfnamefont {J.}~\bibnamefont {{Wehr}}},\ }\bibfield
  {title} {\bibinfo {title} {\emph {{An SDE approximation for stochastic
  differential delay equations with colored state-dependent noise}}},\
  }\href@noop {} {\  (\bibinfo {year} {2014})},\ \Eprint
  {http://arxiv.org/abs/1406.7287} {arXiv:1406.7287} \BibitemShut {NoStop}%
\bibitem [{\citenamefont {Volpe}\ \emph {et~al.}(2010)\citenamefont {Volpe},
  \citenamefont {Helden}, \citenamefont {Brettschneider}, \citenamefont
  {Wehr},\ and\ \citenamefont {Bechinger}}]{Volpe2010}%
  \BibitemOpen
  \bibfield  {author} {\bibinfo {author} {\bibfnamefont {G.}~\bibnamefont
  {Volpe}}, \bibinfo {author} {\bibfnamefont {L.}~\bibnamefont {Helden}},
  \bibinfo {author} {\bibfnamefont {T.}~\bibnamefont {Brettschneider}},
  \bibinfo {author} {\bibfnamefont {J.}~\bibnamefont {Wehr}}, \ and\ \bibinfo
  {author} {\bibfnamefont {C.}~\bibnamefont {Bechinger}},\ }\bibfield  {title}
  {\bibinfo {title} {\emph {Influence of Noise on Force Measurements}},\
  }\href@noop {} {\bibfield  {journal} {\bibinfo  {journal} {Phys. Rev. Lett.}\
  }\textbf {\bibinfo {volume} {104}},\ \bibinfo {pages} {170602} (\bibinfo
  {year} {2010})}\BibitemShut {NoStop}%
\bibitem [{\citenamefont {Brettschneider}\ \emph {et~al.}(2011)\citenamefont
  {Brettschneider}, \citenamefont {Volpe}, \citenamefont {Helden},
  \citenamefont {Wehr},\ and\ \citenamefont {Bechinger}}]{Brettschneider2011}%
  \BibitemOpen
  \bibfield  {author} {\bibinfo {author} {\bibfnamefont {T.}~\bibnamefont
  {Brettschneider}}, \bibinfo {author} {\bibfnamefont {G.}~\bibnamefont
  {Volpe}}, \bibinfo {author} {\bibfnamefont {L.}~\bibnamefont {Helden}},
  \bibinfo {author} {\bibfnamefont {J.}~\bibnamefont {Wehr}}, \ and\ \bibinfo
  {author} {\bibfnamefont {C.}~\bibnamefont {Bechinger}},\ }\bibfield  {title}
  {\bibinfo {title} {\emph {Force measurement in the presence of Brownian
  noise: Equilibrium-distribution method versus drift method}},\ }\href@noop {}
  {\bibfield  {journal} {\bibinfo  {journal} {Phys. Rev. E}\ }\textbf {\bibinfo
  {volume} {83}},\ \bibinfo {pages} {041113} (\bibinfo {year}
  {2011})}\BibitemShut {NoStop}%
\bibitem [{\citenamefont {G.~Pesce}\ and\ \citenamefont
  {Volpe}(2013)}]{Pesce2012}%
  \BibitemOpen
  \bibfield  {author} {\bibinfo {author} {\bibfnamefont {S.~H. J.~W.}\
  \bibnamefont {G.~Pesce}, \bibfnamefont {A.~McDaniel}}\ and\ \bibinfo {author}
  {\bibfnamefont {G.}~\bibnamefont {Volpe}},\ }\bibfield  {title} {\bibinfo
  {title} {\emph {Stratonovich-to-It\^o transition in noisy systems with
  multiplicative feedback}},\ }\href {\doibase 10.1038/ncomms3733} {\bibfield
  {journal} {\bibinfo  {journal} {Nature Communications}\ }\textbf {\bibinfo
  {volume} {4}},\ \bibinfo {pages} {2733} (\bibinfo {year} {2013})}\BibitemShut
  {NoStop}%
\bibitem [{\citenamefont {Hottovy}\ \emph
  {et~al.}(2012{\natexlab{b}})\citenamefont {Hottovy}, \citenamefont {Volpe},\
  and\ \citenamefont {Wehr}}]{Hottovy2012a}%
  \BibitemOpen
  \bibfield  {author} {\bibinfo {author} {\bibfnamefont {S.}~\bibnamefont
  {Hottovy}}, \bibinfo {author} {\bibfnamefont {G.}~\bibnamefont {Volpe}}, \
  and\ \bibinfo {author} {\bibfnamefont {J.}~\bibnamefont {Wehr}},\ }\bibfield
  {title} {\bibinfo {title} {\emph {Thermophoresis of Brownian particles driven
  by coloured noise}},\ }\href@noop {} {\bibfield  {journal} {\bibinfo
  {journal} {EPL}\ }\textbf {\bibinfo {volume} {99}},\ \bibinfo {pages} {60002}
  (\bibinfo {year} {2012}{\natexlab{b}})}\BibitemShut {NoStop}%
\bibitem [{\citenamefont {Haus}\ and\ \citenamefont {Kehr}(1987)}]{Haus1987}%
  \BibitemOpen
  \bibfield  {author} {\bibinfo {author} {\bibfnamefont {J.~W.}\ \bibnamefont
  {Haus}}\ and\ \bibinfo {author} {\bibfnamefont {K.}~\bibnamefont {Kehr}},\
  }\bibfield  {title} {\bibinfo {title} {\emph {Diffusion in regular and
  disordered lattices}},\ }\href@noop {} {\bibfield  {journal} {\bibinfo
  {journal} {Phys. Rep.}\ }\textbf {\bibinfo {volume} {150}},\ \bibinfo {pages}
  {263} (\bibinfo {year} {1987})}\BibitemShut {NoStop}%
\bibitem [{\citenamefont {Havlin}\ and\ \citenamefont
  {Ben-Avraham}(1987)}]{Havlin87}%
  \BibitemOpen
  \bibfield  {author} {\bibinfo {author} {\bibfnamefont {S.}~\bibnamefont
  {Havlin}}\ and\ \bibinfo {author} {\bibfnamefont {D.}~\bibnamefont
  {Ben-Avraham}},\ }\bibfield  {title} {\bibinfo {title} {\emph {Diffusion in
  disordered media}},\ }\href {\doibase 10.1080/00018738700101072} {\bibfield
  {journal} {\bibinfo  {journal} {Advances in Physics}\ }\textbf {\bibinfo
  {volume} {36}},\ \bibinfo {pages} {695} (\bibinfo {year} {1987})}\BibitemShut
  {NoStop}%
\bibitem [{\citenamefont {{Bouchaud}}\ and\ \citenamefont
  {{Georges}}(1990)}]{BG1990}%
  \BibitemOpen
  \bibfield  {author} {\bibinfo {author} {\bibfnamefont {J.-P.}\ \bibnamefont
  {{Bouchaud}}}\ and\ \bibinfo {author} {\bibfnamefont {A.}~\bibnamefont
  {{Georges}}},\ }\bibfield  {title} {\bibinfo {title} {\emph {{Anomalous
  diffusion in disordered media: Statistical mechanisms, models and physical
  applications}}},\ }\href {\doibase 10.1016/0370-1573(90)90099-N} {\bibfield
  {journal} {\bibinfo  {journal} {Physics Reports}\ }\textbf {\bibinfo {volume}
  {195}},\ \bibinfo {pages} {127} (\bibinfo {year} {1990})}\BibitemShut
  {NoStop}%
\bibitem [{\citenamefont {Klafter}\ and\ \citenamefont
  {Sokolov}(2011)}]{Klafter2011}%
  \BibitemOpen
  \bibfield  {author} {\bibinfo {author} {\bibfnamefont {J.}~\bibnamefont
  {Klafter}}\ and\ \bibinfo {author} {\bibfnamefont {I.~M.}\ \bibnamefont
  {Sokolov}},\ }\href@noop {} {\emph {\bibinfo {title} {First Steps in Random
  Walks}}}\ (\bibinfo  {publisher} {Oxford University Press},\ \bibinfo
  {address} {Oxford},\ \bibinfo {year} {2011})\BibitemShut {NoStop}%
\bibitem [{\citenamefont {Metzler}\ and\ \citenamefont
  {Klafter}(2004)}]{Metzler2004}%
  \BibitemOpen
  \bibfield  {author} {\bibinfo {author} {\bibfnamefont {R.}~\bibnamefont
  {Metzler}}\ and\ \bibinfo {author} {\bibfnamefont {J.}~\bibnamefont
  {Klafter}},\ }\bibfield  {title} {\bibinfo {title} {\emph {The restaurant at
  the end of the random walk: recent developments in the description of
  anomalous transport by fractional dynamics}},\ }\href
  {http://stacks.iop.org/0305-4470/37/i=31/a=R01} {\bibfield  {journal}
  {\bibinfo  {journal} {Journal of Physics A: Mathematical and General}\
  }\textbf {\bibinfo {volume} {37}},\ \bibinfo {pages} {R161} (\bibinfo {year}
  {2004})}\BibitemShut {NoStop}%
\bibitem [{\citenamefont {H\"{o}fling}\ and\ \citenamefont
  {Franosch}(2013)}]{Hoefling2013}%
  \BibitemOpen
  \bibfield  {author} {\bibinfo {author} {\bibfnamefont {F.}~\bibnamefont
  {H\"{o}fling}}\ and\ \bibinfo {author} {\bibfnamefont {T.}~\bibnamefont
  {Franosch}},\ }\bibfield  {title} {\bibinfo {title} {\emph {Anomalous
  transport in the crowded world of biological cells}},\ }\href {\doibase
  10.1088/0034-4885/76/4/046602} {\bibfield  {journal} {\bibinfo  {journal}
  {Reports on Progress in Physics}\ }\textbf {\bibinfo {volume} {76}},\
  \bibinfo {pages} {046602} (\bibinfo {year} {2013})}\BibitemShut {NoStop}%
\bibitem [{\citenamefont {Metzler}\ \emph {et~al.}(2014)\citenamefont
  {Metzler}, \citenamefont {Jeon}, \citenamefont {Cherstvy},\ and\
  \citenamefont {Barkai}}]{Metzler2014}%
  \BibitemOpen
  \bibfield  {author} {\bibinfo {author} {\bibfnamefont {R.}~\bibnamefont
  {Metzler}}, \bibinfo {author} {\bibfnamefont {J.-H.}\ \bibnamefont {Jeon}},
  \bibinfo {author} {\bibfnamefont {A.~G.}\ \bibnamefont {Cherstvy}}, \ and\
  \bibinfo {author} {\bibfnamefont {E.}~\bibnamefont {Barkai}},\ }\bibfield
  {title} {\bibinfo {title} {\emph {Anomalous diffusion models and their
  properties: non-stationarity{,} non-ergodicity{,} and ageing at the centenary
  of single particle tracking}},\ }\href {\doibase 10.1039/C4CP03465A}
  {\bibfield  {journal} {\bibinfo  {journal} {Phys. Chem. Chem. Phys.}\
  }\textbf {\bibinfo {volume} {16}},\ \bibinfo {pages} {24128} (\bibinfo {year}
  {2014})}\BibitemShut {NoStop}%
\bibitem [{\citenamefont {Montroll}\ and\ \citenamefont
  {Weiss}(1965)}]{MontrollWeiss65}%
  \BibitemOpen
  \bibfield  {author} {\bibinfo {author} {\bibfnamefont {E.~W.}\ \bibnamefont
  {Montroll}}\ and\ \bibinfo {author} {\bibfnamefont {G.~H.}\ \bibnamefont
  {Weiss}},\ }\bibfield  {title} {\bibinfo {title} {\emph {Random Walks on
  Lattices. II}},\ }\href {\doibase http://dx.doi.org/10.1063/1.1704269}
  {\bibfield  {journal} {\bibinfo  {journal} {Journal of Mathematical Physics}\
  }\textbf {\bibinfo {volume} {6}},\ \bibinfo {pages} {167} (\bibinfo {year}
  {1965})}\BibitemShut {NoStop}%
\bibitem [{\citenamefont {Scher}\ and\ \citenamefont
  {Montroll}(1975)}]{ScherMontroll75}%
  \BibitemOpen
  \bibfield  {author} {\bibinfo {author} {\bibfnamefont {H.}~\bibnamefont
  {Scher}}\ and\ \bibinfo {author} {\bibfnamefont {E.~W.}\ \bibnamefont
  {Montroll}},\ }\bibfield  {title} {\bibinfo {title} {\emph {Anomalous
  transit-time dispersion in amorphous solids}},\ }\href {\doibase
  10.1103/PhysRevB.12.2455} {\bibfield  {journal} {\bibinfo  {journal} {Phys.
  Rev. B}\ }\textbf {\bibinfo {volume} {12}},\ \bibinfo {pages} {2455}
  (\bibinfo {year} {1975})}\BibitemShut {NoStop}%
\bibitem [{\citenamefont {Saxton}(1993)}]{Saxton1993}%
  \BibitemOpen
  \bibfield  {author} {\bibinfo {author} {\bibfnamefont {M.~J.}\ \bibnamefont
  {Saxton}},\ }\bibfield  {title} {\bibinfo {title} {\emph {{Lateral diffusion
  in an archipelago. Single-particle diffusion.}}},\ }\href {\doibase
  10.1016/S0006-3495(93)81548-0} {\bibfield  {journal} {\bibinfo  {journal}
  {Biophysical journal}\ }\textbf {\bibinfo {volume} {64}},\ \bibinfo {pages}
  {1766} (\bibinfo {year} {1993})}\BibitemShut {NoStop}%
\bibitem [{\citenamefont {Saxton}(1997)}]{Saxton1997}%
  \BibitemOpen
  \bibfield  {author} {\bibinfo {author} {\bibfnamefont {M.~J.}\ \bibnamefont
  {Saxton}},\ }\bibfield  {title} {\bibinfo {title} {\emph {{Single-particle
  tracking: the distribution of diffusion coefficients.}}},\ }\href {\doibase
  10.1016/S0006-3495(97)78820-9} {\bibfield  {journal} {\bibinfo  {journal}
  {Biophysical journal}\ }\textbf {\bibinfo {volume} {72}},\ \bibinfo {pages}
  {1744} (\bibinfo {year} {1997})}\BibitemShut {NoStop}%
\bibitem [{\citenamefont {Cherstvy}\ and\ \citenamefont
  {Metzler}(2013)}]{Cherstvy2013a}%
  \BibitemOpen
  \bibfield  {author} {\bibinfo {author} {\bibfnamefont {A.~G.}\ \bibnamefont
  {Cherstvy}}\ and\ \bibinfo {author} {\bibfnamefont {R.}~\bibnamefont
  {Metzler}},\ }\bibfield  {title} {\bibinfo {title} {\emph {Population
  splitting{,} trapping{,} and non-ergodicity in heterogeneous diffusion
  processes}},\ }\href {\doibase 10.1039/C3CP53056F} {\bibfield  {journal}
  {\bibinfo  {journal} {Phys. Chem. Chem. Phys.}\ }\textbf {\bibinfo {volume}
  {15}},\ \bibinfo {pages} {20220} (\bibinfo {year} {2013})}\BibitemShut
  {NoStop}%
\bibitem [{\citenamefont {Cherstvy}\ \emph {et~al.}(2014)\citenamefont
  {Cherstvy}, \citenamefont {Chechkin},\ and\ \citenamefont
  {Metzler}}]{Cherstvy2014}%
  \BibitemOpen
  \bibfield  {author} {\bibinfo {author} {\bibfnamefont {A.~G.}\ \bibnamefont
  {Cherstvy}}, \bibinfo {author} {\bibfnamefont {A.~V.}\ \bibnamefont
  {Chechkin}}, \ and\ \bibinfo {author} {\bibfnamefont {R.}~\bibnamefont
  {Metzler}},\ }\bibfield  {title} {\bibinfo {title} {\emph {Particle
  invasion{,} survival{,} and non-ergodicity in 2D diffusion processes with
  space-dependent diffusivity}},\ }\href {http://dx.doi.org/10.1039/C3SM52846D}
  {\bibfield  {journal} {\bibinfo  {journal} {Soft Matter}\ }\textbf {\bibinfo
  {volume} {10}},\ \bibinfo {pages} {1591} (\bibinfo {year}
  {2014})}\BibitemShut {NoStop}%
\bibitem [{\citenamefont {Toli\'{c}-N\o{}rrelykke}\ \emph
  {et~al.}(2004)\citenamefont {Toli\'{c}-N\o{}rrelykke}, \citenamefont
  {Munteanu}, \citenamefont {Thon}, \citenamefont {Oddershede},\ and\
  \citenamefont {Berg-S\o{}rensen}}]{Tolic04}%
  \BibitemOpen
  \bibfield  {author} {\bibinfo {author} {\bibfnamefont {I.~M.}\ \bibnamefont
  {Toli\'{c}-N\o{}rrelykke}}, \bibinfo {author} {\bibfnamefont {E.-L.}\
  \bibnamefont {Munteanu}}, \bibinfo {author} {\bibfnamefont {G.}~\bibnamefont
  {Thon}}, \bibinfo {author} {\bibfnamefont {L.}~\bibnamefont {Oddershede}}, \
  and\ \bibinfo {author} {\bibfnamefont {K.}~\bibnamefont {Berg-S\o{}rensen}},\
  }\bibfield  {title} {\bibinfo {title} {\emph {Anomalous Diffusion in Living
  Yeast Cells}},\ }\href {\doibase 10.1103/PhysRevLett.93.078102} {\bibfield
  {journal} {\bibinfo  {journal} {Phys. Rev. Lett.}\ }\textbf {\bibinfo
  {volume} {93}},\ \bibinfo {pages} {078102} (\bibinfo {year}
  {2004})}\BibitemShut {NoStop}%
\bibitem [{\citenamefont {Golding}\ and\ \citenamefont
  {Cox}(2006)}]{Golding06}%
  \BibitemOpen
  \bibfield  {author} {\bibinfo {author} {\bibfnamefont {I.}~\bibnamefont
  {Golding}}\ and\ \bibinfo {author} {\bibfnamefont {E.~C.}\ \bibnamefont
  {Cox}},\ }\bibfield  {title} {\bibinfo {title} {\emph {Physical Nature of
  Bacterial Cytoplasm}},\ }\href {\doibase 10.1103/PhysRevLett.96.098102}
  {\bibfield  {journal} {\bibinfo  {journal} {Phys. Rev. Lett.}\ }\textbf
  {\bibinfo {volume} {96}},\ \bibinfo {pages} {098102} (\bibinfo {year}
  {2006})}\BibitemShut {NoStop}%
\bibitem [{\citenamefont {Jeon}\ \emph {et~al.}(2011)\citenamefont {Jeon},
  \citenamefont {Tejedor}, \citenamefont {Burov}, \citenamefont {Barkai},
  \citenamefont {Selhuber-Unkel}, \citenamefont {Berg-S\o{}rensen},
  \citenamefont {Oddershede},\ and\ \citenamefont {Metzler}}]{Jeon2011}%
  \BibitemOpen
  \bibfield  {author} {\bibinfo {author} {\bibfnamefont {J.-H.}\ \bibnamefont
  {Jeon}}, \bibinfo {author} {\bibfnamefont {V.}~\bibnamefont {Tejedor}},
  \bibinfo {author} {\bibfnamefont {S.}~\bibnamefont {Burov}}, \bibinfo
  {author} {\bibfnamefont {E.}~\bibnamefont {Barkai}}, \bibinfo {author}
  {\bibfnamefont {C.}~\bibnamefont {Selhuber-Unkel}}, \bibinfo {author}
  {\bibfnamefont {K.}~\bibnamefont {Berg-S\o{}rensen}}, \bibinfo {author}
  {\bibfnamefont {L.}~\bibnamefont {Oddershede}}, \ and\ \bibinfo {author}
  {\bibfnamefont {R.}~\bibnamefont {Metzler}},\ }\bibfield  {title} {\bibinfo
  {title} {\emph {In Vivo Anomalous Diffusion and Weak Ergodicity Breaking of
  Lipid Granules}},\ }\href {\doibase 10.1103/PhysRevLett.106.048103}
  {\bibfield  {journal} {\bibinfo  {journal} {Phys. Rev. Lett.}\ }\textbf
  {\bibinfo {volume} {106}},\ \bibinfo {pages} {048103} (\bibinfo {year}
  {2011})}\BibitemShut {NoStop}%
\bibitem [{\citenamefont {Weigel}\ \emph {et~al.}(2011)\citenamefont {Weigel},
  \citenamefont {Simon}, \citenamefont {Tamkun},\ and\ \citenamefont
  {Krapf}}]{Weigel2011}%
  \BibitemOpen
  \bibfield  {author} {\bibinfo {author} {\bibfnamefont {A.~V.}\ \bibnamefont
  {Weigel}}, \bibinfo {author} {\bibfnamefont {B.}~\bibnamefont {Simon}},
  \bibinfo {author} {\bibfnamefont {M.~M.}\ \bibnamefont {Tamkun}}, \ and\
  \bibinfo {author} {\bibfnamefont {D.}~\bibnamefont {Krapf}},\ }\bibfield
  {title} {\bibinfo {title} {\emph {Ergodic and nonergodic processes coexist in
  the plasma membrane as observed by single-molecule tracking}},\ }\href
  {\doibase 10.1073/pnas.1016325108} {\bibfield  {journal} {\bibinfo  {journal}
  {Proceedings of the National Academy of Sciences}\ }\textbf {\bibinfo
  {volume} {108}},\ \bibinfo {pages} {6438} (\bibinfo {year}
  {2011})}\BibitemShut {NoStop}%
\bibitem [{\citenamefont {Kusumi}\ \emph {et~al.}(2012)\citenamefont {Kusumi},
  \citenamefont {Fujiwara}, \citenamefont {Chadda}, \citenamefont {Xie},
  \citenamefont {Tsunoyama}, \citenamefont {Kalay}, \citenamefont {Kasai},\
  and\ \citenamefont {Suzuki}}]{Kusumi2012}%
  \BibitemOpen
  \bibfield  {author} {\bibinfo {author} {\bibfnamefont {A.}~\bibnamefont
  {Kusumi}}, \bibinfo {author} {\bibfnamefont {T.~K.}\ \bibnamefont
  {Fujiwara}}, \bibinfo {author} {\bibfnamefont {R.}~\bibnamefont {Chadda}},
  \bibinfo {author} {\bibfnamefont {M.}~\bibnamefont {Xie}}, \bibinfo {author}
  {\bibfnamefont {T.~A.}\ \bibnamefont {Tsunoyama}}, \bibinfo {author}
  {\bibfnamefont {Z.}~\bibnamefont {Kalay}}, \bibinfo {author} {\bibfnamefont
  {R.~S.}\ \bibnamefont {Kasai}}, \ and\ \bibinfo {author} {\bibfnamefont
  {K.~G.}\ \bibnamefont {Suzuki}},\ }\bibfield  {title} {\bibinfo {title}
  {\emph {Dynamic Organizing Principles of the Plasma Membrane that Regulate
  Signal Transduction: Commemorating the Fortieth Anniversary of Singer and
  Nicolson's Fluid-Mosaic Model}},\ }\href {\doibase
  10.1146/annurev-cellbio-100809-151736} {\bibfield  {journal} {\bibinfo
  {journal} {Annual Review of Cell and Developmental Biology}\ }\textbf
  {\bibinfo {volume} {28}},\ \bibinfo {pages} {215} (\bibinfo {year}
  {2012})}\BibitemShut {NoStop}%
\bibitem [{\citenamefont {Bakker}\ \emph {et~al.}(2012)\citenamefont {Bakker},
  \citenamefont {Eich}, \citenamefont {Torreno-Pina}, \citenamefont
  {Diez-Ahedo}, \citenamefont {Perez-Samper}, \citenamefont {van Zanten},
  \citenamefont {Figdor}, \citenamefont {Cambi},\ and\ \citenamefont
  {Garc\'ia-Parajo}}]{Bakker2012}%
  \BibitemOpen
  \bibfield  {author} {\bibinfo {author} {\bibfnamefont {G.~J.}\ \bibnamefont
  {Bakker}}, \bibinfo {author} {\bibfnamefont {C.}~\bibnamefont {Eich}},
  \bibinfo {author} {\bibfnamefont {J.~A.}\ \bibnamefont {Torreno-Pina}},
  \bibinfo {author} {\bibfnamefont {R.}~\bibnamefont {Diez-Ahedo}}, \bibinfo
  {author} {\bibfnamefont {G.}~\bibnamefont {Perez-Samper}}, \bibinfo {author}
  {\bibfnamefont {T.~S.}\ \bibnamefont {van Zanten}}, \bibinfo {author}
  {\bibfnamefont {C.~G.}\ \bibnamefont {Figdor}}, \bibinfo {author}
  {\bibfnamefont {A.}~\bibnamefont {Cambi}}, \ and\ \bibinfo {author}
  {\bibfnamefont {M.~F.}\ \bibnamefont {Garc\'ia-Parajo}},\ }\bibfield  {title}
  {\bibinfo {title} {\emph {Lateral mobility of individual integrin
  nanoclusters orchestrates the onset for leukocyte adhesion}},\ }\href
  {\doibase 10.1073/pnas.1116425109} {\bibfield  {journal} {\bibinfo  {journal}
  {Proceedings of the National Academy of Sciences}\ }\textbf {\bibinfo
  {volume} {109}},\ \bibinfo {pages} {4869} (\bibinfo {year}
  {2012})}\BibitemShut {NoStop}%
\bibitem [{\citenamefont {Cisse}\ \emph {et~al.}(2013)\citenamefont {Cisse},
  \citenamefont {Izeddin}, \citenamefont {Causse}, \citenamefont {Boudarene},
  \citenamefont {Senecal}, \citenamefont {Muresan}, \citenamefont
  {Dugast-Darzacq}, \citenamefont {Hajj}, \citenamefont {Dahan},\ and\
  \citenamefont {Darzacq}}]{Cisse2013}%
  \BibitemOpen
  \bibfield  {author} {\bibinfo {author} {\bibfnamefont {I.~I.}\ \bibnamefont
  {Cisse}}, \bibinfo {author} {\bibfnamefont {I.}~\bibnamefont {Izeddin}},
  \bibinfo {author} {\bibfnamefont {S.~Z.}\ \bibnamefont {Causse}}, \bibinfo
  {author} {\bibfnamefont {L.}~\bibnamefont {Boudarene}}, \bibinfo {author}
  {\bibfnamefont {A.}~\bibnamefont {Senecal}}, \bibinfo {author} {\bibfnamefont
  {L.}~\bibnamefont {Muresan}}, \bibinfo {author} {\bibfnamefont
  {C.}~\bibnamefont {Dugast-Darzacq}}, \bibinfo {author} {\bibfnamefont
  {B.}~\bibnamefont {Hajj}}, \bibinfo {author} {\bibfnamefont {M.}~\bibnamefont
  {Dahan}}, \ and\ \bibinfo {author} {\bibfnamefont {X.}~\bibnamefont
  {Darzacq}},\ }\bibfield  {title} {\bibinfo {title} {\emph {Real-Time Dynamics
  of RNA Polymerase II Clustering in Live Human Cells}},\ }\href {\doibase
  10.1126/science.1239053} {\bibfield  {journal} {\bibinfo  {journal}
  {Science}\ }\textbf {\bibinfo {volume} {341}},\ \bibinfo {pages} {664}
  (\bibinfo {year} {2013})}\BibitemShut {NoStop}%
\bibitem [{\citenamefont {Massignan}\ \emph
  {et~al.}(2014{\natexlab{a}})\citenamefont {Massignan}, \citenamefont {Manzo},
  \citenamefont {Torreno-Pina}, \citenamefont {Garc\'ia-Parajo}, \citenamefont
  {Lewenstein},\ and\ \citenamefont {Lapeyre}}]{Massignan2014}%
  \BibitemOpen
  \bibfield  {author} {\bibinfo {author} {\bibfnamefont {P.}~\bibnamefont
  {Massignan}}, \bibinfo {author} {\bibfnamefont {C.}~\bibnamefont {Manzo}},
  \bibinfo {author} {\bibfnamefont {J.~A.}\ \bibnamefont {Torreno-Pina}},
  \bibinfo {author} {\bibfnamefont {M.~F.}\ \bibnamefont {Garc\'ia-Parajo}},
  \bibinfo {author} {\bibfnamefont {M.}~\bibnamefont {Lewenstein}}, \ and\
  \bibinfo {author} {\bibfnamefont {G.~J.}\ \bibnamefont {Lapeyre}},\
  }\bibfield  {title} {\bibinfo {title} {\emph {Nonergodic Subdiffusion from
  Brownian Motion in an Inhomogeneous Medium}},\ }\href {\doibase
  10.1103/PhysRevLett.112.150603} {\bibfield  {journal} {\bibinfo  {journal}
  {Phys. Rev. Lett.}\ }\textbf {\bibinfo {volume} {112}},\ \bibinfo {pages}
  {150603} (\bibinfo {year} {2014}{\natexlab{a}})}\BibitemShut {NoStop}%
\bibitem [{\citenamefont {{Manzo}}\ \emph {et~al.}(2014)\citenamefont
  {{Manzo}}, \citenamefont {{Torreno-Pina}}, \citenamefont {{Massignan}},
  \citenamefont {{Lapeyre}}, \citenamefont {{Lewenstein}},\ and\ \citenamefont
  {{Garc\'ia-Parajo}}}]{Manzo2014}%
  \BibitemOpen
  \bibfield  {author} {\bibinfo {author} {\bibfnamefont {C.}~\bibnamefont
  {{Manzo}}}, \bibinfo {author} {\bibfnamefont {J.~A.}\ \bibnamefont
  {{Torreno-Pina}}}, \bibinfo {author} {\bibfnamefont {P.}~\bibnamefont
  {{Massignan}}}, \bibinfo {author} {\bibfnamefont {G.~J.}\ \bibnamefont
  {{Lapeyre}}, \bibfnamefont {Jr.}}, \bibinfo {author} {\bibfnamefont
  {M.}~\bibnamefont {{Lewenstein}}}, \ and\ \bibinfo {author} {\bibfnamefont
  {M.~F.}\ \bibnamefont {{Garc\'ia-Parajo}}},\ }\bibfield  {title} {\bibinfo
  {title} {\emph {{Weak ergodicity breaking of receptor motion in living cells
  stemming from random diffusivity}}},\ }\href@noop {} {\  (\bibinfo {year}
  {2014})},\ \Eprint {http://arxiv.org/abs/1407.2552} {arXiv:1407.2552}
  \BibitemShut {NoStop}%
\bibitem [{\citenamefont {Lewenstein}\ \emph {et~al.}(2012)\citenamefont
  {Lewenstein}, \citenamefont {Sanpera},\ and\ \citenamefont
  {Ahufinger}}]{Lewenstein2012}%
  \BibitemOpen
  \bibfield  {author} {\bibinfo {author} {\bibfnamefont {M.}~\bibnamefont
  {Lewenstein}}, \bibinfo {author} {\bibfnamefont {A.}~\bibnamefont {Sanpera}},
  \ and\ \bibinfo {author} {\bibfnamefont {V.}~\bibnamefont {Ahufinger}},\
  }\href@noop {} {\emph {\bibinfo {title} {Ultracold atoms in optical lattices:
  Simulating quantum many-body systems}}}\ (\bibinfo  {publisher} {OUP},\
  \bibinfo {address} {Oxford},\ \bibinfo {year} {2012})\BibitemShut {NoStop}%
\bibitem [{\citenamefont {{Krinner}}\ \emph {et~al.}(2013)\citenamefont
  {{Krinner}}, \citenamefont {{Stadler}}, \citenamefont {{Meineke}},
  \citenamefont {{Brantut}},\ and\ \citenamefont {{Esslinger}}}]{Krinner2013}%
  \BibitemOpen
  \bibfield  {author} {\bibinfo {author} {\bibfnamefont {S.}~\bibnamefont
  {{Krinner}}}, \bibinfo {author} {\bibfnamefont {D.}~\bibnamefont
  {{Stadler}}}, \bibinfo {author} {\bibfnamefont {J.}~\bibnamefont
  {{Meineke}}}, \bibinfo {author} {\bibfnamefont {J.-P.}\ \bibnamefont
  {{Brantut}}}, \ and\ \bibinfo {author} {\bibfnamefont {T.}~\bibnamefont
  {{Esslinger}}},\ }\bibfield  {title} {\bibinfo {title} {\emph {{Direct
  Observation of Fragmentation in a Disordered, Strongly Interacting Fermi
  Gas}}},\ }\href@noop {} {\  (\bibinfo {year} {2013})},\ \Eprint
  {http://arxiv.org/abs/1311.5174} {arXiv:1311.5174} \BibitemShut {NoStop}%
\bibitem [{\citenamefont {Schirotzek}\ \emph {et~al.}(2009)\citenamefont
  {Schirotzek}, \citenamefont {Wu}, \citenamefont {Sommer},\ and\ \citenamefont
  {Zwierlein}}]{Schirotzek2009}%
  \BibitemOpen
  \bibfield  {author} {\bibinfo {author} {\bibfnamefont {A.}~\bibnamefont
  {Schirotzek}}, \bibinfo {author} {\bibfnamefont {C.-H.}\ \bibnamefont {Wu}},
  \bibinfo {author} {\bibfnamefont {A.}~\bibnamefont {Sommer}}, \ and\ \bibinfo
  {author} {\bibfnamefont {M.~W.}\ \bibnamefont {Zwierlein}},\ }\bibfield
  {title} {\bibinfo {title} {\emph {Observation of Fermi Polarons in a Tunable
  Fermi Liquid of Ultracold Atoms}},\ }\href {\doibase
  10.1103/PhysRevLett.102.230402} {\bibfield  {journal} {\bibinfo  {journal}
  {Phys. Rev. Lett.}\ }\textbf {\bibinfo {volume} {102}},\ \bibinfo {pages}
  {230402} (\bibinfo {year} {2009})}\BibitemShut {NoStop}%
\bibitem [{\citenamefont {Kohstall}\ \emph {et~al.}(2012)\citenamefont
  {Kohstall}, \citenamefont {Zaccanti}, \citenamefont {Jag}, \citenamefont
  {Trenkwalder}, \citenamefont {Massignan}, \citenamefont {Bruun},
  \citenamefont {Schreck},\ and\ \citenamefont {Grimm}}]{Kohstall2012}%
  \BibitemOpen
  \bibfield  {author} {\bibinfo {author} {\bibfnamefont {C.}~\bibnamefont
  {Kohstall}}, \bibinfo {author} {\bibfnamefont {M.}~\bibnamefont {Zaccanti}},
  \bibinfo {author} {\bibfnamefont {M.}~\bibnamefont {Jag}}, \bibinfo {author}
  {\bibfnamefont {A.}~\bibnamefont {Trenkwalder}}, \bibinfo {author}
  {\bibfnamefont {P.}~\bibnamefont {Massignan}}, \bibinfo {author}
  {\bibfnamefont {G.~M.}\ \bibnamefont {Bruun}}, \bibinfo {author}
  {\bibfnamefont {F.}~\bibnamefont {Schreck}}, \ and\ \bibinfo {author}
  {\bibfnamefont {R.}~\bibnamefont {Grimm}},\ }\bibfield  {title} {\bibinfo
  {title} {\emph {Metastability and coherence of repulsive polarons in a
  strongly interacting Fermi mixture}},\ }\href
  {http://dx.doi.org/10.1038/nature11065} {\bibfield  {journal} {\bibinfo
  {journal} {Nature}\ }\textbf {\bibinfo {volume} {485}},\ \bibinfo {pages}
  {615} (\bibinfo {year} {2012})}\BibitemShut {NoStop}%
\bibitem [{\citenamefont {{Koschorreck}}\ \emph {et~al.}(2012)\citenamefont
  {{Koschorreck}}, \citenamefont {{Pertot}}, \citenamefont {{Vogt}},
  \citenamefont {{Fr{\"o}hlich}}, \citenamefont {{Feld}},\ and\ \citenamefont
  {{K{\"o}hl}}}]{Koschorreck2012}%
  \BibitemOpen
  \bibfield  {author} {\bibinfo {author} {\bibfnamefont {M.}~\bibnamefont
  {{Koschorreck}}}, \bibinfo {author} {\bibfnamefont {D.}~\bibnamefont
  {{Pertot}}}, \bibinfo {author} {\bibfnamefont {E.}~\bibnamefont {{Vogt}}},
  \bibinfo {author} {\bibfnamefont {B.}~\bibnamefont {{Fr{\"o}hlich}}},
  \bibinfo {author} {\bibfnamefont {M.}~\bibnamefont {{Feld}}}, \ and\ \bibinfo
  {author} {\bibfnamefont {M.}~\bibnamefont {{K{\"o}hl}}},\ }\bibfield  {title}
  {\bibinfo {title} {\emph {{Attractive and repulsive Fermi polarons in two
  dimensions}}},\ }\href {\doibase 10.1038/nature11151} {\bibfield  {journal}
  {\bibinfo  {journal} {Nature}\ }\textbf {\bibinfo {volume} {485}},\ \bibinfo
  {pages} {619} (\bibinfo {year} {2012})}\BibitemShut {NoStop}%
\bibitem [{\citenamefont {Massignan}\ \emph
  {et~al.}(2014{\natexlab{b}})\citenamefont {Massignan}, \citenamefont
  {Zaccanti},\ and\ \citenamefont {Bruun}}]{MassignanPolRev2014}%
  \BibitemOpen
  \bibfield  {author} {\bibinfo {author} {\bibfnamefont {P.}~\bibnamefont
  {Massignan}}, \bibinfo {author} {\bibfnamefont {M.}~\bibnamefont {Zaccanti}},
  \ and\ \bibinfo {author} {\bibfnamefont {G.~M.}\ \bibnamefont {Bruun}},\
  }\bibfield  {title} {\bibinfo {title} {\emph {Polarons, dressed molecules and
  itinerant ferromagnetism in ultracold Fermi gases}},\ }\href
  {http://stacks.iop.org/0034-4885/77/i=3/a=034401} {\bibfield  {journal}
  {\bibinfo  {journal} {Reports on Progress in Physics}\ }\textbf {\bibinfo
  {volume} {77}},\ \bibinfo {pages} {034401} (\bibinfo {year}
  {2014}{\natexlab{b}})}\BibitemShut {NoStop}%
\bibitem [{\citenamefont {{Lan}}\ and\ \citenamefont {{Lobo}}(2014)}]{Lan2014}%
  \BibitemOpen
  \bibfield  {author} {\bibinfo {author} {\bibfnamefont {Z.}~\bibnamefont
  {{Lan}}}\ and\ \bibinfo {author} {\bibfnamefont {C.}~\bibnamefont {{Lobo}}},\
  }\bibfield  {title} {\bibinfo {title} {\emph {{A single impurity in an ideal
  atomic Fermi gas: current understanding and some open problems}}},\
  }\href@noop {} {\bibfield  {journal} {\bibinfo  {journal} {J. Indian I.
  Sci.}\ }\textbf {\bibinfo {volume} {94}},\ \bibinfo {pages} {179} (\bibinfo
  {year} {2014})}\BibitemShut {NoStop}%
\bibitem [{\citenamefont {{Levinsen}}\ and\ \citenamefont
  {{Parish}}(2015)}]{Levinsen2014}%
  \BibitemOpen
  \bibfield  {author} {\bibinfo {author} {\bibfnamefont {J.}~\bibnamefont
  {{Levinsen}}}\ and\ \bibinfo {author} {\bibfnamefont {M.~M.}\ \bibnamefont
  {{Parish}}},\ }\bibfield  {title} {\bibinfo {title} {\emph {{Strongly
  interacting two-dimensional Fermi gases}}},\ }\href
  {http://www.worldscientific.com/worldscibooks/10.1142/9561} {\bibfield
  {journal} {\bibinfo  {journal} {Annual Review of Cold Atoms and Molecules}\
  }\textbf {\bibinfo {volume} {3}},\ \bibinfo {pages} {1} (\bibinfo {year}
  {2015})}\BibitemShut {NoStop}%
\bibitem [{\citenamefont {C\^ot\'e}\ \emph {et~al.}(2002)\citenamefont
  {C\^ot\'e}, \citenamefont {Kharchenko},\ and\ \citenamefont
  {Lukin}}]{Cote2002}%
  \BibitemOpen
  \bibfield  {author} {\bibinfo {author} {\bibfnamefont {R.}~\bibnamefont
  {C\^ot\'e}}, \bibinfo {author} {\bibfnamefont {V.}~\bibnamefont
  {Kharchenko}}, \ and\ \bibinfo {author} {\bibfnamefont {M.~D.}\ \bibnamefont
  {Lukin}},\ }\bibfield  {title} {\bibinfo {title} {\emph {Mesoscopic Molecular
  Ions in Bose-Einstein Condensates}},\ }\href {\doibase
  10.1103/PhysRevLett.89.093001} {\bibfield  {journal} {\bibinfo  {journal}
  {Phys. Rev. Lett.}\ }\textbf {\bibinfo {volume} {89}},\ \bibinfo {pages}
  {093001} (\bibinfo {year} {2002})}\BibitemShut {NoStop}%
\bibitem [{\citenamefont {Massignan}\ \emph {et~al.}(2005)\citenamefont
  {Massignan}, \citenamefont {Pethick},\ and\ \citenamefont
  {Smith}}]{Massignan2005}%
  \BibitemOpen
  \bibfield  {author} {\bibinfo {author} {\bibfnamefont {P.}~\bibnamefont
  {Massignan}}, \bibinfo {author} {\bibfnamefont {C.~J.}\ \bibnamefont
  {Pethick}}, \ and\ \bibinfo {author} {\bibfnamefont {H.}~\bibnamefont
  {Smith}},\ }\bibfield  {title} {\bibinfo {title} {\emph {Static properties of
  positive ions in atomic Bose-Einstein condensates}},\ }\href {\doibase
  10.1103/PhysRevA.71.023606} {\bibfield  {journal} {\bibinfo  {journal} {Phys.
  Rev. A}\ }\textbf {\bibinfo {volume} {71}},\ \bibinfo {pages} {023606}
  (\bibinfo {year} {2005})}\BibitemShut {NoStop}%
\bibitem [{\citenamefont {Cucchietti}\ and\ \citenamefont
  {Timmermans}(2006)}]{Cucchietti2006}%
  \BibitemOpen
  \bibfield  {author} {\bibinfo {author} {\bibfnamefont {F.~M.}\ \bibnamefont
  {Cucchietti}}\ and\ \bibinfo {author} {\bibfnamefont {E.}~\bibnamefont
  {Timmermans}},\ }\bibfield  {title} {\bibinfo {title} {\emph {Strong-Coupling
  Polarons in Dilute Gas Bose-Einstein Condensates}},\ }\href {\doibase
  10.1103/PhysRevLett.96.210401} {\bibfield  {journal} {\bibinfo  {journal}
  {Phys. Rev. Lett.}\ }\textbf {\bibinfo {volume} {96}},\ \bibinfo {pages}
  {210401} (\bibinfo {year} {2006})}\BibitemShut {NoStop}%
\bibitem [{\citenamefont {Palzer}\ \emph {et~al.}(2009)\citenamefont {Palzer},
  \citenamefont {Zipkes}, \citenamefont {Sias},\ and\ \citenamefont
  {K\"ohl}}]{Palzer2009}%
  \BibitemOpen
  \bibfield  {author} {\bibinfo {author} {\bibfnamefont {S.}~\bibnamefont
  {Palzer}}, \bibinfo {author} {\bibfnamefont {C.}~\bibnamefont {Zipkes}},
  \bibinfo {author} {\bibfnamefont {C.}~\bibnamefont {Sias}}, \ and\ \bibinfo
  {author} {\bibfnamefont {M.}~\bibnamefont {K\"ohl}},\ }\bibfield  {title}
  {\bibinfo {title} {\emph {Quantum Transport through a Tonks-Girardeau Gas}},\
  }\href {\doibase 10.1103/PhysRevLett.103.150601} {\bibfield  {journal}
  {\bibinfo  {journal} {Phys. Rev. Lett.}\ }\textbf {\bibinfo {volume} {103}},\
  \bibinfo {pages} {150601} (\bibinfo {year} {2009})}\BibitemShut {NoStop}%
\bibitem [{\citenamefont {Catani}\ \emph {et~al.}(2012)\citenamefont {Catani},
  \citenamefont {Lamporesi}, \citenamefont {Naik}, \citenamefont {Gring},
  \citenamefont {Inguscio}, \citenamefont {Minardi}, \citenamefont {Kantian},\
  and\ \citenamefont {Giamarchi}}]{Catani2012}%
  \BibitemOpen
  \bibfield  {author} {\bibinfo {author} {\bibfnamefont {J.}~\bibnamefont
  {Catani}}, \bibinfo {author} {\bibfnamefont {G.}~\bibnamefont {Lamporesi}},
  \bibinfo {author} {\bibfnamefont {D.}~\bibnamefont {Naik}}, \bibinfo {author}
  {\bibfnamefont {M.}~\bibnamefont {Gring}}, \bibinfo {author} {\bibfnamefont
  {M.}~\bibnamefont {Inguscio}}, \bibinfo {author} {\bibfnamefont
  {F.}~\bibnamefont {Minardi}}, \bibinfo {author} {\bibfnamefont
  {A.}~\bibnamefont {Kantian}}, \ and\ \bibinfo {author} {\bibfnamefont
  {T.}~\bibnamefont {Giamarchi}},\ }\bibfield  {title} {\bibinfo {title} {\emph
  {Quantum dynamics of impurities in a one-dimensional Bose gas}},\ }\href
  {\doibase 10.1103/PhysRevA.85.023623} {\bibfield  {journal} {\bibinfo
  {journal} {Phys. Rev. A}\ }\textbf {\bibinfo {volume} {85}},\ \bibinfo
  {pages} {023623} (\bibinfo {year} {2012})}\BibitemShut {NoStop}%
\bibitem [{\citenamefont {Rath}\ and\ \citenamefont
  {Schmidt}(2013)}]{Rath2013}%
  \BibitemOpen
  \bibfield  {author} {\bibinfo {author} {\bibfnamefont {S.~P.}\ \bibnamefont
  {Rath}}\ and\ \bibinfo {author} {\bibfnamefont {R.}~\bibnamefont {Schmidt}},\
  }\bibfield  {title} {\bibinfo {title} {\emph {Field-theoretical study of the
  Bose polaron}},\ }\href {\doibase 10.1103/PhysRevA.88.053632} {\bibfield
  {journal} {\bibinfo  {journal} {Phys. Rev. A}\ }\textbf {\bibinfo {volume}
  {88}},\ \bibinfo {pages} {053632} (\bibinfo {year} {2013})}\BibitemShut
  {NoStop}%
\bibitem [{\citenamefont {{Fukuhara}}\ \emph {et~al.}(2013)\citenamefont
  {{Fukuhara}}, \citenamefont {{Kantian}}, \citenamefont {{Endres}},
  \citenamefont {{Cheneau}}, \citenamefont {{Schau{\ss}}}, \citenamefont
  {{Hild}}, \citenamefont {{Bellem}}, \citenamefont {{Schollw{\"o}ck}},
  \citenamefont {{Giamarchi}}, \citenamefont {{Gross}}, \citenamefont
  {{Bloch}},\ and\ \citenamefont {{Kuhr}}}]{Fukuhara2013}%
  \BibitemOpen
  \bibfield  {author} {\bibinfo {author} {\bibfnamefont {T.}~\bibnamefont
  {{Fukuhara}}}, \bibinfo {author} {\bibfnamefont {A.}~\bibnamefont
  {{Kantian}}}, \bibinfo {author} {\bibfnamefont {M.}~\bibnamefont {{Endres}}},
  \bibinfo {author} {\bibfnamefont {M.}~\bibnamefont {{Cheneau}}}, \bibinfo
  {author} {\bibfnamefont {P.}~\bibnamefont {{Schau{\ss}}}}, \bibinfo {author}
  {\bibfnamefont {S.}~\bibnamefont {{Hild}}}, \bibinfo {author} {\bibfnamefont
  {D.}~\bibnamefont {{Bellem}}}, \bibinfo {author} {\bibfnamefont
  {U.}~\bibnamefont {{Schollw{\"o}ck}}}, \bibinfo {author} {\bibfnamefont
  {T.}~\bibnamefont {{Giamarchi}}}, \bibinfo {author} {\bibfnamefont
  {C.}~\bibnamefont {{Gross}}}, \bibinfo {author} {\bibfnamefont
  {I.}~\bibnamefont {{Bloch}}}, \ and\ \bibinfo {author} {\bibfnamefont
  {S.}~\bibnamefont {{Kuhr}}},\ }\bibfield  {title} {\bibinfo {title} {\emph
  {{Quantum dynamics of a mobile spin impurity}}},\ }\href {\doibase
  10.1038/nphys2561} {\bibfield  {journal} {\bibinfo  {journal} {Nature
  Physics}\ }\textbf {\bibinfo {volume} {9}},\ \bibinfo {pages} {235} (\bibinfo
  {year} {2013})}\BibitemShut {NoStop}%
\bibitem [{\citenamefont {Shashi}\ \emph {et~al.}(2014)\citenamefont {Shashi},
  \citenamefont {Grusdt}, \citenamefont {Abanin},\ and\ \citenamefont
  {Demler}}]{Shashi2014}%
  \BibitemOpen
  \bibfield  {author} {\bibinfo {author} {\bibfnamefont {A.}~\bibnamefont
  {Shashi}}, \bibinfo {author} {\bibfnamefont {F.}~\bibnamefont {Grusdt}},
  \bibinfo {author} {\bibfnamefont {D.~A.}\ \bibnamefont {Abanin}}, \ and\
  \bibinfo {author} {\bibfnamefont {E.}~\bibnamefont {Demler}},\ }\bibfield
  {title} {\bibinfo {title} {\emph {Radio-frequency spectroscopy of polarons in
  ultracold Bose gases}},\ }\href {\doibase 10.1103/PhysRevA.89.053617}
  {\bibfield  {journal} {\bibinfo  {journal} {Phys. Rev. A}\ }\textbf {\bibinfo
  {volume} {89}},\ \bibinfo {pages} {053617} (\bibinfo {year}
  {2014})}\BibitemShut {NoStop}%
\bibitem [{\citenamefont {{Grusdt}}\ \emph
  {et~al.}(2014{\natexlab{a}})\citenamefont {{Grusdt}}, \citenamefont
  {{Shashi}}, \citenamefont {{Abanin}},\ and\ \citenamefont
  {{Demler}}}]{Grusdt2014a}%
  \BibitemOpen
  \bibfield  {author} {\bibinfo {author} {\bibfnamefont {F.}~\bibnamefont
  {{Grusdt}}}, \bibinfo {author} {\bibfnamefont {A.}~\bibnamefont {{Shashi}}},
  \bibinfo {author} {\bibfnamefont {D.}~\bibnamefont {{Abanin}}}, \ and\
  \bibinfo {author} {\bibfnamefont {E.}~\bibnamefont {{Demler}}},\ }\bibfield
  {title} {\bibinfo {title} {\emph {{Bloch oscillations of bosonic lattice
  polarons}}},\ }\href@noop {} {\  (\bibinfo {year} {2014}{\natexlab{a}})},\
  \Eprint {http://arxiv.org/abs/1410.1513} {arXiv:1410.1513} \BibitemShut
  {NoStop}%
\bibitem [{\citenamefont {{Grusdt}}\ \emph
  {et~al.}(2014{\natexlab{b}})\citenamefont {{Grusdt}}, \citenamefont
  {{Shchadilova}}, \citenamefont {{Rubtsov}},\ and\ \citenamefont
  {{Demler}}}]{Grusdt2014b}%
  \BibitemOpen
  \bibfield  {author} {\bibinfo {author} {\bibfnamefont {F.}~\bibnamefont
  {{Grusdt}}}, \bibinfo {author} {\bibfnamefont {Y.~E.}\ \bibnamefont
  {{Shchadilova}}}, \bibinfo {author} {\bibfnamefont {A.~N.}\ \bibnamefont
  {{Rubtsov}}}, \ and\ \bibinfo {author} {\bibfnamefont {E.}~\bibnamefont
  {{Demler}}},\ }\bibfield  {title} {\bibinfo {title} {\emph {{Renormalization
  group approach to the Fr\"ohlich polaron model: application to impurity-BEC
  problem}}},\ }\href@noop {} {\  (\bibinfo {year} {2014}{\natexlab{b}})},\
  \Eprint {http://arxiv.org/abs/1410.2203} {arXiv:1410.2203} \BibitemShut
  {NoStop}%
\bibitem [{\citenamefont {Devreese}\ and\ \citenamefont
  {Alexandrov}(2009)}]{Devreese2009}%
  \BibitemOpen
  \bibfield  {author} {\bibinfo {author} {\bibfnamefont {J.~T.}\ \bibnamefont
  {Devreese}}\ and\ \bibinfo {author} {\bibfnamefont {A.~S.}\ \bibnamefont
  {Alexandrov}},\ }\bibfield  {title} {\bibinfo {title} {\emph {Fr{\"o}hlich
  polaron and bipolaron: recent developments}},\ }\href
  {http://stacks.iop.org/0034-4885/72/i=6/a=066501} {\bibfield  {journal}
  {\bibinfo  {journal} {Reports on Progress in Physics}\ }\textbf {\bibinfo
  {volume} {72}},\ \bibinfo {pages} {066501} (\bibinfo {year}
  {2009})}\BibitemShut {NoStop}%
\bibitem [{\citenamefont {Alexandrov}\ and\ \citenamefont
  {Devreese}(2009)}]{Alexandrov2009}%
  \BibitemOpen
  \bibfield  {author} {\bibinfo {author} {\bibfnamefont {A.}~\bibnamefont
  {Alexandrov}}\ and\ \bibinfo {author} {\bibfnamefont {J.}~\bibnamefont
  {Devreese}},\ }\href {http://books.google.es/books?id=EI0Hql-9oY8C} {\emph
  {\bibinfo {title} {Advances in Polaron Physics}}},\ Springer Series in
  Solid-State Sciences\ (\bibinfo  {publisher} {Springer},\ \bibinfo {year}
  {2009})\BibitemShut {NoStop}%
\bibitem [{\citenamefont {Lieb}\ and\ \citenamefont
  {Yamazaki}(1958)}]{Lieb1958}%
  \BibitemOpen
  \bibfield  {author} {\bibinfo {author} {\bibfnamefont {E.~H.}\ \bibnamefont
  {Lieb}}\ and\ \bibinfo {author} {\bibfnamefont {K.}~\bibnamefont
  {Yamazaki}},\ }\bibfield  {title} {\bibinfo {title} {\emph {Ground-State
  Energy and Effective Mass of the Polaron}},\ }\href@noop {} {\bibfield
  {journal} {\bibinfo  {journal} {Phys. Rev.}\ }\textbf {\bibinfo {volume}
  {111}},\ \bibinfo {pages} {728} (\bibinfo {year} {1958})}\BibitemShut
  {NoStop}%
\bibitem [{\citenamefont {Lieb}\ and\ \citenamefont {Thomas}(1997)}]{Lieb1997}%
  \BibitemOpen
  \bibfield  {author} {\bibinfo {author} {\bibfnamefont {E.~H.}\ \bibnamefont
  {Lieb}}\ and\ \bibinfo {author} {\bibfnamefont {L.~E.}\ \bibnamefont
  {Thomas}},\ }\bibfield  {title} {\bibinfo {title} {\emph {Exact ground state
  energy of the strong-coupling polaron}},\ }\href@noop {} {\bibfield
  {journal} {\bibinfo  {journal} {Comm. Math. Phys.}\ }\textbf {\bibinfo
  {volume} {183}},\ \bibinfo {pages} {511} (\bibinfo {year}
  {1997})}\BibitemShut {NoStop}%
\bibitem [{\citenamefont {Frank}\ \emph {et~al.}(2010)\citenamefont {Frank},
  \citenamefont {Lieb}, \citenamefont {Seiringer},\ and\ \citenamefont
  {Thomas}}]{Frank2010}%
  \BibitemOpen
  \bibfield  {author} {\bibinfo {author} {\bibfnamefont {R.~L.}\ \bibnamefont
  {Frank}}, \bibinfo {author} {\bibfnamefont {E.~H.}\ \bibnamefont {Lieb}},
  \bibinfo {author} {\bibfnamefont {R.}~\bibnamefont {Seiringer}}, \ and\
  \bibinfo {author} {\bibfnamefont {L.~E.}\ \bibnamefont {Thomas}},\ }\bibfield
   {title} {\bibinfo {title} {\emph {Bi-polaron and N-polaron binding
  energies}},\ }\href@noop {} {\bibfield  {journal} {\bibinfo  {journal} {Phys.
  Rev. Lett.}\ }\textbf {\bibinfo {volume} {104}},\ \bibinfo {pages} {210402}
  (\bibinfo {year} {2010})}\BibitemShut {NoStop}%
\bibitem [{\citenamefont {Anapolitanos}\ and\ \citenamefont
  {Landon}(2013)}]{Anapolitanos2013}%
  \BibitemOpen
  \bibfield  {author} {\bibinfo {author} {\bibfnamefont {I.}~\bibnamefont
  {Anapolitanos}}\ and\ \bibinfo {author} {\bibfnamefont {B.}~\bibnamefont
  {Landon}},\ }\bibfield  {title} {\bibinfo {title} {\emph {The Ground State
  Energy of the Multi-Polaron in the Strong Coupling Limit}},\ }\href@noop {}
  {\bibfield  {journal} {\bibinfo  {journal} {Lett. Math. Phys.}\ }\textbf
  {\bibinfo {volume} {103}},\ \bibinfo {pages} {1347} (\bibinfo {year}
  {2013})}\BibitemShut {NoStop}%
\bibitem [{\citenamefont {Sancho}\ \emph {et~al.}(1982)\citenamefont {Sancho},
  \citenamefont {Miguel},\ and\ \citenamefont {D\"urr}}]{Sancho1982}%
  \BibitemOpen
  \bibfield  {author} {\bibinfo {author} {\bibfnamefont {J.~M.}\ \bibnamefont
  {Sancho}}, \bibinfo {author} {\bibfnamefont {M.~S.}\ \bibnamefont {Miguel}},
  \ and\ \bibinfo {author} {\bibfnamefont {D.}~\bibnamefont {D\"urr}},\
  }\bibfield  {title} {\bibinfo {title} {\emph {{Adiabatic elimination for
  systems of Brownian particles with nonconstant damping coefficients}}},\
  }\href@noop {} {\bibfield  {journal} {\bibinfo  {journal} {J. Stat. Phys.}\
  }\textbf {\bibinfo {volume} {28}},\ \bibinfo {pages} {291} (\bibinfo {year}
  {1982})}\BibitemShut {NoStop}%
\bibitem [{\citenamefont {Freidlin}(2004)}]{Freidlin2004}%
  \BibitemOpen
  \bibfield  {author} {\bibinfo {author} {\bibfnamefont {M.}~\bibnamefont
  {Freidlin}},\ }\bibfield  {title} {\bibinfo {title} {\emph {{Some remarks on
  the Smoluchowski-Kramers approximation}}},\ }\href@noop {} {\bibfield
  {journal} {\bibinfo  {journal} {J. Stat. Phys.}\ }\textbf {\bibinfo {volume}
  {117}},\ \bibinfo {pages} {314} (\bibinfo {year} {2004})}\BibitemShut
  {NoStop}%
\bibitem [{\citenamefont {Cerrai}\ and\ \citenamefont
  {Freidlin}(2011)}]{Cerrai2011}%
  \BibitemOpen
  \bibfield  {author} {\bibinfo {author} {\bibfnamefont {S.}~\bibnamefont
  {Cerrai}}\ and\ \bibinfo {author} {\bibfnamefont {M.}~\bibnamefont
  {Freidlin}},\ }\bibfield  {title} {\bibinfo {title} {\emph {{Small mass
  asymptotics for a charged particle in a magnetic field and long-time
  influence of small perturbations}}},\ }\href@noop {} {\bibfield  {journal}
  {\bibinfo  {journal} {J. Stat. Phys.}\ }\textbf {\bibinfo {volume} {144}},\
  \bibinfo {pages} {101} (\bibinfo {year} {2011})}\BibitemShut {NoStop}%
\bibitem [{\citenamefont {Freidlin}\ and\ \citenamefont
  {Hu}(2011)}]{Freidlin2011}%
  \BibitemOpen
  \bibfield  {author} {\bibinfo {author} {\bibfnamefont {M.}~\bibnamefont
  {Freidlin}}\ and\ \bibinfo {author} {\bibfnamefont {W.}~\bibnamefont {Hu}},\
  }\bibfield  {title} {\bibinfo {title} {\emph {{Smoluchowski--Kramers
  approximation in the case of variable friction}}},\ }\href@noop {} {\bibfield
   {journal} {\bibinfo  {journal} {J. Math. Sci.}\ }\textbf {\bibinfo {volume}
  {179}},\ \bibinfo {pages} {184} (\bibinfo {year} {2011})}\BibitemShut
  {NoStop}%
\bibitem [{\citenamefont {Shi}\ \emph {et~al.}(2012)\citenamefont {Shi},
  \citenamefont {Chen}, \citenamefont {Yuan}, \citenamefont {Yuan},\ and\
  \citenamefont {Ao}}]{Shi2012}%
  \BibitemOpen
  \bibfield  {author} {\bibinfo {author} {\bibfnamefont {J.}~\bibnamefont
  {Shi}}, \bibinfo {author} {\bibfnamefont {T.}~\bibnamefont {Chen}}, \bibinfo
  {author} {\bibfnamefont {R.}~\bibnamefont {Yuan}}, \bibinfo {author}
  {\bibfnamefont {B.}~\bibnamefont {Yuan}}, \ and\ \bibinfo {author}
  {\bibfnamefont {P.}~\bibnamefont {Ao}},\ }\bibfield  {title} {\bibinfo
  {title} {\emph {Relation of a new interpretation of stochastic differential
  equations to Ito process}},\ }\href@noop {} {\bibfield  {journal} {\bibinfo
  {journal} {J. Stat. Phys.}\ }\textbf {\bibinfo {volume} {148}},\ \bibinfo
  {pages} {579} (\bibinfo {year} {2012})}\BibitemShut {NoStop}%
\bibitem [{\citenamefont {Caldeira}\ and\ \citenamefont
  {Leggett}(1983{\natexlab{a}})}]{Caldeira1983a}%
  \BibitemOpen
  \bibfield  {author} {\bibinfo {author} {\bibfnamefont {A.}~\bibnamefont
  {Caldeira}}\ and\ \bibinfo {author} {\bibfnamefont {A.}~\bibnamefont
  {Leggett}},\ }\bibfield  {title} {\bibinfo {title} {\emph {Path integral
  approach to quantum Brownian motion}},\ }\href {\doibase
  http://dx.doi.org/10.1016/0378-4371(83)90013-4} {\bibfield  {journal}
  {\bibinfo  {journal} {Physica A: Statistical Mechanics and its Applications}\
  }\textbf {\bibinfo {volume} {121}},\ \bibinfo {pages} {587 } (\bibinfo {year}
  {1983}{\natexlab{a}})}\BibitemShut {NoStop}%
\bibitem [{\citenamefont {Caldeira}\ and\ \citenamefont
  {Leggett}(1983{\natexlab{b}})}]{Caldeira1983b}%
  \BibitemOpen
  \bibfield  {author} {\bibinfo {author} {\bibfnamefont {A.}~\bibnamefont
  {Caldeira}}\ and\ \bibinfo {author} {\bibfnamefont {A.}~\bibnamefont
  {Leggett}},\ }\bibfield  {title} {\bibinfo {title} {\emph {Quantum tunnelling
  in a dissipative system}},\ }\href {\doibase
  http://dx.doi.org/10.1016/0003-4916(83)90202-6} {\bibfield  {journal}
  {\bibinfo  {journal} {Annals of Physics}\ }\textbf {\bibinfo {volume}
  {149}},\ \bibinfo {pages} {374 } (\bibinfo {year}
  {1983}{\natexlab{b}})}\BibitemShut {NoStop}%
\bibitem [{\citenamefont {Fleming}\ and\ \citenamefont
  {Hu}(2012)}]{Fleming2012}%
  \BibitemOpen
  \bibfield  {author} {\bibinfo {author} {\bibfnamefont {C.}~\bibnamefont
  {Fleming}}\ and\ \bibinfo {author} {\bibfnamefont {B.}~\bibnamefont {Hu}},\
  }\bibfield  {title} {\bibinfo {title} {\emph {Non-Markovian dynamics of open
  quantum systems: Stochastic equations and their perturbative solutions}},\
  }\href {\doibase http://dx.doi.org/10.1016/j.aop.2011.12.006} {\bibfield
  {journal} {\bibinfo  {journal} {Annals of Physics}\ }\textbf {\bibinfo
  {volume} {327}},\ \bibinfo {pages} {1238 } (\bibinfo {year}
  {2012})}\BibitemShut {NoStop}%
\bibitem [{\citenamefont {Rz{\c a}\.zewski}\ and\ \citenamefont
  {\.Zakowicz}(1976)}]{Rzazewski1976}%
  \BibitemOpen
  \bibfield  {author} {\bibinfo {author} {\bibfnamefont {K.}~\bibnamefont
  {Rz{\c a}\.zewski}}\ and\ \bibinfo {author} {\bibfnamefont {W.}~\bibnamefont
  {\.Zakowicz}},\ }\bibfield  {title} {\bibinfo {title} {\emph {Initial value
  problem and causality of radiating oscillator}},\ }\href@noop {} {\bibfield
  {journal} {\bibinfo  {journal} {J. Phys. A}\ }\textbf {\bibinfo {volume}
  {9}},\ \bibinfo {pages} {1159} (\bibinfo {year} {1976})}\BibitemShut
  {NoStop}%
\bibitem [{\citenamefont {Lewenstein}\ and\ \citenamefont {Rz{\c
  a}\.zewski}(1980)}]{Lewenstein1980}%
  \BibitemOpen
  \bibfield  {author} {\bibinfo {author} {\bibfnamefont {M.}~\bibnamefont
  {Lewenstein}}\ and\ \bibinfo {author} {\bibfnamefont {K.}~\bibnamefont {Rz{\c
  a}\.zewski}},\ }\bibfield  {title} {\bibinfo {title} {\emph {Collective
  radiation and the near zone field}},\ }\href@noop {} {\bibfield  {journal}
  {\bibinfo  {journal} {J. Phys. A}\ }\textbf {\bibinfo {volume} {13}},\
  \bibinfo {pages} {743} (\bibinfo {year} {1980})}\BibitemShut {NoStop}%
\bibitem [{\citenamefont {Redfield}(1957)}]{Redfield1957}%
  \BibitemOpen
  \bibfield  {author} {\bibinfo {author} {\bibfnamefont {A.~G.}\ \bibnamefont
  {Redfield}},\ }\bibfield  {title} {\bibinfo {title} {\emph {On the Theory of
  Relaxation Processes}},\ }\href {\doibase 10.1147/rd.11.0019} {\bibfield
  {journal} {\bibinfo  {journal} {IBM Journal of Research and Development}\
  }\textbf {\bibinfo {volume} {1}},\ \bibinfo {pages} {19} (\bibinfo {year}
  {1957})}\BibitemShut {NoStop}%
\bibitem [{\citenamefont {Blum}(1981)}]{Blum1981}%
  \BibitemOpen
  \bibfield  {author} {\bibinfo {author} {\bibfnamefont {K.}~\bibnamefont
  {Blum}},\ }\href@noop {} {\emph {\bibinfo {title} {Density Matrix Theory and
  Applications}}}\ (\bibinfo  {publisher} {Plenum Press},\ \bibinfo {address}
  {New York, London},\ \bibinfo {year} {1981})\BibitemShut {NoStop}%
\bibitem [{\citenamefont {Haake}(1982)}]{Haake-ZfP}%
  \BibitemOpen
  \bibfield  {author} {\bibinfo {author} {\bibfnamefont {F.}~\bibnamefont
  {Haake}},\ }\bibfield  {title} {\bibinfo {title} {\emph {Systematic adiabatic
  elimination for stochastic processes}},\ }\href {\doibase 10.1007/BF02026425}
  {\bibfield  {journal} {\bibinfo  {journal} {Zeitschrift f\"ur Physik B
  Condensed Matter}\ }\textbf {\bibinfo {volume} {48}},\ \bibinfo {pages} {31}
  (\bibinfo {year} {1982})}\BibitemShut {NoStop}%
\bibitem [{\citenamefont {Haake}\ and\ \citenamefont
  {Lewenstein}(1983)}]{MLFH}%
  \BibitemOpen
  \bibfield  {author} {\bibinfo {author} {\bibfnamefont {F.}~\bibnamefont
  {Haake}}\ and\ \bibinfo {author} {\bibfnamefont {M.}~\bibnamefont
  {Lewenstein}},\ }\bibfield  {title} {\bibinfo {title} {\emph {Adiabatic drag
  and initial slip in random processes}},\ }\href {\doibase
  10.1103/PhysRevA.28.3606} {\bibfield  {journal} {\bibinfo  {journal} {Phys.
  Rev. A}\ }\textbf {\bibinfo {volume} {28}},\ \bibinfo {pages} {3606}
  (\bibinfo {year} {1983})}\BibitemShut {NoStop}%
\bibitem [{\citenamefont {Haake}\ \emph {et~al.}(1985)\citenamefont {Haake},
  \citenamefont {Lewenstein},\ and\ \citenamefont {Reibold}}]{Reibold1}%
  \BibitemOpen
  \bibfield  {author} {\bibinfo {author} {\bibfnamefont {F.}~\bibnamefont
  {Haake}}, \bibinfo {author} {\bibfnamefont {M.}~\bibnamefont {Lewenstein}}, \
  and\ \bibinfo {author} {\bibfnamefont {R.}~\bibnamefont {Reibold}},\
  }\href@noop {} {\emph {\bibinfo {title} {Adiabatic drag and initial slip for
  random processes with slow and fast variables}}},\ edited by\ \bibinfo
  {editor} {\bibfnamefont {L.}~\bibnamefont {Accardi}}\ and\ \bibinfo {editor}
  {\bibfnamefont {W.}~\bibnamefont {von Waldenfels}},\ Lecture Note in
  Mathematics: Quantum Probability and Applications II\ (\bibinfo  {publisher}
  {Springer},\ \bibinfo {address} {Heidelberg},\ \bibinfo {year}
  {1985})\BibitemShut {NoStop}%
\bibitem [{\citenamefont {Grabert}\ \emph {et~al.}(1988)\citenamefont
  {Grabert}, \citenamefont {Schramm},\ and\ \citenamefont
  {Ingold}}]{Grabert1988}%
  \BibitemOpen
  \bibfield  {author} {\bibinfo {author} {\bibfnamefont {H.}~\bibnamefont
  {Grabert}}, \bibinfo {author} {\bibfnamefont {P.}~\bibnamefont {Schramm}}, \
  and\ \bibinfo {author} {\bibfnamefont {G.-L.}\ \bibnamefont {Ingold}},\
  }\bibfield  {title} {\bibinfo {title} {\emph {Quantum Brownian motion: The
  functional integral approach}},\ }\href {\doibase
  http://dx.doi.org/10.1016/0370-1573(88)90023-3} {\bibfield  {journal}
  {\bibinfo  {journal} {Physics Reports}\ }\textbf {\bibinfo {volume} {168}},\
  \bibinfo {pages} {115 } (\bibinfo {year} {1988})}\BibitemShut {NoStop}%
\bibitem [{\citenamefont {Fleming}\ \emph {et~al.}(2011)\citenamefont
  {Fleming}, \citenamefont {Roura},\ and\ \citenamefont {Hu}}]{Fleming2011}%
  \BibitemOpen
  \bibfield  {author} {\bibinfo {author} {\bibfnamefont {C.}~\bibnamefont
  {Fleming}}, \bibinfo {author} {\bibfnamefont {A.}~\bibnamefont {Roura}}, \
  and\ \bibinfo {author} {\bibfnamefont {B.}~\bibnamefont {Hu}},\ }\bibfield
  {title} {\bibinfo {title} {\emph {Exact analytical solutions to the master
  equation of quantum Brownian motion for a general environment}},\ }\href
  {\doibase http://dx.doi.org/10.1016/j.aop.2010.12.003} {\bibfield  {journal}
  {\bibinfo  {journal} {Annals of Physics}\ }\textbf {\bibinfo {volume}
  {326}},\ \bibinfo {pages} {1207 } (\bibinfo {year} {2011})}\BibitemShut
  {NoStop}%
\bibitem [{\citenamefont {Ullersma}(1966{\natexlab{a}})}]{Ullersma1}%
  \BibitemOpen
  \bibfield  {author} {\bibinfo {author} {\bibfnamefont {P.}~\bibnamefont
  {Ullersma}},\ }\bibfield  {title} {\bibinfo {title} {\emph {An exactly
  solvable model for Brownian motion: I. Derivation of the Langevin
  equation}},\ }\href {\doibase 10.1016/0031-8914(66)90102-9} {\bibfield
  {journal} {\bibinfo  {journal} {Physica}\ }\textbf {\bibinfo {volume} {32}},\
  \bibinfo {pages} {27} (\bibinfo {year} {1966}{\natexlab{a}})}\BibitemShut
  {NoStop}%
\bibitem [{\citenamefont {Ullersma}(1966{\natexlab{b}})}]{Ullersma2}%
  \BibitemOpen
  \bibfield  {author} {\bibinfo {author} {\bibfnamefont {P.}~\bibnamefont
  {Ullersma}},\ }\bibfield  {title} {\bibinfo {title} {\emph {An exactly
  solvable model for Brownian motion: II. Derivation of the Fokker-Planck
  equation and the master equation}},\ }\href@noop {} {\bibfield  {journal}
  {\bibinfo  {journal} {Physica}\ }\textbf {\bibinfo {volume} {32}},\ \bibinfo
  {pages} {56} (\bibinfo {year} {1966}{\natexlab{b}})}\BibitemShut {NoStop}%
\bibitem [{\citenamefont {Ullersma}(1966{\natexlab{c}})}]{Ullersma3}%
  \BibitemOpen
  \bibfield  {author} {\bibinfo {author} {\bibfnamefont {P.}~\bibnamefont
  {Ullersma}},\ }\bibfield  {title} {\bibinfo {title} {\emph {An exactly
  solvable model for Brownian motion: III. Motion of a heavy mass in a linear
  chain}},\ }\href@noop {} {\bibfield  {journal} {\bibinfo  {journal}
  {Physica}\ }\textbf {\bibinfo {volume} {32}},\ \bibinfo {pages} {74}
  (\bibinfo {year} {1966}{\natexlab{c}})}\BibitemShut {NoStop}%
\bibitem [{\citenamefont {Ullersma}(1966{\natexlab{d}})}]{Ullersma4}%
  \BibitemOpen
  \bibfield  {author} {\bibinfo {author} {\bibfnamefont {P.}~\bibnamefont
  {Ullersma}},\ }\bibfield  {title} {\bibinfo {title} {\emph {An exactly
  solvable model for Brownian motion: IV. Susceptibility and Nyquist's
  theorem}},\ }\href@noop {} {\bibfield  {journal} {\bibinfo  {journal}
  {Physica}\ }\textbf {\bibinfo {volume} {32}},\ \bibinfo {pages} {90}
  (\bibinfo {year} {1966}{\natexlab{d}})}\BibitemShut {NoStop}%
\bibitem [{\citenamefont {Haake}\ and\ \citenamefont
  {Reibold}(1984)}]{Reibold}%
  \BibitemOpen
  \bibfield  {author} {\bibinfo {author} {\bibfnamefont {F.}~\bibnamefont
  {Haake}}\ and\ \bibinfo {author} {\bibfnamefont {R.}~\bibnamefont
  {Reibold}},\ }\bibfield  {title} {\bibinfo {title} {\emph {Strong damping and
  low-temperature anomalies for the harmonic oscillator}},\ }\href@noop {}
  {\bibfield  {journal} {\bibinfo  {journal} {Phys. Rev. A}\ }\textbf {\bibinfo
  {volume} {29}},\ \bibinfo {pages} {3208} (\bibinfo {year}
  {1984})}\BibitemShut {NoStop}%
\bibitem [{\citenamefont {Zurek}(2003)}]{Zurek2003}%
  \BibitemOpen
  \bibfield  {author} {\bibinfo {author} {\bibfnamefont {W.~H.}\ \bibnamefont
  {Zurek}},\ }\bibfield  {title} {\bibinfo {title} {\emph {Decoherence,
  einselection, and the quantum origins of the classical}},\ }\href {\doibase
  10.1103/RevModPhys.75.715} {\bibfield  {journal} {\bibinfo  {journal} {Rev.
  Mod. Phys.}\ }\textbf {\bibinfo {volume} {75}},\ \bibinfo {pages} {715}
  (\bibinfo {year} {2003})}\BibitemShut {NoStop}%
\bibitem [{\citenamefont {Schlosshauer}(2005)}]{Schlosshauer2005}%
  \BibitemOpen
  \bibfield  {author} {\bibinfo {author} {\bibfnamefont {M.}~\bibnamefont
  {Schlosshauer}},\ }\bibfield  {title} {\bibinfo {title} {\emph {Decoherence,
  the measurement problem, and interpretations of quantum mechanics}},\ }\href
  {\doibase 10.1103/RevModPhys.76.1267} {\bibfield  {journal} {\bibinfo
  {journal} {Rev. Mod. Phys.}\ }\textbf {\bibinfo {volume} {76}},\ \bibinfo
  {pages} {1267} (\bibinfo {year} {2005})}\BibitemShut {NoStop}%
\bibitem [{\citenamefont {Fleming}\ and\ \citenamefont
  {Cummings}(2011)}]{FlemPRE2011}%
  \BibitemOpen
  \bibfield  {author} {\bibinfo {author} {\bibfnamefont {C.~H.}\ \bibnamefont
  {Fleming}}\ and\ \bibinfo {author} {\bibfnamefont {N.~I.}\ \bibnamefont
  {Cummings}},\ }\bibfield  {title} {\bibinfo {title} {\emph {Accuracy of
  perturbative master equations}},\ }\href {\doibase
  10.1103/PhysRevE.83.031117} {\bibfield  {journal} {\bibinfo  {journal} {Phys.
  Rev. E}\ }\textbf {\bibinfo {volume} {83}},\ \bibinfo {pages} {031117}
  (\bibinfo {year} {2011})}\BibitemShut {NoStop}%
\bibitem [{\citenamefont {Suba\c{s}i}\ \emph {et~al.}(2012)\citenamefont
  {Suba\c{s}i}, \citenamefont {Fleming}, \citenamefont {Taylor},\ and\
  \citenamefont {Hu}}]{Subasi2012}%
  \BibitemOpen
  \bibfield  {author} {\bibinfo {author} {\bibfnamefont {Y.}~\bibnamefont
  {Suba\c{s}i}}, \bibinfo {author} {\bibfnamefont {C.~H.}\ \bibnamefont
  {Fleming}}, \bibinfo {author} {\bibfnamefont {J.~M.}\ \bibnamefont {Taylor}},
  \ and\ \bibinfo {author} {\bibfnamefont {B.~L.}\ \bibnamefont {Hu}},\
  }\bibfield  {title} {\bibinfo {title} {\emph {Equilibrium states of open
  quantum systems in the strong coupling regime}},\ }\href {\doibase
  10.1103/PhysRevE.86.061132} {\bibfield  {journal} {\bibinfo  {journal} {Phys.
  Rev. E}\ }\textbf {\bibinfo {volume} {86}},\ \bibinfo {pages} {061132}
  (\bibinfo {year} {2012})}\BibitemShut {NoStop}%
\bibitem [{\citenamefont {Gardiner}(2009)}]{Gardiner-cl}%
  \BibitemOpen
  \bibfield  {author} {\bibinfo {author} {\bibfnamefont {C.}~\bibnamefont
  {Gardiner}},\ }\href@noop {} {\emph {\bibinfo {title} {Stochastic Methods: A
  Handbook for the Natural and Social Sciences}}},\ \bibinfo {series} {Springer
  Series in Synergetics}, Vol.~\bibinfo {volume} {13}\ (\bibinfo  {publisher}
  {Springer},\ \bibinfo {address} {Heidelberg},\ \bibinfo {year}
  {2009})\BibitemShut {NoStop}%
\bibitem [{\citenamefont {Risken}(2012)}]{Risken}%
  \BibitemOpen
  \bibfield  {author} {\bibinfo {author} {\bibfnamefont {H.}~\bibnamefont
  {Risken}},\ }\href {http://books.google.es/books?id=ThNnMQEACAAJ} {\emph
  {\bibinfo {title} {The Fokker-Planck Equation: Methods of Solution and
  Applications}}},\ \bibinfo {series} {Springer Series in Synergetics},
  Vol.~\bibinfo {volume} {18}\ (\bibinfo  {publisher} {Springer},\ \bibinfo
  {address} {Heidelberg},\ \bibinfo {year} {2012})\BibitemShut {NoStop}%
\bibitem [{\citenamefont {Papanicolaou}(1977)}]{papa-book}%
  \BibitemOpen
  \bibfield  {author} {\bibinfo {author} {\bibfnamefont {G.}~\bibnamefont
  {Papanicolaou}},\ }\href@noop {} {\emph {\bibinfo {title} {Modern modeling of
  continuum phenomena (Ninth Summer Sem. Appl. Math., Rensselaer Polytech.
  Inst., Troy, N.Y., 1975)}}},\ \bibinfo {series} {Lect. in Appl. Math.},
  Vol.~\bibinfo {volume} {16}\ (\bibinfo  {publisher} {Amer. Math. Soc.},\
  \bibinfo {year} {1977})\ pp.\ \bibinfo {pages} {109--147}\BibitemShut
  {NoStop}%
\bibitem [{\citenamefont {Pavliotis}\ and\ \citenamefont
  {Stuart}(2008)}]{Pavliotis}%
  \BibitemOpen
  \bibfield  {author} {\bibinfo {author} {\bibfnamefont {G.}~\bibnamefont
  {Pavliotis}}\ and\ \bibinfo {author} {\bibfnamefont {A.}~\bibnamefont
  {Stuart}},\ }\href@noop {} {\emph {\bibinfo {title} {Quantum Brownian Motion:
  The Functional Integral Approach}}},\ \bibinfo {series} {Texts in Applied
  Mathematics}, Vol.~\bibinfo {volume} {53}\ (\bibinfo  {publisher}
  {Springer},\ \bibinfo {address} {New York},\ \bibinfo {year}
  {2008})\BibitemShut {NoStop}%
\bibitem [{\citenamefont {Maier}\ and\ \citenamefont
  {Ankerhold}(2010)}]{Maier2010}%
  \BibitemOpen
  \bibfield  {author} {\bibinfo {author} {\bibfnamefont {S.~A.}\ \bibnamefont
  {Maier}}\ and\ \bibinfo {author} {\bibfnamefont {J.}~\bibnamefont
  {Ankerhold}},\ }\bibfield  {title} {\bibinfo {title} {\emph {Quantum
  Smoluchowski equation: A systematic study}},\ }\href {\doibase
  10.1103/PhysRevE.81.021107} {\bibfield  {journal} {\bibinfo  {journal} {Phys.
  Rev. E}\ }\textbf {\bibinfo {volume} {81}},\ \bibinfo {pages} {021107}
  (\bibinfo {year} {2010})}\BibitemShut {NoStop}%
\bibitem [{\citenamefont {Ankerhold}\ \emph {et~al.}(2001)\citenamefont
  {Ankerhold}, \citenamefont {Pechukas},\ and\ \citenamefont
  {Grabert}}]{Ankerhold1}%
  \BibitemOpen
  \bibfield  {author} {\bibinfo {author} {\bibfnamefont {J.}~\bibnamefont
  {Ankerhold}}, \bibinfo {author} {\bibfnamefont {P.}~\bibnamefont {Pechukas}},
  \ and\ \bibinfo {author} {\bibfnamefont {H.}~\bibnamefont {Grabert}},\
  }\bibfield  {title} {\bibinfo {title} {\emph {Strong Friction Limit in
  Quantum Mechanics: The Quantum Smoluchowski Equation}},\ }\href {\doibase
  10.1103/PhysRevLett.87.086802} {\bibfield  {journal} {\bibinfo  {journal}
  {Phys. Rev. Lett.}\ }\textbf {\bibinfo {volume} {87}},\ \bibinfo {pages}
  {086802} (\bibinfo {year} {2001})}\BibitemShut {NoStop}%
\bibitem [{\citenamefont {Ankerhold}\ and\ \citenamefont
  {Grabert}(2008)}]{Ankerhold2}%
  \BibitemOpen
  \bibfield  {author} {\bibinfo {author} {\bibfnamefont {J.}~\bibnamefont
  {Ankerhold}}\ and\ \bibinfo {author} {\bibfnamefont {H.}~\bibnamefont
  {Grabert}},\ }\bibfield  {title} {\bibinfo {title} {\emph {Erratum: Strong
  Friction Limit in Quantum Mechanics: The Quantum Smoluchowski Equation [Phys.
  Rev. Lett. 87, 086802 (2001)]}},\ }\href {\doibase
  10.1103/PhysRevLett.101.119903} {\bibfield  {journal} {\bibinfo  {journal}
  {Phys. Rev. Lett.}\ }\textbf {\bibinfo {volume} {101}},\ \bibinfo {pages}
  {119903} (\bibinfo {year} {2008})}\BibitemShut {NoStop}%
\bibitem [{\citenamefont {Zwanzig}(1961)}]{Zwanzig}%
  \BibitemOpen
  \bibfield  {author} {\bibinfo {author} {\bibfnamefont {R.}~\bibnamefont
  {Zwanzig}},\ }\bibfield  {title} {\bibinfo {title} {\emph {Memory Effects in
  Irreversible Thermodynamics}},\ }\href {\doibase 10.1103/PhysRev.124.983}
  {\bibfield  {journal} {\bibinfo  {journal} {Phys. Rev.}\ }\textbf {\bibinfo
  {volume} {124}},\ \bibinfo {pages} {983} (\bibinfo {year}
  {1961})}\BibitemShut {NoStop}%
\bibitem [{\citenamefont {Gao}(1997)}]{Gao1997}%
  \BibitemOpen
  \bibfield  {author} {\bibinfo {author} {\bibfnamefont {S.}~\bibnamefont
  {Gao}},\ }\bibfield  {title} {\bibinfo {title} {\emph {Dissipative Quantum
  Dynamics with a Lindblad Functional}},\ }\href {\doibase
  10.1103/PhysRevLett.79.3101} {\bibfield  {journal} {\bibinfo  {journal}
  {Phys. Rev. Lett.}\ }\textbf {\bibinfo {volume} {79}},\ \bibinfo {pages}
  {3101} (\bibinfo {year} {1997})}\BibitemShut {NoStop}%
\bibitem [{\citenamefont {Di\'osi}(1993)}]{Diosi1993}%
  \BibitemOpen
  \bibfield  {author} {\bibinfo {author} {\bibfnamefont {L.}~\bibnamefont
  {Di\'osi}},\ }\bibfield  {title} {\bibinfo {title} {\emph {On
  High-Temperature Markovian Equation for Quantum Brownian Motion}},\ }\href
  {http://stacks.iop.org/0295-5075/22/i=1/a=001} {\bibfield  {journal}
  {\bibinfo  {journal} {EPL (Europhysics Letters)}\ }\textbf {\bibinfo {volume}
  {22}},\ \bibinfo {pages} {1} (\bibinfo {year} {1993})}\BibitemShut {NoStop}%
\bibitem [{\citenamefont {Gradshteyn}\ and\ \citenamefont
  {Ryzhik}(2000)}]{Gradshteyn2000}%
  \BibitemOpen
  \bibfield  {author} {\bibinfo {author} {\bibfnamefont {I.~S.}\ \bibnamefont
  {Gradshteyn}}\ and\ \bibinfo {author} {\bibfnamefont {I.~M.}\ \bibnamefont
  {Ryzhik}},\ }\href@noop {} {\emph {\bibinfo {title} {Table of Integrals,
  Series, and Products}}},\ \bibinfo {edition} {sixth}\ ed.\ (\bibinfo
  {publisher} {Academic Press},\ \bibinfo {address} {San Diego},\ \bibinfo
  {year} {2000})\BibitemShut {NoStop}%
\bibitem [{\citenamefont {Knap}\ \emph {et~al.}(2012)\citenamefont {Knap},
  \citenamefont {Shashi}, \citenamefont {Nishida}, \citenamefont {Imambekov},
  \citenamefont {Abanin},\ and\ \citenamefont {Demler}}]{Demler-cata}%
  \BibitemOpen
  \bibfield  {author} {\bibinfo {author} {\bibfnamefont {M.}~\bibnamefont
  {Knap}}, \bibinfo {author} {\bibfnamefont {A.}~\bibnamefont {Shashi}},
  \bibinfo {author} {\bibfnamefont {Y.}~\bibnamefont {Nishida}}, \bibinfo
  {author} {\bibfnamefont {A.}~\bibnamefont {Imambekov}}, \bibinfo {author}
  {\bibfnamefont {D.~A.}\ \bibnamefont {Abanin}}, \ and\ \bibinfo {author}
  {\bibfnamefont {E.}~\bibnamefont {Demler}},\ }\bibfield  {title} {\bibinfo
  {title} {\emph {Time dependent impurity in ultracold fermions: orthogonality
  catastrophe and beyond}},\ }\href {\doibase 10.1103/PhysRevX.2.041020}
  {\bibfield  {journal} {\bibinfo  {journal} {Phys. Rev. X}\ }\textbf {\bibinfo
  {volume} {2}},\ \bibinfo {pages} {041020} (\bibinfo {year}
  {2012})}\BibitemShut {NoStop}%
\bibitem [{\citenamefont {Kim}\ and\ \citenamefont {Huse}(2012)}]{Kim2012}%
  \BibitemOpen
  \bibfield  {author} {\bibinfo {author} {\bibfnamefont {H.}~\bibnamefont
  {Kim}}\ and\ \bibinfo {author} {\bibfnamefont {D.~A.}\ \bibnamefont {Huse}},\
  }\bibfield  {title} {\bibinfo {title} {\emph {Superdiffusive nonequilibrium
  motion of an impurity in a Fermi sea}},\ }\href {\doibase
  10.1103/PhysRevA.85.043603} {\bibfield  {journal} {\bibinfo  {journal} {Phys.
  Rev. A}\ }\textbf {\bibinfo {volume} {85}},\ \bibinfo {pages} {043603}
  (\bibinfo {year} {2012})}\BibitemShut {NoStop}%
\bibitem [{\citenamefont {Pitaevskii}\ and\ \citenamefont
  {Stringari}(2003)}]{Pita-String}%
  \BibitemOpen
  \bibfield  {author} {\bibinfo {author} {\bibfnamefont {L.}~\bibnamefont
  {Pitaevskii}}\ and\ \bibinfo {author} {\bibfnamefont {S.}~\bibnamefont
  {Stringari}},\ }\href@noop {} {\emph {\bibinfo {title} {Bose-Einstein
  Condensation}}}\ (\bibinfo  {publisher} {Oxford University Press},\ \bibinfo
  {address} {Oxford},\ \bibinfo {year} {2003})\BibitemShut {NoStop}%
\bibitem [{\citenamefont {Lahaye}\ \emph {et~al.}(2009)\citenamefont {Lahaye},
  \citenamefont {Menotti}, \citenamefont {Santos}, \citenamefont {Lewenstein},\
  and\ \citenamefont {Pfau}}]{Lahaye2009}%
  \BibitemOpen
  \bibfield  {author} {\bibinfo {author} {\bibfnamefont {T.}~\bibnamefont
  {Lahaye}}, \bibinfo {author} {\bibfnamefont {C.}~\bibnamefont {Menotti}},
  \bibinfo {author} {\bibfnamefont {L.}~\bibnamefont {Santos}}, \bibinfo
  {author} {\bibfnamefont {M.}~\bibnamefont {Lewenstein}}, \ and\ \bibinfo
  {author} {\bibfnamefont {T.}~\bibnamefont {Pfau}},\ }\bibfield  {title}
  {\bibinfo {title} {\emph {The physics of dipolar bosonic quantum gases}},\
  }\href {http://stacks.iop.org/0034-4885/72/i=12/a=126401} {\bibfield
  {journal} {\bibinfo  {journal} {Reports on Progress in Physics}\ }\textbf
  {\bibinfo {volume} {72}},\ \bibinfo {pages} {126401} (\bibinfo {year}
  {2009})}\BibitemShut {NoStop}%
\bibitem [{\citenamefont {Giamarchi}(2004)}]{GiamarchiBook}%
  \BibitemOpen
  \bibfield  {author} {\bibinfo {author} {\bibfnamefont {T.}~\bibnamefont
  {Giamarchi}},\ }\href@noop {} {\emph {\bibinfo {title} {Quantum Physics in
  One Dimension}}}\ (\bibinfo  {publisher} {Oxford University Press, Oxford},\
  \bibinfo {year} {2004})\BibitemShut {NoStop}%
\bibitem [{\citenamefont {Gogolin}\ \emph {et~al.}(2004)\citenamefont
  {Gogolin}, \citenamefont {Nersesyan},\ and\ \citenamefont
  {Tsvelik}}]{GogolinBook}%
  \BibitemOpen
  \bibfield  {author} {\bibinfo {author} {\bibfnamefont {A.~O.}\ \bibnamefont
  {Gogolin}}, \bibinfo {author} {\bibfnamefont {A.~A.}\ \bibnamefont
  {Nersesyan}}, \ and\ \bibinfo {author} {\bibfnamefont {A.~M.}\ \bibnamefont
  {Tsvelik}},\ }\href@noop {} {\emph {\bibinfo {title} {Bosonization and
  Strongly Correlated Systems}}}\ (\bibinfo  {publisher} {Cambridge University
  Press, Cambridge},\ \bibinfo {year} {2004})\BibitemShut {NoStop}%
\bibitem [{\citenamefont {Mahan}(1993)}]{MahanBook}%
  \BibitemOpen
  \bibfield  {author} {\bibinfo {author} {\bibfnamefont {G.~D.}\ \bibnamefont
  {Mahan}},\ }\href@noop {} {\emph {\bibinfo {title} {Many-Particle Physics}}}\
  (\bibinfo  {publisher} {Plenum Press, New York},\ \bibinfo {year}
  {1993})\BibitemShut {NoStop}%
\end{thebibliography}%

\end{document}